\documentclass[12pt]{iopjournal}
\usepackage{amsmath,amssymb,amsfonts}
\usepackage{textcomp}
\usepackage{multirow}
\usepackage{url}
\usepackage{placeins}

\newcommand{\capm}[1]{{\mathversion{normal}$#1$}}
\makeatletter
\renewcommand{\@makecaption}[2]{\vskip\abovecaptionskip
\sbox\@tempboxa{\fontsize{10}{12}\selectfont {\bfseries #1.} #2}%
\ifdim \wd\@tempboxa >\hsize
\raggedright \fontsize{10}{12}\selectfont {\bfseries #1.} #2\par
\else
\global \@minipagefalse
\hb@xt@\hsize{\hfil\box\@tempboxa\hfil}%
\fi
\vskip\belowcaptionskip}
\makeatother
\begin{document}

\articletype{Paper}

\title{VNA-Only Calibration of Waveguide Thermistor Mounts with Self-Consistent Determination of the Thermistor-Port Impedance}

\author{Patricio Greco$^{1,*}$\orcid{0000-0002-4255-7558}}

\affil{$^1$Independent researcher, Buenos Aires, Argentina}

\affil{$^*$Author to whom any correspondence should be addressed.}

\email{patricio\_greco@me.com}


\begin{abstract}
A method is presented for determining the calibration factor and the
effective efficiency of waveguide thermistor mounts from
vector-network-analyzer reflection data alone, without a microcalorimeter,
extending the three reflection-coefficient method of the coaxial case. In
waveguide the thermistor is coupled to the line through a matching
structure and the reference impedance of the thermistor port, $Z_0$, is
not known a priori; it is inferred through a power-consistency
relation between pairs of states, under the convention
that it is real. The method reproduces the factory certificates ($U=2\%$)
of four K486A mounts within $1.0\%$ rms and the PTB-traceable values
($U=0.85$--$0.98\%$) of a fifth within $0.34\%$ rms; its expanded
measurement uncertainty is $0.5$--$1.16\%$ ($k=2$) for the best specimen
and below $2\%$ for the five. A mount with its waveguide-to-coaxial
adapter is calibrated directly, without characterizing the adapter, from
the model factor of the mount and the geometric factor measured through
the adapter. The scattering matrix implied by the solution is not
passive;
a minimum-norm correction of the thermistor impedances along the two
directions invisible to the calibration factor -- the ratios of the
efficiencies of the three states -- brings it to the floor of passivity,
where the efficiency depends on the load only through its mismatch
factor, leaves the result unchanged, and gives the sensitivity of the
efficiency to the bias resistance in closed form: below
$4\times10^{-5}~\Omega^{-1}$ in band mean, proportional to the mismatch
of the operating state and to the loss of the mount.
\end{abstract}

\section{Introduction}
\label{sec:intro}
Waveguide bolometric sensors are a working standard in metrology
laboratories worldwide and serve as highly stable transfer references. The
microcalorimetric method is the primary route for their calibration and
provides a direct link to the physical quantities involved; however, such
services are increasingly scarce and costly, and are available only at a
limited number of frequencies. Moreover, since these mounts are no longer
manufactured, each unit becomes more valuable within a laboratory, and the
possibility of an in situ calibration -- at lower cost and with
uncertainties adequate for use as a transfer standard -- is attractive.
National laboratories quote effective-efficiency uncertainties of the order
of $1\%$ or somewhat below ($k=2$) for these sensors when calibrated in
WR-42 waveguide; the
NIST service reports an absolute expanded uncertainty of $1.1\times10^{-2}$
over 18--26.5~GHz \cite{nist_cmc}, and comparable
figures are declared by other national metrology institutes \cite{kcdb}.

\begin{figure}[!t]
\centering
\includegraphics[width=0.56\textwidth]{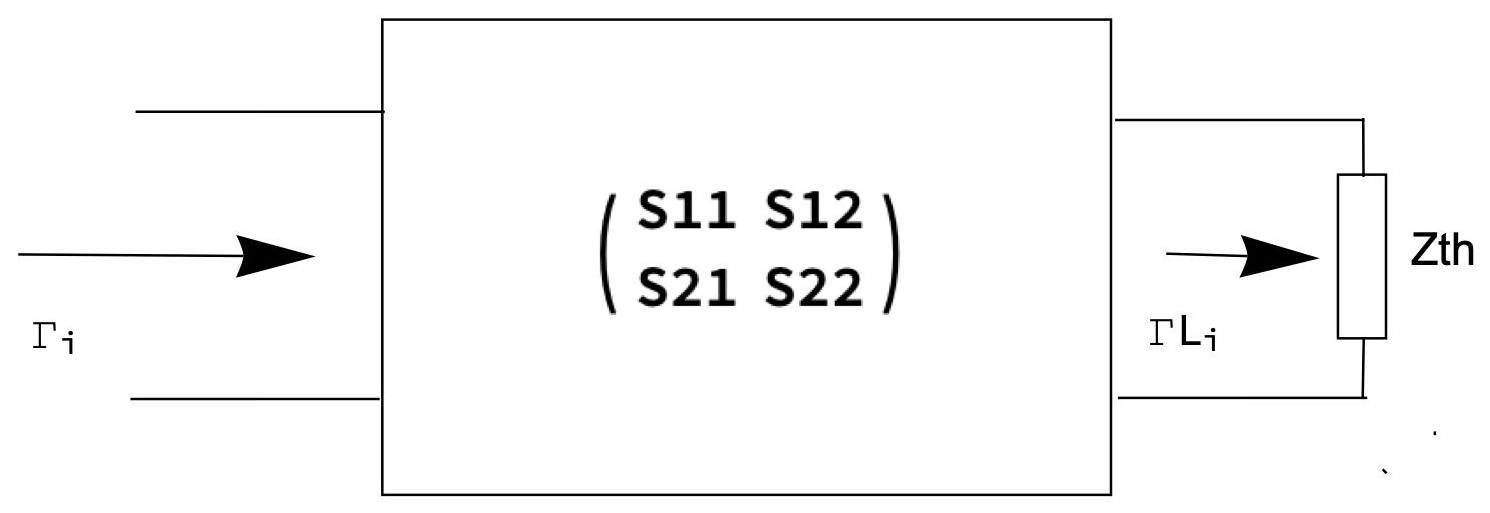}
\caption{Two-port representation of the sensor. \capm{\Gamma_i} is the reflection
coefficient measured at the sensor port; \capm{\Gamma_{Li}} is the reflection
coefficient at the thermistor terminals; $Z_{th,i}$ is the thermistor
impedance in state $i$. The subscript $i$ denotes the bridge resistance
setting $R_{b,i}$.}
\label{fig:twoport}
\end{figure}

The methodology developed here is the one introduced in
\cite{greco_coaxial}, the three reflection-coefficient method,
whose origin is the impedance method of Beatty and Reggia \cite{beatty1955};
we restate it briefly. The thermistor together with the sensor body can be
represented by the two-port of Fig.~\ref{fig:twoport}. It is essential to
note that the thermistor and the scattering matrix of the body cannot be
physically separated: whenever the thermistor is connected to any source,
an associated two-port $S$ is present.

On the waveguide side, the reference plane of this two-port is well defined:
it coincides with the plane of the waveguide flange. On the thermistor side
the port is not a cross-section of a uniform line but a surface enclosing
the thermistor, of the kind introduced by Kerns in his formulation of the
efficiency \cite{kerns1949}, whose shape varies with frequency. Its position
is characterized by an electrical delay obtained from the scattering
parameters (Section~\ref{sec:passivity}).

Here $Z_{th,i}$ is the thermistor impedance, described by the model of
Section~\ref{sec:circuit}; $\Gamma_{Li}$ is its reflection coefficient, and
$\Gamma_i$ that at the sensor input, which is the quantity measured with
the vector network analyzer (VNA). The subscript
$i$ refers to the resistance settings of the Wheatstone bridge in the
power meter, denoted $R_{b,i}$; each setting, with the thermistor held at
that resistance, is called a state of the sensor, and the method uses
three states, $i=1,2,3$. The scattering parameters correspond to the
mechanical structure of the sensor and are the same in the three states.
We adopt $i=2$, the operating state, as the value the manufacturer
specifies as optimal for sensor operation.

The aim of this work is twofold. First, to show that the three
reflection-coefficient method of \cite{greco_coaxial} can be applied to
waveguide mounts despite the reference impedance of the thermistor
port not being known a priori, and that both the calibration factor and
the effective efficiency can be obtained from VNA reflection data alone,
without a microcalorimeter. Second, to establish an uncertainty budget
adequate for the use of these mounts as transfer standards, and to show
that a mount together with its waveguide-to-coaxial adapter can be
calibrated directly as an assembly, without characterizing the adapter or
de-embedding it. The method is
applied to five K486A mounts: four reproduced within $1.0\%$ rms of their
manufacturer certificates ($U=2\%$, $k=2$; two of the reference points, one
each of specimens~2 and~3, are ratiometric comparisons made by the author,
Section~\ref{sec:results}), and a fifth reproduced within
$0.34\%$ rms of reference values traceable to PTB (Physikalisch-Technische Bundesanstalt) ($U=0.85$--$0.98\%$, $k=2$) -- the best
agreement of the set falling on the mount whose reference is traceable to
a primary standard. The expanded measurement uncertainty is
$0.5$--$1.16\%$ ($k=2$) over 18--26.5~GHz for the best specimen and below
$2\%$ for the five.

\section{Measurement Procedure}
\label{sec:measproc}
\begin{figure}[!t]
\centering
\includegraphics[width=0.56\textwidth]{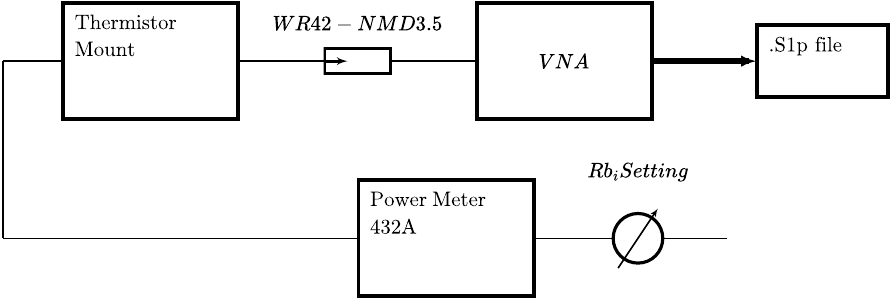}
\caption{Measurement arrangement. The reflection coefficient at the sensor
port is measured with the VNA through a waveguide-to-coaxial adapter
installed at the test port, while the power-meter bridge sets the state
$R_{b,i}$; the one-port calibration is performed in WR-42 at the mount
flange.}
\label{fig:setup}
\end{figure}

The input quantity of the method is the reflection coefficient $\Gamma_i$
of the mount, measured at its waveguide port in each of the three bridge
states $R_{b,i}$. No power, substituted or otherwise, is read: the
measurement is purely a one-port vector reflection measurement, and the
only role of the power-meter bridge is to set the thermistor to the three
operating resistances. The arrangement is shown in Fig.~\ref{fig:setup}
and is described in full below so that the procedure can be reproduced
without reference to other work. Five K486A mounts were measured in this
way; they are referred to below as specimens~1--5, and their data are
presented in Section~\ref{sec:results}.

The reflectometer is a vector network analyzer operated over
18--26.5~GHz, the full WR-42 band, on its internal frequency grid of 801
points at a step of 11.875~MHz; the sweep starts at 17~GHz, and the 716
points within 18--26.5~GHz are used (specimen~5 of Section~\ref{sec:results}
was measured on a 501-point sweep of the same band). Because the analyzer test ports are
coaxial, an HP~K281C waveguide-to-coaxial adapter is installed directly at
the measurement port, with no intervening cable, so that the measurement
reference is carried into WR-42 at the adapter output flange. Removing the
cable eliminates its flexure and phase drift from the connection and makes
the reference plane a mechanical flange to which the mount is repeatably
mated.

A one-port calibration is then performed in WR-42 at that flange. The
standards are a flush waveguide short, a set of offset shorts realized as
precision shims of 1/8, 1/4 and 3/8 wavelength placed ahead of the short, and
a matched waveguide termination measured directly and through a
quarter-wave shim (load and offset load); they are the components of the
Pasternack PEWCK1003 WR-42 kit \cite{pewck1003} -- flush shorts, offsets of
1/8, 1/4 and 3/8 wavelength (2.17, 4.35 and 6.52~mm) and a fixed
termination specified at VSWR~$\le1.03$ ($|\Gamma|\le0.015$) -- acquired
individually rather than as a kit. The offset shorts
provide the phase diversity of the reflection standards, and the offset load
separates the residual directivity from the reflection of the termination,
as a two-position sliding load does, so that the directivity is not limited
by the return loss of the termination. Waveguide standards are simple
pieces whose quality can be verified in the laboratory: the short is a flat
plate, and the shims were checked dimensionally, with the length-measuring
equipment of the author's workshop (resolution of about $10~\mu$m), and by
the phase shift they produce on the analyzer, as the manufacturer
recommends; the termination and the offset load were inspected in the same
way. The residual directivity, source match and reflection tracking of the
calibrated system are the quantities that enter the uncertainty budget of
Section~\ref{sec:budget}; they are properties of the calibrated
reflectometer at the mount flange, not of the analyzer alone. The datasheet
of the kit \cite{pewck1003} specifies the standards but not the residuals
of a calibration made with them; the figures used are those adopted for
this type of calibration (offset shorts, load and offset load), consistent
with the specified reflection of the termination, which the offset load
removes from the directivity, and they are what a commercial kit carries,
since a kit is delivered against the same specification and without a
measurement report unless one is requested.
The analyzer, an HP~8510C, is verified beforehand with the manufacturer's
performance test, which uses the verification kit of attenuators and
airlines, so that it is known to operate within its specification; it is
operated with its averaging factor at the maximum, $4096$.

With the reference plane established, the thermistor mount is connected at
the flange and its bias leads are connected to the Wheatstone bridge of the
power meter. The bridge is set in turn to the three resistances $R_{b,i}$,
and at each setting, once the thermistor has reached thermal equilibrium at
the new bias, the full-band reflection coefficient $\Gamma_i$ is recorded.
The stabilization follows \cite{greco_coaxial}: the meter is switched off
before the setting is changed, and about 30~s are allowed, six times the
settling time measured for the reflection coefficient to reach a level
indistinguishable from the noise of the analyzer.
The three states are measured without disturbing the connection, so a
single mechanical mating produces the three traces and any residual error
of the calibrated system is common to them. The values $R_{b,i}$ are a
property of the bridge and its leads, not of the mount, and are therefore
common to every specimen measured on the same instrument. The bridge of the
power meter provides $100$ and $200~\Omega$ from its front panel; the
$300~\Omega$ state was obtained by adding a precision resistor to the bridge
arm through a switch fitted to the instrument. After the modification the
three resulting bridge-arm resistances were measured with an 8.5-digit
multimeter at the instrument and again at the sensor end of the
interconnecting cable; since the cable lies in the thermistor arm of the
bridge, the resistance held by the thermistor at balance is the bridge-arm
value less the cable and connector resistance obtained from the difference
of the two readings. These are the values $R_{b,i}$ reported with
the parameters in Section~\ref{sec:results}, and the same
combination of
front-panel setting and added resistor is used for every state, since
$200~\Omega$ can be formed in two ways and only the measured combination is
valid.

To characterize the connection itself the mount is disconnected and
remated, and the three-state measurement is repeated; this is done for a
total of three independent connections, and the three resulting sets of
$\Gamma_i$ are averaged. Repeated disconnection and reconnection is the way
connection repeatability is evaluated in the EURAMET guidelines for vector
network analysers \cite{euramet_cg12}, whose principles, stated for coaxial
lines, extend to waveguide; the effect of an imperfect connection on the
calibration factor is bounded in Appendix~\ref{app:flange} from the
published repeatability of waveguide flange connections. The laboratory is held
at $23\,^{\circ}\mathrm{C}\pm2\,^{\circ}\mathrm{C}$; since the measured quantity is a
reflection coefficient and not a substituted power, the thermal drifts that
dominate bolometric \emph{power} measurement do not enter, and apart from
the short stabilization after each change of bridge setting the procedure
is an ordinary one-port vector measurement.

\section{The Method}
\label{sec:method}
\begin{figure}[!t]
\centering
\includegraphics[width=0.56\textwidth]{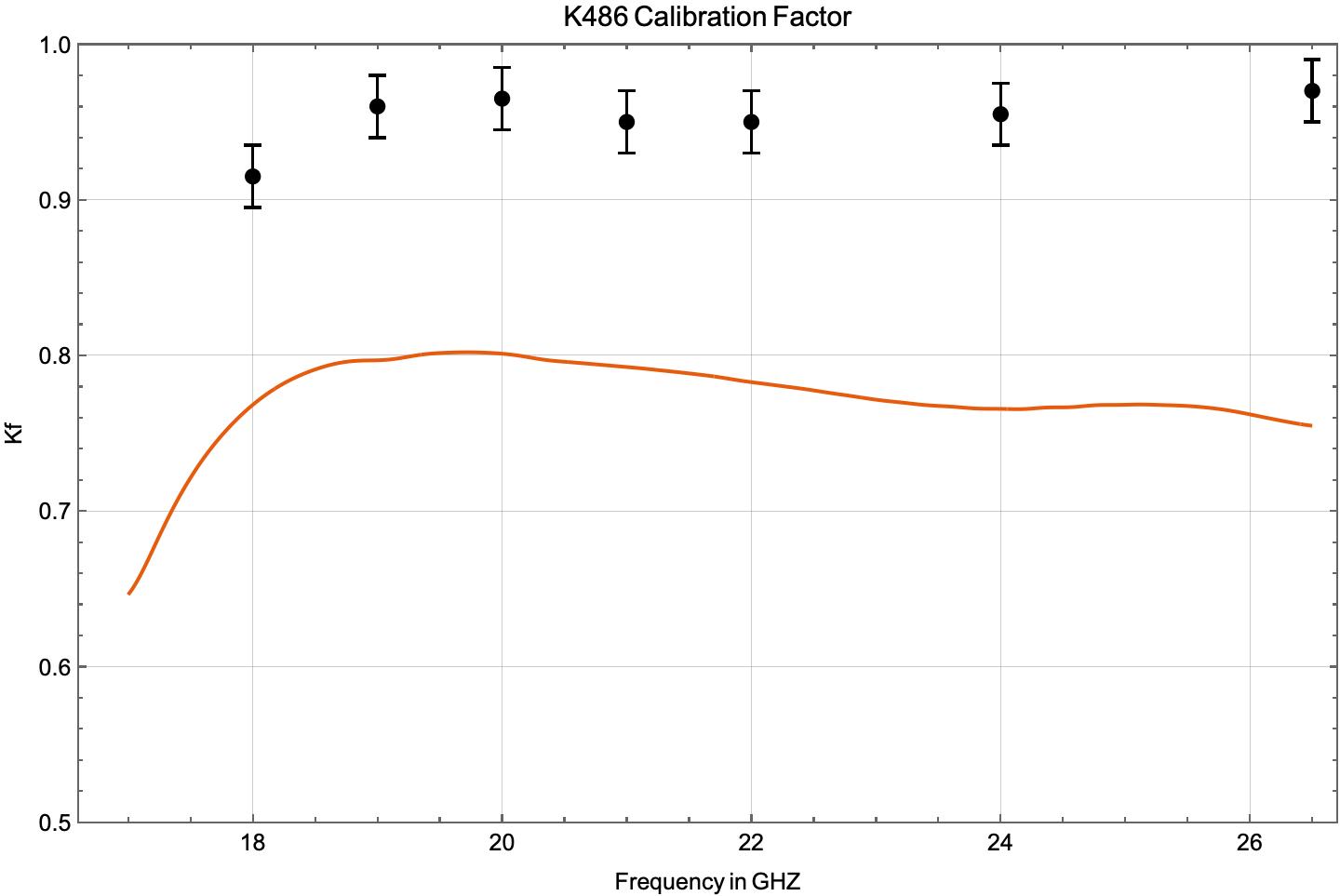}
\caption{Trivial case: $K_f$ obtained with $Z_{th,i}=R_{b,i}$ (solid line),
compared with the K486A certificate (points with $k=2$ bars). The departure
of 15--20\% in $K_f$ is incompatible with any metrological use.}
\label{fig:trivial}
\end{figure}

The input reflection coefficient of the two-port terminated in the
thermistor is
\begin{equation}
\Gamma_i = S_{11} + \frac{S_{21}^2\,\Gamma_{Li}}{1 - S_{22}\Gamma_{Li}} ,
\label{eq:gamma}
\end{equation}
where the two-port, a passive structure containing no anisotropic material,
is taken as reciprocal, $S_{12}=S_{21}$, each port being normalized to its
own real characteristic impedance, with reflection coefficients in the
pseudo-wave convention $(Z-Z_0)/(Z+Z_0)$ -- reciprocity holds for any real
normalization at each port, so the two impedances need not be equal, as
they were in the coaxial case \cite{greco_coaxial}; this is assumed throughout, and
$S_{21}^2$ stands for $S_{12}S_{21}$. The efficiency may be written
\begin{equation}
\eta_i = \frac{|S_{21}|^2\,(1-|\Gamma_{Li}|^2)}
{|1-S_{22}\Gamma_{Li}|^2 - |(S_{21}^2-S_{11}S_{22})\Gamma_{Li}+S_{11}|^2}.
\label{eq:eta}
\end{equation}
The calibration factor of state $i$ is $K_{f,i}=\eta_i\,(1-|\Gamma_i|^2)$,
the product of the efficiency and the mismatch factor at the sensor port;
$K_f$ without further index denotes that of the operating state $i=2$. The
efficiency determined from reflection data is the efficiency of the mount in
the sense of Kerns \cite{kerns1949}, the fraction of the power accepted at
the flange that is delivered to the thermistor element. Its identification
with the effective efficiency of a bolometric calibration requires thermal
equivalence between the RF power absorbed in the bead and the dc power
substituted \cite{keysight2001,weidman1981}. The classical literature
regards the intrinsic substitution error of a thermistor bead as small
\cite{rumfelt1967}; the error of the dual-element coaxial detector
\cite{engen1964,clague1993} does not apply to the configuration of one
detecting and one compensating thermistor of the 486A family
\cite{hp486a}. The comparisons of this work with calorimetric references
(Section~\ref{sec:results}) support the identification at the level of the
uncertainties involved, without determining the substitution error
separately. Throughout, the efficiency is called effective efficiency, as
is customary.
Working with \eqref{eq:gamma} and \eqref{eq:eta}, the efficiency can be
expressed in terms of the measured coefficients and of a factor $C_i$ that
depends on the thermistor-port coefficients only,
\begin{equation}
\eta_i = \frac{C_i}{1-|\Gamma_i|^2}
\left| \frac{(\Gamma_i-\Gamma_j)(\Gamma_i-\Gamma_k)}{\Gamma_j-\Gamma_k} \right|,
\label{eq:etameas}
\end{equation}
with $i,j,k \in \{1,2,3\}$ all different,
where the coefficients $C_i$ are defined as
\begin{equation}
C_i = (1-|\Gamma_{Li}|^2)
\left| \frac{(\Gamma_{Lj}-\Gamma_{Lk})}
{(\Gamma_{Li}-\Gamma_{Lj})(\Gamma_{Li}-\Gamma_{Lk})} \right|.
\label{eq:Ci}
\end{equation}
Replacing the reflection coefficients by their impedance expression, and
using
\begin{equation}
1-|\Gamma_{Li}|^2 = 4Z_0\,\frac{\mathrm{Re}[Z_{th,i}]}{|Z_{th,i}+Z_0|^2},
\label{eq:oneminusgamma}
\end{equation}
with $Z_0$ the reference impedance of the thermistor port (Section~\ref{sec:zoref}),
expression \eqref{eq:Ci} transforms into
\begin{equation}
C_i = 2\,\mathrm{Re}[Z_{th,i}]
\left| \frac{(Z_{th,j}-Z_{th,k})}
{(Z_{th,i}-Z_{th,j})(Z_{th,i}-Z_{th,k})} \right|,
\label{eq:CiZ}
\end{equation}
which is real and positive by construction and independent of the
characteristic impedance of the system. In an
initial computation the thermistor impedance is taken directly as the bridge
resistance of each state, $Z_{th,i}=R_{b,i}$; since \eqref{eq:CiZ} is
independent of the characteristic impedance, $Z_0$ plays no role here. We
call this the trivial case, as in the coaxial treatment \cite{greco_coaxial}.
It leads to a catastrophic error, shown in Fig.~\ref{fig:trivial}: the
departure from the certificate value is enormous and entirely incompatible
with any metrological application. The discrepancy is precisely what the
thermistor model is introduced to resolve.

\subsection{The two reference impedances}
\label{sec:zoref}
\begin{figure}[!t]
\centering
\includegraphics[width=0.62\textwidth]{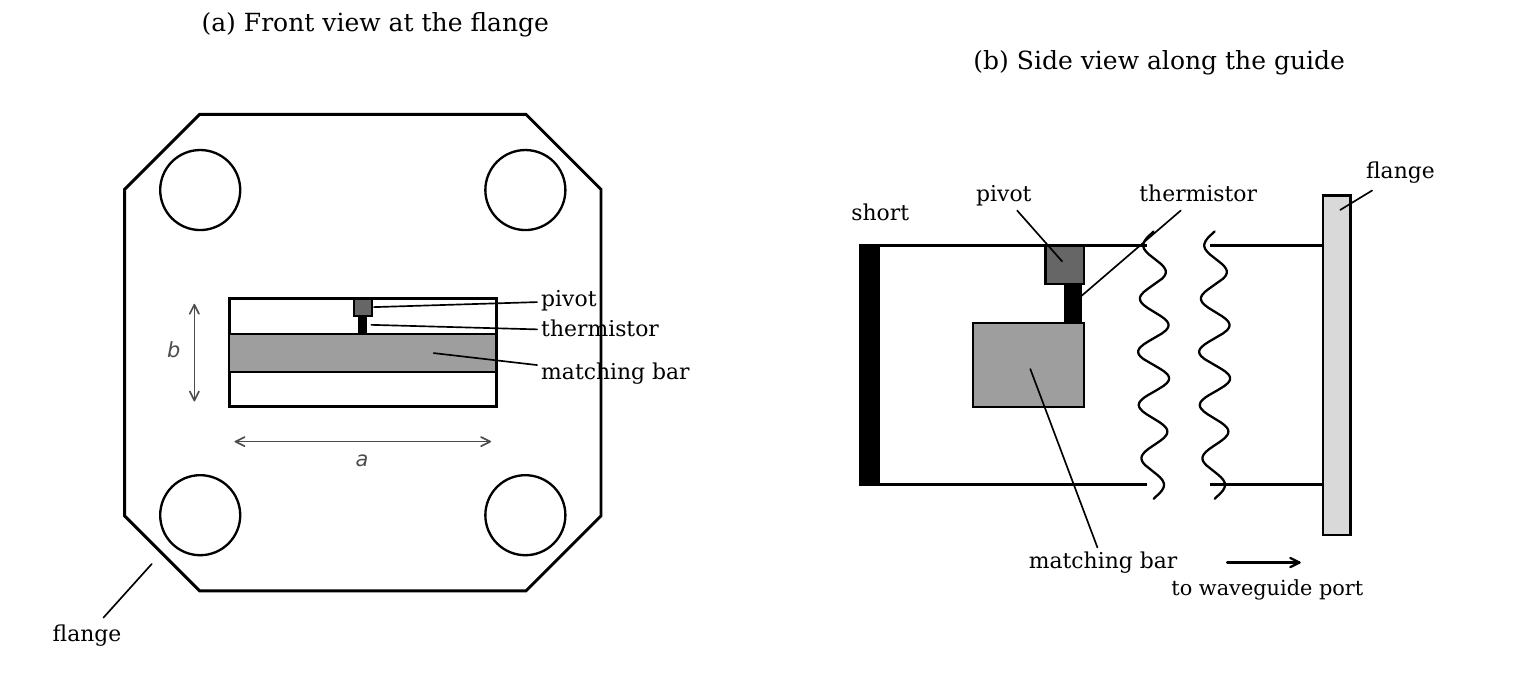}
\caption{Schematic of the thermistor mount in WR-42 (not to scale).
(a)~Front view at the flange: the matching bar spans the full width $a_g$ and
is centered in the height $b_g$; the inductive pivot hangs from the top wall at
the center of the width; the thermistor bridges the gap between the top face
of the bar and the pivot tip. (b)~Side view along the guide: the bar sits
near the short, and the flange defines the reference plane at which
\capm{\Gamma_i} is measured.}
\label{fig:mount}
\end{figure}

At this point the two-port represented by $S$ must be examined more
closely. Waveguide sensors require an impedance-matching structure between
the waveguide itself and the thermistor, which is essentially a two-terminal
lumped device. The matching is produced by a matching bar and an inductive
pivot. The bar bridges the full width of the guide along the dimension known
as $a_g$. The thermistor is mounted between the pivot -- which is capacitively
coupled to the sensor body -- and the center of the bar; the pivot carries
the bias connection to the bridge of the power meter and is itself part of
the matching system. The thermistor is connected to the
points just described by two fine wires. Figure~\ref{fig:mount} shows a
schematic of the mount. We denote by $Z_{0}^{(g)}$ the characteristic
impedance at the sensor (waveguide) port and by $Z_0$, without superscript,
the reference impedance of the thermistor port, a real quantity whose
meaning is made precise in Section~\ref{sec:convention}; $Z_0$ is the
impedance used throughout the rest of this work.

It is therefore evident that the characteristic impedance in the two-port
$S$ is not the same at both ports. It can be verified from
\eqref{eq:etameas} and \eqref{eq:CiZ} that it is not necessary to compute these impedances to obtain the results:
the efficiency, and hence the calibration factor, is independent of them.
While the impedance at the thermistor port is thus not required for the direct
evaluation of $\eta_i$ and $K_f$, it is needed to build the functional of Section~\ref{sec:determination}
from which the sensor parameters are obtained -- specifically for the
relative-curvature condition of Section~\ref{sec:zo}, which is evaluated at
the thermistor port --
and its self-consistent determination is the subject of
Section~\ref{sec:zo}.

\subsection{Neutral transformation of the thermistor impedance}
\label{sec:neutral}
A condition that is particularly important, and that will be examined
towards the end of this work, can be stated now. The thermistor
impedances may be transformed according to
\begin{equation}
Z_{th,k}' = a\,Z_{th,k} + j\,b, \qquad a,b\in\mathbb{R},
\label{eq:neutral}
\end{equation}
with $a$ and $b$ common to the three states. A common real scale factor $a$
cancels in the ratio of differences in \eqref{eq:CiZ}, and a common
imaginary additive $j\,b$ cancels in the differences $(Z_{th,i}-Z_{th,j})$
while leaving $\mathrm{Re}[Z_{th,i}]$ unchanged; both therefore leave $C_i$
-- and hence $\eta_i$ -- invariant. A real additive, by contrast, shifts
$\mathrm{Re}[Z_{th,i}]$, which enters \eqref{eq:CiZ} directly, and is thus
not a free direction: it changes the result. Within the class of thermistor models admitted here,
the method therefore determines the thermistor impedances up to the common
scale ($a>0$) and the common reactance of \eqref{eq:neutral}, with no real
additive freedom. This condition also applies to the
coaxial case, where it was of no consequence; here, the change of
characteristic impedance across the two-port leads us to add further
information to the model in order to determine the thermistor impedances
through corrections consistent with \eqref{eq:neutral}
(Section~\ref{sec:scalingA}).

\section{Equivalent Circuit}
\label{sec:circuit}
\begin{figure}[!t]
\centering
\includegraphics[width=0.53\textwidth]{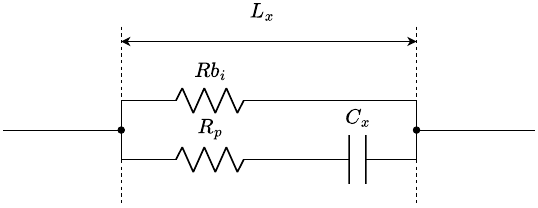}
\caption{Equivalent circuit of the thermistor seen from its port: the
bolometric resistance $R_{b,i}$ in parallel with the series branch
$R_p$--$C_x$. The skin impedance of \eqref{eq:skin} applies to each resistor,
with $L_x$ the length of the thermistor body.}
\label{fig:circuit}
\end{figure}

As follows from Fig.~\ref{fig:trivial}, where the calibration factor
computed with $Z_{th,i}=R_{b,i}$ falls $15$--$20\%$ below the certificate,
a mechanism must exist that delivers
additional heat to the thermistor and thereby raises its efficiency inside
the guide, consistent with \eqref{eq:eta}. The resistance $R_{b,i}$ is the
bolometric resistance; the pair $C_x$, $R_p$ of Fig.~\ref{fig:circuit} represents the dielectric
losses of the thermistor, which are essentially responsible for that extra
heating, and $L_x$ is the length of the body of the thermistor bead. The
thermistor-impedance equations include the skin effect \cite{ramo1994},
\begin{equation}
Z_{\mathrm{skin}}(R_{dc}) = \frac{1-j}{2}\,
\sqrt{R_{dc}\,L_x\mu_0 f}\;
\frac{J_0\!\left((1-j)\sqrt{L_x\mu_0 f/R_{dc}}\right)}
     {J_1\!\left((1-j)\sqrt{L_x\mu_0 f/R_{dc}}\right)},
\label{eq:skin}
\end{equation}
where $f$ is the frequency, $R_{dc}$ the direct-current resistance, and
$J_0$, $J_1$ the Bessel functions of the first kind of orders zero and one.
The quantity $L_x = L\,\mu_r$ has units of length, $L$ being the physical
length of the conductor; we take $\mu_r = 1$, so that $L_x = L$. For the
values involved \eqref{eq:skin} is well represented by its series expansion,
\begin{equation}
Z_{\mathrm{skin}}(R_{dc}) \simeq R_{dc}
+ \frac{L_x^2\mu_0^2 f^2}{48\,R_{dc}}
+ j\,\frac{L_x\mu_0 f}{4}.
\label{eq:skinseries}
\end{equation}
The accuracy of the expansion was quantified in \cite{greco_coaxial}
(Appendix~C) for an argument of the Bessel functions,
$|u|=\sqrt{2}\sqrt{L_x\mu_0 f/R_{dc}}$, reaching 1.14; here, with
$L_x=0.32$~mm, $|u|$ does not exceed 0.46 over the band and the truncation
error is correspondingly smaller. With
\begin{equation}
z_{b,i} = Z_{\mathrm{skin}}(R_{b,i}),\qquad
z_p = \frac{-j}{\omega C_x} + Z_{\mathrm{skin}}(R_p),
\label{eq:zbzp}
\end{equation}
the thermistor impedance in each state is
\begin{equation}
Z_{th,i} = \frac{z_{b,i}\,z_p}{z_{b,i}+z_p},
\label{eq:zth}
\end{equation}
that is, the bolometric branch $R_{b,i}$ in parallel with the loss branch
$C_x$--$R_p$, as shown in Fig.~\ref{fig:circuit}. The branch $C_x$--$R_p$
is a dielectric loss: the reactance of $C_x$ is comparable to $R_p$ over the
band, and the branch carries the frequency dependence of \eqref{eq:rpf}
below.

\begin{figure}[!t]
\centering
\includegraphics[width=0.47\textwidth]{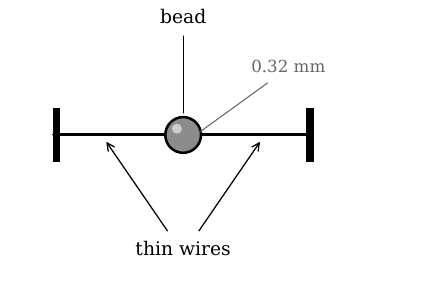}
\caption{Structure of the thermistor: a vitrified sintered bead,
$0.32$~mm in diameter, supported by two fine wires of $25~\mu$m diameter.}
\label{fig:thermistor}
\end{figure}

As in \cite{greco_coaxial}, the loss resistance $R_p$ in general decreases
with frequency; here it is written as
\begin{equation}
R_p = R_{p0}\left(1 - \kappa\,\sqrt{\frac{f}{f_{\max}}}\right),
\label{eq:rpf}
\end{equation}
where $R_{p0}$ is the initial loss resistance, $f_{\max}$ the upper edge of
the band in use (26.5~GHz for WR-42, 40~GHz for WR-28), and $\kappa\ge0$ a constant, so that $R_p$ does not
increase with frequency over the band. The law
\eqref{eq:rpf} is empirical. It is suggested by the structure of the
thermistor material, a vitrified sintered composite of diverse fragments
(Fig.~\ref{fig:thermistor}): the fragments do not share a single relaxation
time constant, and a distribution of time constants produces a dispersion
whose frequency dependence is of the $\omega^{-1/2}$ type of the Warburg
impedance \cite{warburg1899,barsoukov2018}. The square root in
\eqref{eq:rpf} takes that dependence for the real loss resistance; in the
present band the five specimens follow it, and the
linear law used for the coaxial mount \cite{greco_coaxial} is the simpler
approximation of the same behavior. The analogy motivates the form of
\eqref{eq:rpf}; the exponent is adopted on that basis, not derived from a
transport model of the bead, and it is not a fitted quantity.
Over a single waveguide band the laws with exponents $1/2$ and $1$ differ
by less than $0.3\%$ rms in $R_p(f)$ over 18--26.5~GHz once $R_{p0}$ and
$\kappa$ are re-determined by least squares to the other law, so a free
exponent would be degenerate with $\kappa$; and
every parameter added without a condition of its own degrades the
conditioning of the others (Appendix~\ref{app:ident}). The parameter set is
therefore kept to the elements that the three-state measurement can
determine, and the exponent is fixed at $1/2$ for consistency with the
dispersive behavior of the material.

\begin{figure}[!t]
\centering
\includegraphics[width=0.62\textwidth]{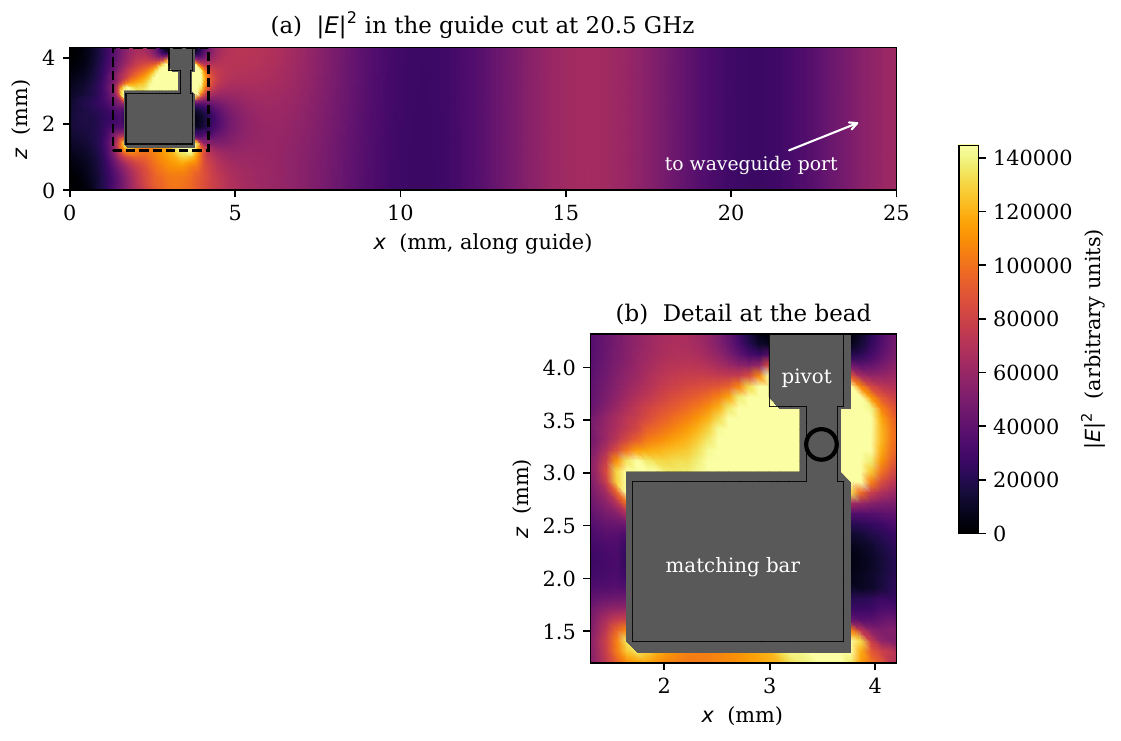}
\caption{Squared electric-field magnitude $|E|^2$ in the longitudinal cut of
the mount at 20.5~GHz, from the finite-element (FEM) model, proportional to the electric
energy density. (a)~Full guide cut; the energy is confined to the thermistor
region while a low-level traveling wave fills the guide towards the waveguide
port. (b)~Detail at the bead; the black circle marks the thermistor bead and
the metal parts (matching bar, pivot) are shown in gray.}
\label{fig:efield}
\end{figure}

It is worth pausing here on two points. The first is the mechanism by which
power flows into the body of the thermistor. The element sits in a region of
intense electric field produced by the design of the mount,
illustrated in Fig.~\ref{fig:efield}; in this field the thermistor material
absorbs power by induction. The supporting wires act as a boundary, but the
mechanism differs from that of the coaxial sensor, and for this reason a
reader comparing the two will not find here the equivalent circuit of the
leads. The topology is otherwise that validated in \cite{greco_coaxial}
against calorimetric references, and it obeys the rule stated there: a
reactance common to the three states and in series with the thermistor
cancels in the differences of \eqref{eq:CiZ}; a series resistance survives
only through $\mathrm{Re}\,Z_{th,2}$, as a heat source
(Section~\ref{sec:budget}); and the elements that leave a trace in the
relation between the states, and can therefore be determined by the method,
are those in parallel with $R_{b,i}$ or state dependent through it -- the
dielectric branch, and the skin impedance of each branch. The second point is the structure of the thermistor itself,
Fig.~\ref{fig:thermistor}: a vitrified sintered sphere $0.32$~mm in diameter
with two fine wires $25~\mu$m in diameter. The dimensions of the sensor and
of the thermistor are accessible by direct measurement; the bead size fixes
$L_x = 0.32$~mm $\pm\,10~\mu$m, a value which, as shown in
Section~\ref{sec:budget}, is not critical in the uncertainty analysis. In
\cite{greco_coaxial} $L_x$ was left free, as an effective electrical length
of the coaxial lead structure, and was the least-determined direction of the
fit, consistent with the insensitivity found here; in the waveguide mount
the bead geometry fixes it.

Since the sintered material is a composite of diverse fragments, there is a
dependence of the thermistor impedance on frequency, because the heating of
those fragments is produced directly by the electric field incident on the
material -- as if further thermistor elements were brought into conduction
at microwave frequencies. The thermistor impedance as a whole -- not only
the loss branch -- is therefore expected to be lower at microwave frequencies
than the dc resistance set by the bridge. The idea is old: the equivalent
circuit proposed for the bead in the Radiation Laboratory treatment of
thermistor mounts \cite{montgomery1947} places a second thermistor
resistance in parallel with the first through a capacitance whose reactance
is negligible at microwave frequencies, attributed to the finite particle
size of the sintered oxides and to the capacitive gaps between imperfectly
contacting particles: at dc only the first path conducts, at microwave
frequencies both do, and the microwave resistance is lower than the dc
value. That element is not the branch $C_x$--$R_p$ of
Fig.~\ref{fig:circuit}, whose capacitance is not a short at these
frequencies and which represents a loss; it is the mechanism behind the
scale of the whole impedance, which the present model does not represent
by an element because it does not need to. The determination of $K_f$ and
$\eta$ is immune to this, since the method works with $Z_{th,i}$ up to the
common scale of \eqref{eq:neutral} and $Z_0$ is determined in that same
scale (Section~\ref{sec:zo}); the expectation becomes observable only when
the scale is fixed from outside, which is what the comparison with the
finite-element impedance of the mount does in Section~\ref{sec:scalingA},
where the scaled bridge resistances come out below their dc values in every
specimen.

The parameters $C_x$, $R_{p0}$ and $\kappa$ are taken common to the three
states, although the bead operates at a different temperature in each of
them: the bridge sets the bolometric resistance precisely through that
temperature. A state dependence of the thermistor impedance beyond $R_b$,
modeled for coaxial mounts in \cite{kazemipour2010,ziade2011}, is not
represented by an element of the circuit; the skin impedance of each
branch is state dependent through $R_{b,i}$, and the part of any further
dependence that the three-state measurement requires is determined in
Section~\ref{sec:passivity}, along the directions that the calibration
factor does not see (Sections~\ref{sec:neutral} and~\ref{sec:invariant}).

The problem that follows is to determine the parameter set
\begin{equation}
X = \{C_x,\,R_{p0},\,\kappa\}
\label{eq:paramset}
\end{equation}
that characterizes the sensor. A parameter search against the known
calibration data of a given specimen could be of interest, but it is not the
object of this work; the aim is to obtain these parameters directly from the
measured set of reflection coefficients,
\begin{equation}
\{\Gamma_i\} \;\Rightarrow\; X \;\Rightarrow\; \{\eta_i,\,K_{f,i}\}.
\label{eq:pipeline}
\end{equation}
For this, a set of properties is required that will form a functional to be
minimized, as shown in what follows.

\section{Determination of $X$ from the Reflection Coefficients}
\label{sec:determination}
\begin{figure}[!t]
\centering
\includegraphics{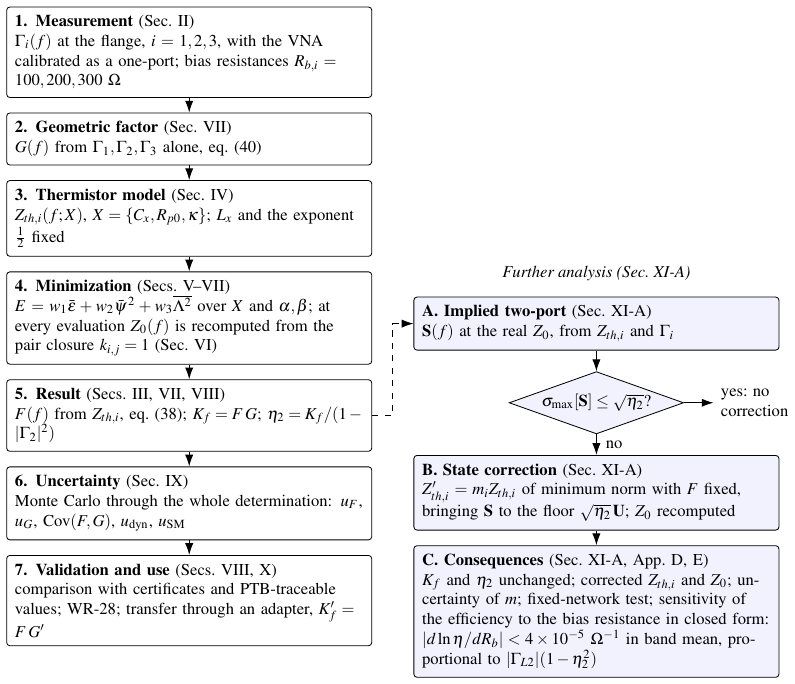}
\caption{Computation flow of the method: the main path (left) and the
further analysis of Section~\ref{sec:passivity} (right), with the
sections and appendices in which each step is developed.}
\label{fig:flow}
\end{figure}
Fig.~\ref{fig:flow} summarizes the steps that follow and the sections in
which each is developed. The parameter set $X$ is obtained by minimizing a functional built from
several properties of the three-state measurement. The bridge resistances
$R_{b,i}$ are not arbitrary; they must satisfy the symmetry condition
\begin{equation}
R_{b,i}=\{R_0-\delta,\;R_0,\;R_0+\delta\},\qquad \delta\le R_0/2 .
\label{eq:rbsym}
\end{equation}
Their extreme values are bounded by two considerations. Small resistances
imply a larger power dissipated in the thermistor, limited by the thermistor
specifications, which set the admissible operating range. Large resistances
leave a small bias power in the thermistor, comparable to the power to be
measured, which compromises the dynamic range of the measurement.
The symmetry of \eqref{eq:rbsym} is what makes the combination
$\Gamma_1+\Gamma_3-2\Gamma_2$ of the curvature condition of
Section~\ref{sec:zo} a second difference: unequal steps bias $\psi$, the
curvature residual defined there, even when the model is evaluated at the
resistances actually held, so the steps are kept equal to the tolerance
with which the bridge resistances are known. The calibration factor itself is unaffected by a departure
from symmetry provided the model uses the resistances actually held by the
bridge \cite{greco_coaxial} (Appendix~A); that result carries over, since
the asymmetry enters $F$ only through $\Gamma_i$ and $Z_{th,i}(R_{b,i})$.
The set used here, $\{100.415, 200.366, 300.345\}~\Omega$
(Section~\ref{sec:results}), has steps of $99.951$ and $99.979~\Omega$, a
difference of $3\times10^{-4}$, within that tolerance.

The first property is the proportionality of the efficiency ratios within the
operating band, established in \cite{greco_coaxial}. The efficiencies of the
three states need not be equal to one another -- each state $R_{b,i}$
presents a different load to the matching structure -- but their ratios
are taken as constant on average over the band, as in the coaxial case.
The ratios can be written
with two auxiliary variables $\alpha,\beta$ as
\begin{equation}
\frac{\eta_2}{\eta_1}=\frac{2-\alpha}{\beta},\quad
\frac{\eta_2}{\eta_3}=\frac{\alpha}{\beta},\quad
\frac{\eta_3}{\eta_1}=\frac{2-\alpha}{\alpha}.
\label{eq:etaratios}
\end{equation}
with $\alpha=\beta=1$ corresponding to three equal efficiencies.
It is this difference between the three efficiencies, and the constancy of
their ratios across the band, that carries the information, whatever
their value; Section~\ref{sec:passivity} shows that the value is unity
once the matching structure is taken at the floor of passivity, at which
the efficiencies of the three states coincide up to their mismatch
factors. From \eqref{eq:etaratios}, each ratio gives one
relation among the efficiencies at a given frequency; collecting the two
independent ones as squared residuals, the quantity to be minimized at
each frequency is
\begin{equation}
\epsilon(\alpha,\beta)=\big(\eta_2\beta-\eta_1(2-\alpha)\big)^2
+\big(\eta_2\beta-\eta_3\alpha\big)^2 ,
\label{eq:eps}
\end{equation}
the third relation of \eqref{eq:etaratios} following from the other two
by transitivity. Since the
property holds on average over the useful band of the sensor, with $N_f$
measured frequencies the first term of the functional is the band average
\begin{equation}
E(\alpha,\beta)=\frac{1}{N_f}\sum_{n=1}^{N_f}\epsilon_n(\alpha,\beta),
\label{eq:Eeps}
\end{equation}
with $\epsilon_n$ the value of \eqref{eq:eps} at the $n$th frequency.
This condition alone does not determine $X$; a second property is required.
It is the relative curvature of the reflection coefficient, also established
in \cite{greco_coaxial}. Its evaluation at the thermistor port requires the
reference impedance $Z_0$, which is not known a priori in waveguide;
this is the subject of the next section, after which the complete functional
is assembled.

\section{Self-Consistent Determination of the Thermistor-Port Impedance}
\label{sec:zo}
The relative-curvature property established in \cite{greco_coaxial}, for the
measured reflection coefficients,
\begin{equation}
\mathrm{Re}[C_v] = \mathrm{Re}\!\left[\frac{\Delta^2\Gamma}{\Gamma_2}\right]
= \mathrm{Re}\!\left[\frac{\Gamma_1+\Gamma_3-2\Gamma_2}{\Gamma_2}\right],
\label{eq:cv}
\end{equation}
and its counterpart at the thermistor port,
\begin{equation}
\mathrm{Re}[C_{vL}] = \mathrm{Re}\!\left[\frac{\Delta^2\Gamma_L}{\Gamma_{L2}}\right],
\qquad \psi = \mathrm{Re}[C_v - C_{vL}],
\label{eq:cvl}
\end{equation}
completes the functional to be minimized. The property states that the
real part of the relative curvature is carried from the thermistor port to
the sensor input, so that the band mean of $\psi$ vanishes; it holds
exactly when the matching structure acts on the three states as a phase
rotation (Section~\ref{sec:convention}). Its evaluation, however, requires
$\Gamma_{Li}$ and therefore the reference impedance $Z_0$ of the
thermistor port. In the coaxial case \cite{greco_coaxial} this impedance was
homogeneous along the sensor; in waveguide it is not, and $Z_0$ at the
thermistor port is not known a priori.

It might be supposed that $Z_0$ could be obtained by FEM simulation. This
procedure is not metrologically valid: it yields a value that does not
represent the specimen under study, since small mechanical differences
produce variations in every parameter of the sensor, including the
reference impedance that the method requires to evaluate the relative
curvature.

As established in Section~\ref{sec:neutral}, the method works with values
that are exact up to the common transformation \eqref{eq:neutral}. The same
must hold for $Z_0$; only its real part is retained, $Z_0\in\mathbb{R}$,
since a reactive part of the reference is indistinguishable, at fixed $X$,
from the common reactance $b$ of \eqref{eq:neutral}; the curvature
condition, which fixes $X$, is not invariant to it
(Section~\ref{sec:scalingA}). The
reflection coefficient at the thermistor port is
\begin{equation}
\Gamma_{Li}(Z_0) = \frac{Z_{th,i}-Z_0}{Z_{th,i}+Z_0}.
\label{eq:gammaLzo}
\end{equation}
The cross-relation between the power accepted at the sensor input and the
power dissipated in the thermistor is expected to take similar values in the
three states; accordingly we impose, for each pair $(i,j)$ with $i\neq j$,
\begin{equation}
\left(\frac{1-|\Gamma_i|^2}{1-|\Gamma_j|^2}\right)
\left(\frac{1-|\Gamma_{Lj}(Z_{0,ij})|^2}{1-|\Gamma_{Li}(Z_{0,ij})|^2}\right)
= k_{i,j},\qquad k_{i,j}\approx 1.
\label{eq:zocond}
\end{equation}
where $Z_{0,ij}$, the pair impedance, is the value of the thermistor-port
impedance that satisfies the relation for the pair $(i,j)$. The left side
compares, between the two states, the mismatch factor $1-|\Gamma|^2$ at
the sensor input with that at the thermistor port; its exact content is
given by \eqref{eq:kidentity} below, and the reference case in which
$k_{i,j}=1$ holds identically is described in Section~\ref{sec:convention}.
The calibration factor is invariant under a common scaling of $Z_0$ and
$Z_{th,i}$ (Section~\ref{sec:scalingA} and Appendix~\ref{app:sens}), so the
scale of $Z_0$ is an internal quantity of the algorithm. With three states,
two independent combinations exist, and for convenience we take the pairs
$(1,2)$ and $(3,2)$.

We set $k_{i,j}=1$. This is a convention: it defines the real reference
impedance $Z_0$ as the one at which the power-consistency relation
\eqref{eq:zocond} closes exactly for each pair, and it reduces the
condition to a closed quadratic for $Z_{0,ij}$. What the convention leaves
undetermined is examined in Section~\ref{sec:scalingA} and
Appendix~\ref{app:sens}, and the calibration factor does not depend on it
except through the curvature condition.
Writing $Z_{th,i}=r_i + j\,x_i$ and
\begin{equation}
p_{i,2}=\frac{1-|\Gamma_i|^2}{1-|\Gamma_2|^2},\qquad
\phi_i = p_{i,2}\,\frac{r_2}{r_i},
\label{eq:phi}
\end{equation}
the condition \eqref{eq:zocond} becomes
\begin{equation}
Z_{0,i2}^2
+ 2\,Z_{0,i2}\left(\frac{r_i\phi_i - r_2}{\phi_i-1}\right)
+ \left(\frac{(r_i^2+x_i^2)\phi_i-(r_2^2+x_2^2)}{\phi_i-1}\right)=0,
\label{eq:zoquad}
\end{equation}
whose physical root satisfies
\begin{equation}
Z_{0,i2}\in\mathbb{R},\quad Z_{0,i2}>0,\quad
Z_{-}\le Z_{0,i2}\le Z_{+},
\label{eq:zoroot}
\end{equation}
with
\begin{equation}
\begin{gathered}
Z_{\pm}=\Big(\nu_c\pm\sqrt{\nu_c^2-1}\Big)\,r_2,\\
\nu_c=\frac{1+K_c^2|\Gamma_2|^2}{1-K_c^2|\Gamma_2|^2},\qquad K_c=2.
\end{gathered}
\label{eq:zobound}
\end{equation}
The band \eqref{eq:zobound} is the set of real $Z_0$ for which a purely
resistive load equal to the resistive part of the central-state thermistor
impedance, $r_2=\mathrm{Re}\,Z_{th,2}$, which includes the effect of $C_x$
and $L_x$ at each frequency, would be matched to $Z_0$ no worse than
$K_c$ times the mismatch of the sensor to the guide,
$|\Gamma_{L2}|\le K_c|\Gamma_2|$: solving
$|r_2-Z_0|/(r_2+Z_0)\le K_c|\Gamma_2|$ for $Z_0$ gives
$Z_{\pm}=r_2\,(1\pm K_c|\Gamma_2|)/(1\mp K_c|\Gamma_2|)$, which is
\eqref{eq:zobound}. The factor $K_c$ is a coverage factor: the resistive
test ignores the reactance of the load, which makes the actual mismatch at
the thermistor port larger than that of a resistive load, and $K_c=2$
admits every solution with a margin. It is a criterion of admissibility,
not a constraint on the solution: the two roots of \eqref{eq:zoquad}
differ markedly, the spurious one falling well below the lower limit
$Z_-$, and the test \eqref{eq:zoroot} selects the physical root. The
thermistor-port impedance used in the functional is
$Z_0 = (Z_{0,12}+Z_{0,32})/2$.

\begin{figure}[!t]
\centering
\includegraphics[width=0.62\textwidth]{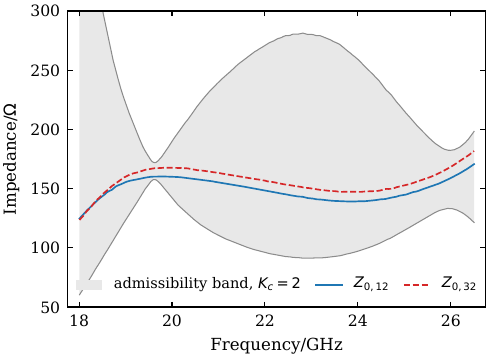}
\caption{Pair impedances $Z_{0,12}(f)$ (solid, lower curve) and
$Z_{0,32}(f)$ (dashed, upper curve) of specimen~4 at the converged solution,
and the admissibility band \eqref{eq:zobound} with $K_c=2$ (shaded). Both
lie within the band over the full range; the band narrows where the central
state is matched (near 19.6~GHz) and widens toward the band edges.}
\label{fig:zoband}
\end{figure}

Figure~\ref{fig:zoband} shows the result for specimen~4: the two pair
impedances $Z_{0,12}(f)$ and $Z_{0,32}(f)$ and the admissibility band
\eqref{eq:zobound} within which the physical root must lie. Both lie inside
the band across the whole range, as do those of the other four specimens;
$Z_{0,32}$ exceeds $Z_{0,12}$ by $3$ to $10\%$ on average, with the same sign in the
five mounts, and this difference is quantified in Section~\ref{sec:results}.
The band narrows sharply near 19.6~GHz, where the
central state is matched and $|\Gamma_2|$ passes through its minimum, so
that $\nu_c\to1$ and the half-width $\sqrt{\nu_c^2-1}\to0$ in
\eqref{eq:zobound}; there $Z_0$ is strongly constrained. At the band edges,
where $|\Gamma_2|$ is largest, the band is wider.

The factor $k_{i,j}$ introduced in \eqref{eq:zocond} is not arbitrary. By
the Schwarz--Pick identity of Appendix~\ref{app:schwarzpick}, the
efficiency of each state is
$\eta_i=|T'(\Gamma_{Li})|\,(1-|\Gamma_{Li}|^2)/(1-|\Gamma_i|^2)$, with
$T'$ the derivative \eqref{eq:Tprime} of the bilinear map $T$ of
\eqref{eq:gamma}, $\Gamma_i=T(\Gamma_{Li})$, at the load point;
the ratio that \eqref{eq:zocond} forces to unity can therefore be written
in closed form as
\begin{equation}
k_{i,j} = \frac{\eta_j}{\eta_i}\,
\frac{|T'(\Gamma_{Li})|}{|T'(\Gamma_{Lj})|},
\label{eq:kidentity}
\end{equation}
that is, the ratio of the efficiencies of the two states multiplied by the
inverse ratio of the local magnifications of the map at the two load
points. Setting $k_{i,j}=1$ selects the real reference at which the two
factors compensate -- the efficiency per unit magnification,
$\eta_i/|T'(\Gamma_{Li})|$, is the same in the three states. The
convention is admissible because the calibration factor is insensitive to
it (Appendix~\ref{app:sens}), whatever the individual values of the two
factors; these are the two directions of the thermistor-impedance space
that the calibration factor does not see, identified in
Section~\ref{sec:invariant}, and
Section~\ref{sec:passivity} shows that, once they are fixed by the
passivity of the matching structure, both factors are within $1\%$ of
unity and the convention is exact to that level. The two pairs $(1,2)$ and
$(3,2)$ yield two such impedances, and their disagreement -- the closure
defect, reported in Section~\ref{sec:results} -- is the margin that the
convention leaves on $Z_0$; its effect on the calibration factor is bounded
in Appendix~\ref{app:sens}.

\subsection{Meaning of the Convention}
\label{sec:convention}
The convention has an exact meaning in a reference case. Let the matching
structure between the mount port and the thermistor port be a lossless,
reciprocal two-port whose output reflection coefficient $S_{22}$, referred to
the real reference $R_{\mathrm{ref}}$ of the thermistor port, is real. Its output impedance
with the mount port matched,
\begin{equation}
Z_M = R_{\mathrm{ref}}\,\frac{1+S_{22}}{1-S_{22}},
\label{eq:zm}
\end{equation}
is then real. Re-referencing the thermistor port to $Z_M$ makes
$S'_{22}=0$; unitarity then forces $|S'_{21}|=1$ and $S'_{11}=0$, so that the
two-port reduces to a phase rotation, $\Gamma_i=e^{j\varphi}\,\Gamma_{Li}(Z_M)$
for every load, real or complex. Hence $|\Gamma_i|=|\Gamma_{Li}(Z_M)|$ in
every state: the condition \eqref{eq:zocond} closes with $k_{i,j}=1$ for every
pair at $Z_0=Z_M$, with no closure defect, and the relative curvature is
reproduced exactly, $C_v=C_{vL}(Z_M)$, because \eqref{eq:cv} is invariant
under a common rotation of the three coefficients. In this case the convention
returns the real output impedance of the mount and the curvature condition
is satisfied identically. In general, $Z_0$ is best described as the
\emph{effective reference impedance} of the thermistor port: the real
reference at which the lossless description of the matching structure is
closest to the measured data.

Table~\ref{tab:synth} verifies the statement on synthetic data: a two-port
of known scattering matrix in $R_{\mathrm{ref}}=170~\Omega$, a reference of the order of
the thermistor-port impedances found in Section~\ref{sec:results}, resistive
loads of 100, 200 and $300~\Omega$ standing for the three states,
$\Gamma_i$ computed from the network, and $Z_{0,12}$, $Z_{0,32}$ obtained
from \eqref{eq:zoquad}. With a lossless network and real $S_{22}$, the two
pairs return $Z_M$ to machine precision and $\psi=0$. The identity is broken
by a complex $S_{22}$ -- the two pairs then disagree and $\psi\neq0$ -- and,
more strongly, by dissipation: a matched attenuator at the mount port, which
leaves $Z_M$ unchanged, displaces the two roots by different amounts, the
disagreement between the pairs and $\psi$ both growing in proportion to the
loss.
The measured mounts are dissipative, $\eta_2\approx0.92$--$0.98$
(Section~\ref{sec:results}), and the closure defect they show, $3$--$10\%$
rms between the two pairs (Section~\ref{sec:results}), is consistent with this
picture: it measures the
departure of the actual structure from the lossless reference, and its
effect on the calibration factor is what Appendix~\ref{app:sens} bounds.
The convention is thus suited to a low-loss matching structure, as these
mounts are; for a strongly dissipative device the pair closure would not
define a useful reference.

\begin{table}[!t]
\caption{Synthetic test of the convention $k_{i,j}=1$. Lossless reciprocal
two-port in $R_{\mathrm{ref}}=170$~\capm{\Omega} with $S_{21}=\sqrt{1-|S_{22}|^2}\,e^{j0.7}$;
loads 100, 200, 300~\capm{\Omega}. The lossy rows cascade a matched attenuator of
the stated power loss at the mount port of the $S_{22}=-0.10$ network. $Z_M$
from \eqref{eq:zm}; $Z_{0,12}$, $Z_{0,32}$ from \eqref{eq:zoquad};
$C_{vL}$ at $Z_0=(Z_{0,12}+Z_{0,32})/2$. Impedances in \capm{\Omega}.}
\label{tab:synth}
\centering
\small
\setlength{\tabcolsep}{2.2pt}
\begin{tabular}{l c c c c c}
\hline
Network & $Z_M$ & $Z_{0,12}$ & $Z_{0,32}$ & $\mathrm{Re}\,C_v$ & $\mathrm{Re}\,C_{vL}$\\
\hline
$S_{22}=-0.10$ & 139.09 & 139.09 & 139.09 & $-0.870$ & $-0.870$\\
$S_{22}=0.20$ & 255.00 & 255.00 & 255.00 & $0.941$ & $0.941$\\
$S_{22}=-0.1\,e^{j0.5}$ & $142.0{-}j13.8$ & 142.63 & 142.45 & $-0.843$ & $-0.925$\\
$-0.10$, 2\% loss & 139.09 & 139.19 & 142.70 & $-0.870$ & $-0.899$\\
$-0.10$, 5\% loss & 139.09 & 139.32 & 148.06 & $-0.870$ & $-0.944$\\
\hline
\end{tabular}
\end{table}

With $Z_0$ determined, the relative-curvature residual
$\psi=\mathrm{Re}[C_v-C_{vL}]$ of \eqref{eq:cvl} can be evaluated. The two
conditions established so far combine into
\begin{equation}
E=w_1\,\frac{1}{N_f}\sum_{n=1}^{N_f}\epsilon_n(\alpha,\beta)
+ w_2\left(\frac{1}{N_f}\sum_{n=1}^{N_f}\psi_n\right)^2 ,
\label{eq:Etwo}
\end{equation}
with $w_1$, $w_2$ the weights of the two terms (their values are given in
Section~\ref{sec:invariant}); $\psi$ enters as the square of its band mean, and not as a mean of
squares like $\epsilon$, because $\psi$ changes sign across the band and
the condition is that its mean vanish. The minimization of \eqref{eq:Etwo}
over $X$ (and the auxiliary $\alpha,\beta$) yields the
parameter set that characterizes the sensor. A third term is added
to \eqref{eq:Etwo} in Section~\ref{sec:invariant} for the stability of the
minimization.

\section{The Hyperbolic Invariant}
\label{sec:invariant}
A third term is added to the functional for the stability of the
minimization: the two conditions of Sections~\ref{sec:determination}
and~\ref{sec:zo} fix $X$ firmly along one direction only and leave it
almost free along $C_x$ and along a combination of $R_{p0}$ and $\kappa$
(Appendix~\ref{app:ident}), and the term supplies the curvature that is
missing there. It
rests on a classical property of the reflection coefficient that is not often
used in this context, so we state it from the beginning.

A passive reflection coefficient lies in the unit disk, $|\Gamma|<1$. On that
disk one can define, between any two points $\Gamma_a$ and $\Gamma_b$, the
\emph{pseudo-hyperbolic distance}
\begin{equation}
\rho(\Gamma_a,\Gamma_b)=\left|\frac{\Gamma_a-\Gamma_b}
{1-\Gamma_a^{*}\,\Gamma_b}\right|,
\label{eq:pseudohyp}
\end{equation}
which takes values in $[0,1)$ and reduces to the ordinary distance
$|\Gamma_a-\Gamma_b|$ near the center. It is the natural distance of the
Poincar\'e disk model of hyperbolic geometry, and its defining property is
the one we need here: it is unchanged by any conformal map of the disk onto
itself (a disk automorphism), a result known as the invariant form of the
Schwarz lemma, or the Schwarz--Pick lemma \cite{garnett,ahlfors}. A lossless,
reciprocal two-port transforms its load reflection coefficient into its input
reflection coefficient precisely by such a map; the quantity \eqref{eq:pseudohyp}
is therefore conserved across a lossless matching network.

With the three states we form, at each frequency, the scalar
\begin{equation}
Q_h = \rho(\Gamma_1,\Gamma_2)^2 + \rho(\Gamma_3,\Gamma_2)^2 ,
\label{eq:qh}
\end{equation}
the sum of the squared pseudo-hyperbolic distances of the two outer states to
the central one. The matching two-port relates $\Gamma_{Li}$ and $\Gamma_i$ by
the bilinear map \eqref{eq:gamma}; if it were lossless that map would be a disk
automorphism, under which $Q_h$ is exactly invariant, and the value computed
from the measured coefficients would equal the value computed from the
thermistor-port coefficients,
\begin{equation}
Q_h(\Gamma_1,\Gamma_2,\Gamma_3) = Q_h(\Gamma_{L1},\Gamma_{L2},\Gamma_{L3}),
\label{eq:qhequality}
\end{equation}
frequency by frequency. The matching bar and pivot are lossy only to the
extent of the efficiency, so \eqref{eq:qhequality} holds up to a contraction
that is accounted for below, and imposing it adds a further constraint to
the fit. Unlike the conditions of
Sections~\ref{sec:determination} and~\ref{sec:zo}, which enter through moduli
and through the real part of the curvature ratio, the pseudo-hyperbolic distance uses the full complex
coefficients and therefore carries phase information that the other terms do
not see directly; this is what improves the conditioning of the fit.

In practice the matching structure is not exactly lossless, and a lossy map
does not conserve the pseudo-hyperbolic distance but contracts it; the model
value in \eqref{eq:qhequality} must therefore be scaled by that contraction
before it is compared with the measured value. Writing the measured quantity
as $q_{h}^{\mathrm{meas}}=Q_h(\Gamma_1,\Gamma_2,\Gamma_3)$ and the model
quantity as $q_{h}^{\mathrm{mod}}=Q_h(\Gamma_{L1},\Gamma_{L2},\Gamma_{L3})$,
with the thermistor-port coefficients evaluated at the real reference $Z_0$ of
\eqref{eq:gammaLzo}, the residual used is
\begin{equation}
\Lambda = q_{h}^{\mathrm{meas}} - q_{h}^{\mathrm{mod}}\,\eta_2^{2} ,
\label{eq:qhresidual}
\end{equation}
where $\eta_2$ is the efficiency of the central state. The origin of this
scaling factor is set out below and proved in
Appendix~\ref{app:schwarzpick}. The functional of \eqref{eq:Etwo} is then
extended with a third term,
\begin{equation}
E=w_1\,\frac{1}{N_f}\sum_n\epsilon_n
+ w_2\left(\frac{1}{N_f}\sum_n\psi_n\right)^2
+ w_3\,\frac{1}{N_f}\sum_n \Lambda_n^2 ,
\label{eq:Ethree}
\end{equation}
and minimized over $X$ as before. The runs reported here used
$w_1=2.5\times10^{-4}$, $w_2=0.74$ and $w_3=1$ for the five WR-42 specimens
($w_1=2.5\times10^{-5}$ for the WR-28 specimen of Section~\ref{sec:wr28}). The weights
govern the speed of convergence and not the solution: for $w_2$ between
$10^{-2}$ and $1$ the parameters reached are the same within the stopping
tolerance, and the calibration factor is insensitive to them within a broad
range.

The form of the scaling factor follows from the geometry of the map. The local
contraction of the pseudo-hyperbolic metric under the bilinear map $T$ of
\eqref{eq:gamma} is the Schwarz--Pick factor, which at a point $z$ of the disk
is
\begin{equation}
\lambda(z)=|T'(z)|\,\frac{1-|z|^2}{1-|T(z)|^2},
\label{eq:lambdabody}
\end{equation}
so that for two nearby states
$\rho\big(T(\Gamma_{La}),T(\Gamma_{Lb})\big)\simeq\lambda\,
\rho(\Gamma_{La},\Gamma_{Lb})$. Appendix~\ref{app:schwarzpick} shows that this
factor has an exact physical meaning: for a reciprocal two-port it equals the
efficiency, $\lambda=\eta$. The contraction that the mount imposes on the
pseudo-hyperbolic distance is therefore its efficiency, and since $Q_h$ is a
sum of squared distances it is scaled by $\eta^2$. This is what allows a
quantity formed from the measured coefficients to be compared with one formed
from the thermistor model: the measured side carries the contraction, and the
efficiency, which the method determines independently through
\eqref{eq:etameas}, supplies it exactly.

The quantity $Q_h$ is built from \emph{finite} distances between the three
states, not from the differential metric element, and $\lambda=\eta$ varies
from point to point across the disk because the three states have different
efficiencies. The scaling in \eqref{eq:qhresidual} is the zeroth-order
approximation in which the contraction is represented by its value at the
central state, $\lambda_2=\eta_2$, applied to both distances; it becomes exact
as the separation between the states vanishes. Its exact form is stated
next, and what follows from it defines the role of the term.

It may be asked whether \eqref{eq:qhresidual} set to zero is not simply
another expression for the efficiency, the one of \eqref{eq:eta}. It is,
and exactly so. In the limit of vanishing separation, by
Appendix~\ref{app:schwarzpick} the contraction of each distance is the
efficiency of \eqref{eq:eta}. At finite separation the exact relation, for
any reciprocal two-port and any two states $i,j$, is
\begin{equation}
\frac{\rho_{\mathrm{in}}^2}{1-\rho_{\mathrm{in}}^2}
=\eta_i\,\eta_j\,\frac{\rho_L^2}{1-\rho_L^2},
\label{eq:hypexact}
\end{equation}
with $\rho_{\mathrm{in}}=\rho(\Gamma_i,\Gamma_j)$ and
$\rho_L=\rho(\Gamma_{Li},\Gamma_{Lj})$; it follows from \eqref{eq:gamma}
and the identity of Appendix~\ref{app:schwarzpick}. Since the efficiencies of
\eqref{eq:etameas} are those of the very bilinear map that carries the
$\Gamma_{Li}(X)$ to the measured $\Gamma_i$, \eqref{eq:hypexact} holds
identically for every $X$: in its exact form the hyperbolic condition
imposes nothing, which is the precise statement that $Q_h$ is not a third
determination of $\eta$.

The transformation \eqref{eq:neutral} of Section~\ref{sec:neutral} is a
particular case of a wider invariance, which can now be stated with the
distance \eqref{eq:pseudohyp} in hand, and which delimits what the method
can determine. Let
\begin{equation}
\rho_{ij}=\left|\frac{Z_{th,i}-Z_{th,j}}{Z_{th,i}+Z_{th,j}^{*}}\right|,
\qquad
q_{ij}=\frac{\rho_{ij}^2}{1-\rho_{ij}^2}
=\frac{|Z_{th,i}-Z_{th,j}|^2}{4\,\mathrm{Re}Z_{th,i}\,\mathrm{Re}Z_{th,j}}
\label{eq:rhoq}
\end{equation}
where $\rho_{ij}$ is the pseudo-hyperbolic distance \eqref{eq:pseudohyp}
between the thermistor-port reflection coefficients of the two states,
written in terms of the impedances, and $q_{ij}$ the quantity associated
with it (the asterisk denotes the complex conjugate); $\rho_{ij}$ is the
same at any real reference, and does not depend on that reference. Substituting $|Z_{th,i}-Z_{th,j}|^2=4q_{ij}\,\mathrm{Re}Z_{th,i}\,\mathrm{Re}Z_{th,j}$
in \eqref{eq:CiZ}, all the real parts cancel and
\begin{equation}
C_i^2=\frac{q_{jk}}{q_{ij}\,q_{ik}} ,
\label{eq:Fq}
\end{equation}
exactly, for each state $i$ with $j,k$ the other two; for the operating
state, $F^2=C_2^2=q_{13}/(q_{12}q_{32})$. We write $F$ for $C_2$ because it
is the factor of the calibration factor that carries the thermistor model;
the measured coefficients enter through a second factor, $G$, obtained
below, and $K_f=F\,G$. The model factor depends on the
three thermistor impedances only through their mutual hyperbolic
distances, and on these only through one combination. The distances are invariant under the isometries of the
disk, which are precisely the load maps of lossless two-ports: a real
transformer and a series reactance -- the transformation
\eqref{eq:neutral}, the subgroup that fixes $\Gamma=1$ -- a shunt
susceptance, a length of line, and their cascades, a group of dimension
three. Any such element common to the three states is invisible to
$C_i$, and so is any change of real reference. The count is then the
following. The three impedances are six real quantities; $F=C_2$ fixes one;
the isometries account for three; two remain -- the shape of the
hyperbolic triangle formed by the three loads beyond the combination
\eqref{eq:Fq}. The set
\begin{equation}
\mathcal M_F=\{(Z_{th,1},Z_{th,2},Z_{th,3}):\ q_{13}=F^2\,q_{12}\,q_{32}\}
\label{eq:MF}
\end{equation}
is the part of the space of thermistor impedances that the calibration
factor does not see; it is a hypersurface of dimension five, and its
quotient by the isometries is a surface of dimension two. The two
coordinates are identified next.

Relation \eqref{eq:hypexact}, combined with \eqref{eq:Fq}, closes the
count. Writing $q^{\mathrm{L}}_{ij}$ and
$q^{\mathrm{in}}_{ij}$ for $\rho^2/(1-\rho^2)$ on the load and on the
input side, \eqref{eq:hypexact} reads $q^{\mathrm{L}}_{ij}=q^{\mathrm{in}}_{ij}/(\eta_i\eta_j)$,
and substituting in \eqref{eq:Fq},
\begin{equation}
F^2=\eta_2^2\,\frac{q^{\mathrm{in}}_{13}}{q^{\mathrm{in}}_{12}\,q^{\mathrm{in}}_{32}},
\qquad
G=(1-|\Gamma_2|^2)\sqrt{\frac{q^{\mathrm{in}}_{12}\,q^{\mathrm{in}}_{32}}{q^{\mathrm{in}}_{13}}},
\label{eq:FGq}
\end{equation}
the second, the geometric factor $G$, being the same identity applied to
the measured coefficients;
hence $FG=\eta_2(1-|\Gamma_2|^2)=K_f$, and $\eta_1$, $\eta_3$ cancel.
With the measured coefficients given and $\eta_2$ fixed by $F$, the two
coordinates of $\mathcal M_F$ are therefore $\eta_1/\eta_2$ and
$\eta_3/\eta_2$, the auxiliary variables of \eqref{eq:etaratios}: what
the calibration factor does not see is exactly the spread of the
efficiencies between the states. The first property of
Section~\ref{sec:determination} constrains those ratios to be constant
across the band, and the calibration factor is invariant to their value;
Section~\ref{sec:passivity} shows how that value is determined. The form \eqref{eq:qhresidual} differs from
\eqref{eq:hypexact} by the replacement of $\eta_i\eta_2$ by $\eta_2^2$ and
by the finite-separation factor; $\Lambda$ is therefore a smooth function
of $X$ that vanishes where the three efficiencies coincide and the states
are close, and not a property of the data. That is what makes it useful as a
stability term and nothing more: it carries no information on the
calibration factor, and the result stated below confirms it.

The term is independent of the reference impedance. For any reference, real or
complex, the pseudo-hyperbolic distance between two thermistor-port
coefficients reduces to
$\rho(\Gamma_{La},\Gamma_{Lb})=|Z_{th,a}-Z_{th,b}|/|Z_{th,a}^{*}+Z_{th,b}|$,
so $q_{h}^{\mathrm{mod}}$ depends on the thermistor impedances alone, and
$\eta_2$ is independent of $Z_0$ by \eqref{eq:etameas} and \eqref{eq:CiZ}. The term thus
neither depends on nor constrains the thermistor-port impedance; it acts on the
thermistor parameters only, along $C_x$ and a second direction combining
$R_{p0}$ and $\kappa$ (Appendix~\ref{app:ident}). With the term the
functional has a proper minimum, with a positive-definite Hessian
(Appendix~\ref{app:ident}), and the calibration factor is insensitive to
those directions: with the term ($w_3=1$) and without it ($w_3=0$) the
calibration factor and its expanded uncertainty are the same -- over
18--26.5~GHz the mean $U$ ($k=2$) differs by about $1\times10^{-3}\%$,
with no systematic sign. The term therefore gives the parameters a unique
value without determining the result; that is the sense in which it is a
stability term.

\subsection{Minimization}
\label{sec:minimization}
The functional \eqref{eq:Ethree} is minimized over five unknowns: the three
parameters $X=\{C_x,R_{p0},\kappa\}$ and the auxiliary variables $\alpha$,
$\beta$ of \eqref{eq:etaratios}; the bead length $L_x=0.32$~mm is fixed from
the measured geometry (Section~\ref{sec:circuit}). With five unknowns the
search is simpler than in the coaxial case \cite{greco_coaxial}: no coarse
grid is required, and the parameters are obtained by direct minimization
within the bounds $5\le C_x\le40$~fF, $300\le R_{p0}\le3000~\Omega$,
$0\le\kappa\le0.99$ and $0.9\le\alpha,\beta\le1.1$; every WR-42
solution lies in the interior of these ranges. The
procedure is iterative, updating $\{C_x,R_{p0},\kappa\}$ and
$\{\alpha,\beta\}$ alternately; each update is a bounded
minimization performed with the \textsf{NMinimize} function of
Mathematica \cite{wolfram_nminimize}, Nelder--Mead method, which starts
from a set of points generated within the bounded region and does not
require a starting estimate; the nominal values serve only as the
reference point of the alternation. As a test of the minimization, the
run for specimen~5 was repeated with four different random seeds of the
Nelder--Mead method and with the Differential Evolution method under the
same bounds; the five runs converge to the same minimum, with $E$ agreeing
to seven digits and $K_f$ within $10^{-5}\%$ at every frequency. The
alternation converges in a few rounds. The reference impedance $Z_0$ is not an
independent unknown; it is computed from the current parameters at each
evaluation through the closed form of Section~\ref{sec:zo}, so the whole
determination reduces to this low-dimensional minimization. The five
specimens were processed with the same code, the same weights and bounds,
and no specimen-specific adjustment.

\section{Results}
\label{sec:results}
The method was applied to five K486A thermistor mounts, identified here as
specimens~1--5. The bias resistances were common to all measurements,
$R_{b}=\{100.415,\,200.366,\,300.345\}~\Omega$, determined to a relative
uncertainty of $2.4\times10^{-4}$. For each specimen the parameters $X=\{C_x,R_{p0},\kappa\}$
were obtained from the measured reflection coefficients as described in
Sections~\ref{sec:determination}--\ref{sec:invariant}, with no specimen-specific
adjustment of the procedure, and the calibration factor was computed at every
point of the analyzer frequency grid over 18--26.5~GHz. The parameters
obtained for each specimen are listed in Table~\ref{tab:params}, together with
the bridge resistances.

\begin{table}[!t]
\caption{Parameters obtained for the five specimens: the set $X$ of the
thermistor model and the auxiliary variables $\alpha$, $\beta$ of
\eqref{eq:etaratios}. The bridge resistances are common to all
specimens.}
\label{tab:params}
\centering
\begin{tabular}{c c c c c c}
\hline
Specimen & $C_x$ (fF) & $R_{p0}$ ($\Omega$) & $\kappa$ & $\alpha$ & $\beta$\\
\hline
1 & 18.24 & 1441.7 & 0.821 & 0.9565 & 0.9798\\
2 & 18.94 & 1045.9 & 0.790 & 0.9577 & 0.9794\\
3 & 24.03 & 1218.8 & 0.474 & 0.9623 & 0.9817\\
4 & 16.45 & 1431.3 & 0.660 & 0.9675 & 0.9838\\
5 & 17.03 & 1546.5 & 0.478 & 0.9740 & 0.9880\\
\hline
\multicolumn{6}{c}{Bridge: $R_{b,1}=100.415$, $R_{b,2}=200.366$,
$R_{b,3}=300.345~\Omega$}\\
\hline
\end{tabular}
\end{table}

Specimens~1--4 are compared with the calibration tables supplied by the
manufacturer with each mount ($U=2\%$, $k=2$), which give $K_f$ at a set of
frequencies within the band; their stated uncertainty limits what the
comparison can resolve, which is why the weight of the verification falls on
the specimen with a traceable reference. Specimen~5 is compared with reference values traceable to PTB
($U=0.85$--$0.98\%$, $k=2$), covering the ten frequencies 18--26.5~GHz. The agreement
between the model and each reference set is summarized in
Table~\ref{tab:results}: the root-mean-square and the largest deviation of
the model $K_f$ from the reference points, taken at the grid point of the
analyzer nearest to each reference frequency, without smoothing or
interpolation, and the largest normalized error, defined below. The model reproduces the four manufacturer tables
within $1.0\%$ rms, and the PTB-traceable reference within $0.34\%$ rms; the
closest agreement of the set is obtained against the most accurate reference,
and on the specimen whose own expanded uncertainty is also the lowest
(Section~\ref{sec:budget}).

Three quantities must be kept distinct in reading Table~\ref{tab:results}.
The rms deviation measures the departure of the central values of the model
from an external reference: it is a property of the model and of the
specimen, and it is the only one of the three that tests the method against
something it did not produce. The expanded uncertainty $U$ of
Section~\ref{sec:budget} is the uncertainty of the result given the model:
it propagates the errors of the reflectometer through the determination;
the adequacy of the model is tested by the comparison with the reference,
not by $U$. What connects the two is the normalized error
\begin{equation}
E_n=\frac{|K_f-K_{f,\mathrm{ref}}|}{\sqrt{U^2+U_{\mathrm{ref}}^2}} ,
\label{eq:en}
\end{equation}
the criterion of consistency between a result and a reference in
interlaboratory comparisons \cite{iso13528,iso17043}: $E_n\le1$ means that the
deviation is accounted for by the two uncertainties together, and it is the
quantity that establishes whether the budget of the method is representative
of its actual departure from the reference. The last column of
Table~\ref{tab:results} gives, for each specimen, the largest $E_n$ over its
reference points, with $U$ taken from the budget of Section~\ref{sec:budget}
at the nearest tabulated frequency. Over the 38 reference points of the five
WR-42 specimens at which the budget is tabulated $E_n$ does not exceed
$0.6$, and for the PTB-traceable specimen it does not exceed $0.3$: the budget of the method is consistent
with the deviations actually observed, against both the manufacturer tables
and the traceable reference. The departures carry no systematic sign:
their signed mean is between $-0.9$ and $+0.3\%$ for the four specimens
referred to manufacturer tables and $+0.08\%$ for the PTB-traceable one,
with five points of each sign.

The set of specimens spans the efficiency range of the mounts used as
traveling standards. The two K486A mounts circulated in the recent key comparison
CCEM.RF-K28.W in WR-42 \cite{ccem_k28} have efficiencies of 0.963--0.981 at
18--26.5~GHz, within the range 0.92--0.98 of the five specimens (Table~\ref{tab:etaall}), and each has
a counterpart in the set: Table~\ref{tab:k28eta} places them beside the two
closest specimens, the NIM (National Institute of Metrology, China) standard NIM-2 matched by specimen~4 within $0.3\%$ in level and
frequency dependence, and NIM-1 by specimen~5 within $0.6\%$. This
comparison serves only to show that the set spans that range; the
quantities are not the same (the efficiency of the central state here, the
key comparison reference value there) and no metrological equivalence is
implied.

\begin{table}[!t]
\caption{Agreement of the model calibration factor with the reference data
for the five WR-42 specimens and the WR-28 specimen of
Section~\ref{sec:wr28}: rms and largest deviation at the reference
frequencies, and largest normalized error \eqref{eq:en}. Model values at
the nearest analyzer grid point, without interpolation; $E_n$ with the
budget $U$ at the nearest tabulated frequency, hence over 9 of the 10
points for specimen~3.}
\label{tab:results}
\centering
\small
\setlength{\tabcolsep}{2pt}
\begin{tabular}{c l c c c c}
\hline
Spec. & Reference & $N$ & rms (\%) & max (\%) & max $E_n$\\
\hline
1 & \multirow{4}{*}{\shortstack[l]{manufacturer\\ $U=2\%$}} & 6  & 0.84 & 1.11 & 0.44\\
2$^{\mathrm a}$ &                                           & 6  & 0.98 & 1.59 & 0.54\\
3$^{\mathrm a}$ &                                           & 10 & 0.97 & 1.81 & 0.60\\
4 &                                                         & 7  & 0.86 & 1.43 & 0.58\\
\hline
5 & \shortstack[l]{PTB-traceable\\ $U=0.85$--$0.98\%$}                 & 10 & 0.34 & 0.53 & 0.29\\
\hline
6, WR-28 & \shortstack[l]{manufacturer\\ $U=3\%$}                  & 6  & 0.92 & 1.51 & 0.38\\
\hline
\multicolumn{6}{@{}p{6.8cm}@{}}{$^{\mathrm a}$Two points were measured in
the author's laboratory by ratiometric comparison of effective efficiencies
against specimen~4: specimen~2 at 18~GHz, where the manufacturer value was
not available, $K_f=0.930$, and specimen~3 at 26.5~GHz, $K_f=0.935$; for
both, $U=2.2\%$ (the manufacturer uncertainty of specimen~4 combined with
that of the ratio measurement).}\\
\end{tabular}
\end{table}

\begin{figure}[!t]
\centering
\includegraphics[width=0.95\textwidth]{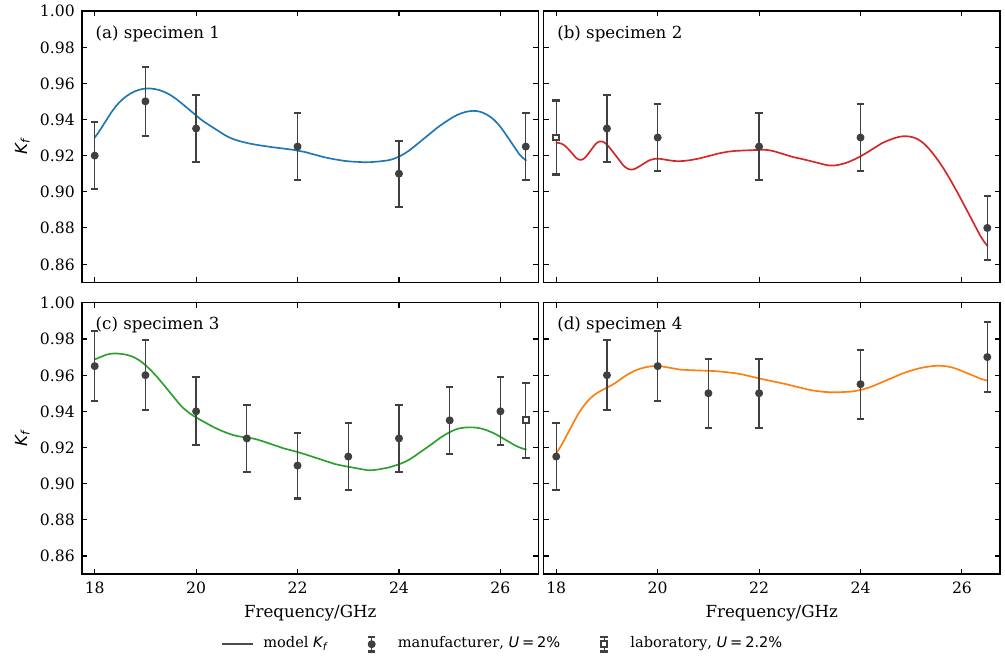}
\caption{Calibration factor of specimens~1--4 (manufacturer reference). Solid
curve: model $K_f$ over the analyzer grid, smoothed with a Gaussian filter
($\sigma=0.17$~GHz). Points: manufacturer table with
$U=2\%$ ($k=2$) bars; the 18~GHz point of specimen~2 and the 26.5~GHz
point of specimen~3 (open squares) are the laboratory comparisons of
Table~\ref{tab:results}, with their own $U=2.2\%$ bars. Same scale
(0.85--1.00) in the four panels.}
\label{fig:kf1234}
\end{figure}

\begin{figure}[!t]
\centering
\includegraphics[width=0.62\textwidth]{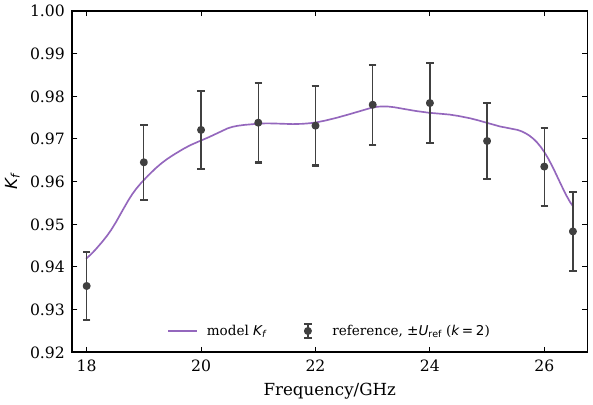}
\caption{Calibration factor of specimen~5 against the PTB-traceable reference.
Solid curve: model $K_f$ over the analyzer grid, smoothed as in
Fig.~\ref{fig:kf1234}; points: reference values with their stated
$U=0.85$--$0.98\%$ ($k=2$) bars. The vertical scale is expanded (0.92--1.00)
relative to Fig.~\ref{fig:kf1234} to resolve the agreement.}
\label{fig:kf5}
\end{figure}

The measured calibration factor of specimens~1--4 is shown against the
manufacturer points in Fig.~\ref{fig:kf1234}. Specimen~5, compared
with the PTB-traceable reference, is shown in Fig.~\ref{fig:kf5} on an
expanded vertical scale, the reference uncertainty being small enough that the
full-band scale of Fig.~\ref{fig:kf1234} would not resolve it. The bars in
these figures are those of the reference alone; the consistency of model and
reference is the $E_n$ of Table~\ref{tab:results}, which also carries the
measurement uncertainty of the method.

The reference of specimen~5 states the effective efficiency separately
from the calibration factor, which allows a second comparison,
complementary to the first -- the two quantities differ by the measured
mismatch factor, so it is a check of consistency rather than an
independent measurement: the efficiency of the central state, $\eta_2$,
which is the quantity the model determines, against the reference
efficiency. The uncertainty of $\eta_2=K_f/(1-|\Gamma_2|^2)$ adds to that
of $K_f$ the term of the mismatch factor,
\begin{equation}
\frac{u(M)}{M}=\frac{2|\Gamma_2|\,u(|\Gamma_2|)}{1-|\Gamma_2|^2},
\qquad M=1-|\Gamma_2|^2,
\label{eq:um}
\end{equation}
with $u(|\Gamma_2|)=0.0033$ from the residuals of the calibrated
reflectometer (Section~\ref{sec:budget}) taken as rectangular limits, dominated by the
directivity. Since $u(K_f)$ already contains the reflectometer channel,
the two are not independent and the expanded uncertainties are added
linearly, $U(\eta)=U(K_f)+U(M)$, which is a bound. Table~\ref{tab:eta5}
gives the result: the model efficiency agrees with the reference within
$0.39\%$ rms, $0.62\%$ maximum, with $E_n\le0.45$ over the ten points, at
the same level as the calibration factor. The efficiency is where the
model acts -- the mismatch factor is measured -- so this is the direct test
of the thermistor model against a traceable reference. The measured
$|\Gamma_2|$ agrees with the reference value within $0.008$ at every
frequency, consistent with the $u(|\Gamma_2|)$ above. Table~\ref{tab:etaall}
gives $\eta_2$ with its uncertainty for the five specimens at the reference
frequencies; it is the quantity that a laboratory holding one of these mounts
would compare with its own calibration, and the form in which the results of
the method are offered for comparison (Section~\ref{sec:conclusion}).

\begin{table}[!t]
\caption{Efficiency of the central state, $\eta_2$, of the five WR-42
specimens at nine frequencies of the band, with the expanded uncertainty
$U(\eta)$ ($k=2$, per cent, in parentheses) from \eqref{eq:um} and the
budgets of Section~\ref{sec:budget}. These are the values a laboratory
compares with its own measurement of the same mount.}
\label{tab:etaall}
\centering
\normalsize
\setlength{\tabcolsep}{6pt}
\begin{tabular}{c c c c c c}
\hline
$f$ (GHz) & 1 & 2 & 3 & 4 & 5\\
\hline
18.0 & 0.9572 (1.53) & 0.9446 (1.09) & 0.9729 (1.49) & 0.9646 (1.67) & 0.9823 (1.46)\\
19.0 & 0.9611 (0.86) & 0.9498 (1.14) & 0.9741 (1.41) & 0.9579 (1.03) & 0.9764 (1.09)\\
20.0 & 0.9628 (1.41) & 0.9532 (1.70) & 0.9631 (1.94) & 0.9667 (0.73) & 0.9787 (0.98)\\
21.0 & 0.9672 (1.76) & 0.9647 (1.71) & 0.9719 (1.83) & 0.9710 (1.12) & 0.9808 (0.78)\\
22.0 & 0.9692 (1.60) & 0.9652 (1.40) & 0.9746 (1.75) & 0.9735 (1.01) & 0.9788 (0.68)\\
23.0 & 0.9620 (1.84) & 0.9608 (1.81) & 0.9657 (2.26) & 0.9697 (1.19) & 0.9803 (0.70)\\
24.0 & 0.9574 (1.81) & 0.9585 (1.61) & 0.9620 (1.91) & 0.9670 (1.28) & 0.9765 (0.54)\\
25.0 & 0.9616 (1.20) & 0.9622 (1.23) & 0.9652 (1.29) & 0.9681 (0.86) & 0.9758 (0.66)\\
26.5 & 0.9231 (1.24) & 0.9371 (2.25) & 0.9494 (1.93) & 0.9597 (0.95) & 0.9792 (1.24)\\
\hline
\end{tabular}
\end{table}

\begin{table}[!t]
\caption{Efficiency of the two traveling standards of CCEM.RF-K28.W (key
comparison reference values, \cite{ccem_k28}) and of the two closest
specimens of this work ($\eta_2$ of the converged solution), at the
frequencies of the comparison. For the range of the set only.}
\label{tab:k28eta}
\centering
\begin{tabular}{l c c c c}
\hline
 & 18 & 21 & 24 & 26.5~GHz\\
\hline
NIM-1 (K28.W)  & 0.976 & 0.981 & 0.978 & 0.973\\
Specimen~5     & 0.982 & 0.981 & 0.976 & 0.979\\
\hline
NIM-2 (K28.W)  & 0.967 & 0.969 & 0.967 & 0.963\\
Specimen~4     & 0.965 & 0.971 & 0.967 & 0.960\\
\hline
\end{tabular}
\end{table}

\begin{table}[!t]
\caption{Efficiency of the central state of specimen~5, $\eta_2$, against
the PTB-traceable reference, with the expanded uncertainty $U(\eta)$ ($k=2$)
from \eqref{eq:um} and the normalized error \eqref{eq:en}; the reference
uncertainty is that stated for its calibration factor, $0.85$--$0.98\%$.}
\label{tab:eta5}
\centering
\small
\setlength{\tabcolsep}{3.5pt}
\begin{tabular}{c c c r c c}
\hline
$f$ (GHz) & $\eta_2$ & $\eta_{\mathrm{ref}}$ & dev. (\%) & $U(\eta)$ (\%)
& $E_n$\\
\hline
18.0 & 0.9823 & 0.9763 & $+0.62$ & 1.46 & 0.36\\
19.0 & 0.9764 & 0.9806 & $-0.43$ & 1.09 & 0.30\\
20.0 & 0.9787 & 0.9804 & $-0.17$ & 0.98 & 0.12\\
21.0 & 0.9808 & 0.9802 & $+0.06$ & 0.78 & 0.05\\
22.0 & 0.9788 & 0.9782 & $+0.06$ & 0.68 & 0.05\\
23.0 & 0.9803 & 0.9805 & $-0.02$ & 0.70 & 0.02\\
24.0 & 0.9765 & 0.9787 & $-0.23$ & 0.54 & 0.21\\
25.0 & 0.9758 & 0.9714 & $+0.45$ & 0.66 & 0.40\\
26.0 & 0.9821 & 0.9761 & $+0.61$ & 1.08 & 0.43\\
26.5 & 0.9792 & 0.9738 & $+0.55$ & 1.24 & 0.35\\
\hline
\end{tabular}
\end{table}

The values of the ratio $k_{i,j}$ of \eqref{eq:zocond}, computed a
posteriori from the converged solution of each specimen with both pairs
evaluated at the common $Z_0=(Z_{0,12}+Z_{0,32})/2$, depart from unity by
$0.4$--$1.3\%$ rms, with the same signature in the five units: the departure
is a systematic property of the method and the mount geometry, not a
per-unit artifact. It is the pair-closure defect analyzed in
Appendix~\ref{app:sens}, and it is the quantity that bounds the
uncertainty of the thermistor-port impedance. Its direct expression is the
difference between the two pair impedances of Section~\ref{sec:zo},
Table~\ref{tab:pairs}: $Z_{0,32}$ exceeds $Z_{0,12}$ in the five mounts, by
$2.6\%$ of $Z_0$ on average for specimen~5 and up to $9.7\%$ for
specimen~2, so that each pair impedance departs from their mean by half
those amounts. By \eqref{eq:kidentity} each pair closes \eqref{eq:zocond}
with the efficiency ratio it has to compensate, so the two pair impedances
coincide only where the three efficiencies are equal; their difference is
an expression of the spread of the efficiencies between states, the two
coordinates of Section~\ref{sec:invariant}. Section~\ref{sec:passivity},
which determines those coordinates, returns to this table: its last
columns anticipate that result, the difference reduced to $1.0$--$4.5\%$
with the mean, which is the $Z_0$ of the method, changed by about
$1~\Omega$ or less. The effect of the closure defect on the
calibration factor is bounded in Appendix~\ref{app:sens}.

\begin{table}[!t]
\caption{Difference between the two pair impedances of \eqref{eq:zoquad},
$(Z_{0,32}-Z_{0,12})/Z_0$ with $Z_0$ their mean, over 18--26.5~GHz: band
mean, rms and largest magnitude, in per cent, with the thermistor model of
Section~\ref{sec:circuit} and with the state correction of
Section~\ref{sec:passivity}, both on the grid of that section (every
second point of the analyzer grid); last column, band mean of $Z_0$ in the
two cases (\capm{\Omega}).}
\label{tab:pairs}
\centering
\normalsize
\setlength{\tabcolsep}{4pt}
\begin{tabular}{c c c c c c c c}
\hline
Spec. & \multicolumn{3}{c}{model} & \multicolumn{3}{c}{corrected} & $Z_0$\\
 & mean & rms & max & mean & rms & max & model / corr.\\
\hline
1 & $+6.9$ & 7.0 & 11.7 & $+3.7$ & 3.7 & 6.3 & 152.4 / 152.0\\
2 & $+9.7$ & 10.0 & 17.1 & $+4.0$ & 4.1 & 6.7 & 203.1 / 202.0\\
3 & $+6.6$ & 6.7 & 9.6 & $+4.5$ & 4.7 & 7.0 & 165.7 / 165.7\\
4 & $+4.7$ & 4.9 & 6.3 & $+2.5$ & 2.8 & 3.8 & 153.1 / 152.9\\
5 & $+2.6$ & 2.9 & 4.0 & $+1.0$ & 1.7 & 3.7 & 188.3 / 188.5\\
\hline
\end{tabular}
\end{table}

\subsection{Application to WR-28}
\label{sec:wr28}
The procedure was applied, with the same functional and the same stopping
rule, to an HP R486A mount of the same family in WR-28 (26.5--40~GHz),
specimen~6; the weight $w_1$ was set to $2.5\times10^{-5}$
(Section~\ref{sec:invariant}), to which the calibration factor is
insensitive within a broad range, and $L_x$ as stated below. The HP 8510C configuration available to
the author does not cover this band, and the reflection coefficients were measured with an experimental
reflectometer assembled from two WR-28 directional couplers, an HP 83630A
swept synthesizer with an HP 83554A millimeter-wave source module, and an
HP 70820A microwave transition analyzer, calibrated at the mount flange, on a
grid of 401 points. Its
measured residuals are a directivity of $-40$~dB, a source match of $-30$~dB,
a tracking of 0.05~dB and a random noise of $1\times10^{-3}$; the dynamic
accuracy was taken from the table of Section~\ref{sec:budget}, as in WR-42. The bias resistances were those of the other
specimens. $L_x$ was fixed at 0.16~mm: the guide is smaller and so is the
thermistor, as observed in a mount of the same model opened for repair, and
the insensitivity of $K_f$ to $L_x$ (Section~\ref{sec:budget}) makes this
estimate sufficient. The fit gives $C_x=12.60$~fF, $R_{p0}=361.6~\Omega$ and $\kappa=0$: the
loss resistance is
constant over the band, the decrease of $R_p$ seen in WR-42 having
reached its limit at these frequencies.

\begin{figure}[!t]
\centering
\includegraphics[width=0.62\textwidth]{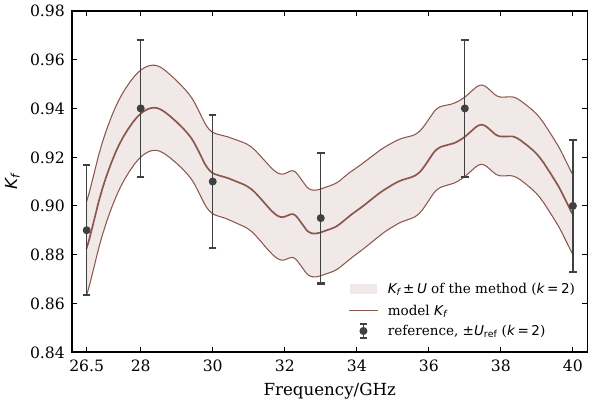}
\caption{Calibration factor of the WR-28 specimen~6. Solid curve: model
$K_f$ over the reflectometer grid, smoothed as in Fig.~\ref{fig:kf1234};
shaded band: $K_f\pm U$, the expanded uncertainty of the method ($k=2$) at
every point of the grid. Points: manufacturer calibration with $U=3\%$ ($k=2$)
bars.}
\label{fig:kf6}
\end{figure}

Figure~\ref{fig:kf6} shows $K_f$ with its expanded measurement uncertainty,
which lies between $1.4\%$ and $2.8\%$ ($k=2$) over the band, together with the six
values of the manufacturer's calibration, whose stated uncertainty is $3\%$.
The rms deviation is $0.92\%$ (Table~\ref{tab:results}), the largest
deviation, $1.5\%$, occurs at the lower band edge, and the normalized error
$E_n$ is at most 0.4 at the six frequencies. The result in this band rests
on one specimen; it shows that the method carries over
to another waveguide size, and to a reflectometer other than the analyzer used
in WR-42, without change.

\section{Uncertainty Budget}
\label{sec:budget}
The calibration factor factors as $K_f = F(X)\,G(\Gamma_1,\Gamma_2,\Gamma_3)$,
where $F$ carries the model parameters and $G$ the measured reflection
coefficients, and the uncertainty is propagated through both channels as in
\cite{greco_coaxial}. The combined standard uncertainty at each frequency is
\begin{equation}
u(K_f) = \sqrt{u_F^2 + u_G^2 + 2\,\mathrm{Cov}(F,G)
+ u_{\mathrm{dyn}}^2 + u_{\mathrm{SM}}^2},
\label{eq:budget}
\end{equation}
with $u_F$ the contribution of the model parameters, $u_G$ that of the
residuals of the calibrated reflectometer propagated through $G$,
$u_{\mathrm{SM}}$ the source-match contribution evaluated per state,
$u_{\mathrm{dyn}}$ the dynamic accuracy of the analyzer, and
$\mathrm{Cov}(F,G)=r\,u_F u_G$ the covariance between the two channels, with $r$ their correlation coefficient. The
expanded measurement uncertainty is $U = 2\,u(K_f)$ ($k=2$). 
The contributions are evaluated by Monte Carlo propagation: the measured
reflection coefficients are perturbed by the error sources of the calibrated
reflectometer and the whole determination -- parameter fit, $Z_0$ recovery and
$K_f$ -- is repeated over $N=400$ replicas, from whose dispersion the standard
uncertainty at each frequency is obtained. In each replica every source of
Table~\ref{tab:sources} is drawn independently and the three measured states
are perturbed consistently, so that terms common to the states retain their
correlation and only the per-state components contribute as independent. This
carries the covariance between $F$ and $G$ automatically, since both are
recomputed from the same perturbed data in each replica. The source
match enters in two parts: its common part is drawn in the replicas as a
term common to the three states, as corresponds to a single calibration,
and its per-state part, $e_{S,i}\Gamma_i^2$ with $e_{S,i}$ independent
between states, is the column $u_{\mathrm{SM}}$ of
Table~\ref{tab:budget}, evaluated outside the Monte Carlo and added in
quadrature in \eqref{eq:budget}. The per-state part has a different
origin from the residual of the calibration: the three sweeps are taken
in sequence, with the bridge re-set between them, and it covers the
drift of the effective source match between the sweeps; its magnitude is
taken at the specification limit, a deliberate margin.

The reflectometer is an HP~8510C with the WR-42 one-port calibration of
Section~\ref{sec:measproc}. After calibration the measured reflection
coefficient is still mapped from the true one by the residual one-port error
model \cite{rytting}
\begin{equation}
\Gamma_m = e_D + \frac{e_R\,\Gamma}{1-e_S\,\Gamma},
\label{eq:errmodel}
\end{equation}
where $e_D$ is the residual directivity, $e_R$ the reflection tracking and
$e_S$ the residual source match; the measured $\Gamma_m$, not $\Gamma$, is what
enters $G$ and $F$. The effective figures of the calibrated system are
those specified by the manufacturer for the HP~8510C with a load/offset-load
calibration \cite{hp8510c}: $|e_D|=-45$~dB, $|e_S|=-35$~dB, $e_R$ a tracking of
$0.05$~dB (a common multiplicative term), and a random trace noise of
$\sigma_\Gamma=3\times10^{-4}$ per point. These are specification values
referred to the mount flange (the WR-28 case of Section~\ref{sec:wr28}
uses measured residuals); as a control of the calibrated reflectometer,
the reflection coefficient of specimen~5 measured with it agrees with the
value reported with its PTB-traceable calibration (Section~\ref{sec:results}).
In the replicas the three specification terms -- directivity, tracking
and the common part of the source match -- are drawn as complex Gaussian
variables of uniform phase whose rms magnitude equals the specification
value, one realization per replica, common to the three states and to all
frequencies; the trace noise
is drawn as a complex Gaussian variable of standard deviation
$\sigma_\Gamma$, independent between points and between states. In each
replica the parameters are re-determined from the perturbed coefficients
by a local minimization started at the nominal solution, with $\alpha$,
$\beta$ held at their nominal values: the calibration factor does not
depend on them (Appendix~\ref{app:ident}), and re-determining them in
every replica only lengthens the computation without changing the
results; $Z_0$, $F$ and $G$ are recomputed. With $N=400$ replicas the relative standard
error of each standard uncertainty so obtained is about
$1/\sqrt{2N}=3.5\%$ of its value, that is, about $0.02\%$ on the entries
of Table~\ref{tab:budget}. These are the inputs of
the propagation; a realization poorer than the specification would enlarge $U$
without affecting the central values, which depend on the measured
coefficients only, and the adequacy of the resulting coverage is what the
normalized errors of Table~\ref{tab:results} test against independent
references. The way each term of
\eqref{eq:errmodel} propagates is fixed by the structure of $G$: the
directivity $e_D$ cancels identically in $G$, which depends only on differences
of the $\Gamma_i$; the tracking $e_R$, a scale $c$ common to the three
states, propagates in full into $G$, which is homogeneous of degree one in
the $\Gamma_i$: its rms magnitude, $0.4\%$ for $0.05$~dB, is the floor of
$u_G$ in Table~\ref{tab:budget}. It is not, however, the floor of
$u(K_f)$. The conditions of Sections~\ref{sec:determination}
to~\ref{sec:invariant} are not exactly invariant under the scaling,
because the efficiencies of \eqref{eq:etameas} carry the factors
$(1-|c\Gamma_i|^2)$, different for each state, so the parameters respond
and $F$ moves with the scale in the direction opposite to $G$. Propagated
alone, with every other source set to zero, the tracking gives
$u_G=0.39\%$, $u_F=0.26$--$0.29\%$ and $r=-0.995$ to $-0.9999$ across
the band, that is, $u(K_f)=0.10$--$0.13\%$: the scale cancels almost
entirely in the product $F\,G$, and this is why the expanded uncertainties
of Table~\ref{tab:budget} can lie below twice the tracking floor of
$u_G$. In \cite{greco_coaxial} this common scale was reported apart from
$U$; here it is inside the replicas, with its compensation. The source
match $e_S$ survives per state, scaling as $|\Gamma_i|^2$, and forms
$u_{\mathrm{SM}}$; the trace noise forms the remainder of $u_G$.
These figures are collected in Table~\ref{tab:sources}.

\begin{table}[!t]
\caption{Residual error sources of the calibrated WR-42 one-port reflectometer
(HP~8510C), and how each enters the calibration factor.}
\label{tab:sources}
\centering
\normalsize
\begin{tabular}{@{}l c p{3.1cm}@{}}
\hline
Source & Value & Enters as\\
\hline
Directivity        & $-45$~dB            & cancels in $G$\\
Reflection tracking& $0.05$~dB           & common scale, in full into $u_G$, compensated by $F$ ($u(K_f)$ 0.1\%)\\
Source match       & $-35$~dB            & common part in the replicas; per-state part $u_{\mathrm{SM}}$\\
Trace noise        & $3\times10^{-4}$    & random per point, $u_G$\\
Dynamic accuracy   & Table~\ref{tab:dynacc} & level-dependent, $u_{\mathrm{dyn}}$\\
\hline
\end{tabular}
\end{table}

\begin{table}[!t]
\caption{Uncertainty budget of specimen~5 (PTB-traceable reference). All
entries in per cent except $r$; $\mathrm{Cov}(F,G)=r\,u_F u_G$ is in
(per cent)$^2$. $u_F$, $u_G$ and $\mathrm{Cov}$ come from the Monte Carlo,
which includes the common part of the source match; $u_{\mathrm{SM}}$
(per-state part) and $u_{\mathrm{dyn}}$ are evaluated outside it. The expanded
measurement uncertainty is $U=2\,u(K_f)$ with $u(K_f)$ from
\eqref{eq:budget}, which the columns reproduce to within rounding.}
\label{tab:budget}
\centering
\normalsize
\begin{tabular}{c c c c c c c c}
\hline
$f$ & $u_{\mathrm{SM}}$ & $u_F$ & $u_G$ & $\mathrm{Cov}$ & $r$
& $u_{\mathrm{dyn}}$ & $U_{k=2}$\\
(GHz) & (\%) & (\%) & (\%) & (\%$^2$) & & (\%) & (\%)\\
\hline
18.00 & 0.292 & 0.341 & 0.653 & $-0.153$ & $-0.686$ & 0.130 & 1.164\\
19.00 & 0.182 & 0.341 & 0.511 & $-0.105$ & $-0.605$ & 0.110 & 0.918\\
20.01 & 0.138 & 0.342 & 0.458 & $-0.090$ & $-0.575$ & 0.123 & 0.849\\
20.99 & 0.120 & 0.344 & 0.450 & $-0.120$ & $-0.772$ & 0.120 & 0.664\\
22.00 & 0.102 & 0.345 & 0.440 & $-0.126$ & $-0.828$ & 0.119 & 0.586\\
23.00 & 0.074 & 0.347 & 0.410 & $-0.104$ & $-0.732$ & 0.119 & 0.631\\
24.00 & 0.031 & 0.349 & 0.394 & $-0.114$ & $-0.832$ & 0.124 & 0.507\\
25.00 & 0.065 & 0.351 & 0.406 & $-0.108$ & $-0.759$ & 0.121 & 0.601\\
26.01 & 0.171 & 0.353 & 0.495 & $-0.101$ & $-0.578$ & 0.112 & 0.913\\
26.50 & 0.236 & 0.355 & 0.572 & $-0.130$ & $-0.642$ & 0.104 & 1.016\\
\hline
\end{tabular}
\end{table}

The dynamic accuracy is treated separately, and differently from the other
sources, because it is a \emph{level-dependent systematic} rather than a random
term. The receiver linearity of the analyzer is a single curve $U_D(L)$ of
error versus input level $L$ (in dB), the same for the three states; its values
are those specified for the instrument \cite{hp8510c}, rising toward low
level, and are listed in Table~\ref{tab:dynacc}. Because the same curve acts
on all three states, their contributions are not combined in quadrature;
$U_D$ being a specification envelope rather than a curve of known shape,
the contributions are added in magnitude, the bound for errors of unknown
sign at each level \cite{gum2008}:
\begin{equation}
u_{\mathrm{dyn}} = \sum_{i=1}^{3} |S_i|\,
\frac{U_D(L_i)\ln 10}{20\sqrt{3}} ,
\qquad L_i = 20\log_{10}|\Gamma_i|,
\label{eq:udyn}
\end{equation}
where the weights $S_i$ are the sensitivities of $K_f$ to each state. These
weights satisfy the identity $S_1+S_2+S_3=1$ across the whole band, which
serves as a numerical check of the propagation, and they are positive over
most of it, so that \eqref{eq:udyn} differs from the signed sum of a
single curve acting on the three levels by less than $0.15\%$ at any
frequency. In \cite{greco_coaxial} this term was combined in quadrature
between states. Equation \eqref{eq:udyn} propagates the level error
through $G$ alone, without the compensation by the parameter channel that
the replicas provide for the other common terms.

\begin{table}[!t]
\caption{Dynamic-accuracy curve $U_D(L)$ of the analyzer: magnitude error
versus input level, used in \eqref{eq:udyn}.}
\label{tab:dynacc}
\centering
\normalsize
\setlength{\tabcolsep}{4pt}
\begin{tabular}{r c c c c c c c}
\hline
$L$ (dB)       & $-60$ & $-50$ & $-40$ & $-30$ & $-20$ & $-10$ & $0$\\
$U_D$ (dB)    & 0.500 & 0.220 & 0.080 & 0.035 & 0.020 & 0.015 & 0.010\\
\hline
\end{tabular}
\end{table}

The remaining sources were considered and are absent from
Table~\ref{tab:sources} for the following reasons. An error $\Delta R$ in
the resistance entered in the model, with the measured $\Gamma_i$ fixed,
acts on $K_f$ through $\mathrm{Re}\,Z_{th,2}$ in \eqref{eq:CiZ}, with
sensitivity $1/\mathrm{Re}\,Z_{th,2}\approx6\times10^{-3}~\Omega^{-1}$
($\mathrm{Re}\,Z_{th,2}=140$--$170~\Omega$ in the five specimens); this
is not the sensitivity of Appendix~\ref{app:rb}, which refers to a change of
the operating point with the measured coefficients following it. The bias
resistances are determined to $2.4\times10^{-4}$
(Section~\ref{sec:results}), that is to $0.05~\Omega$ at $R_{b,2}$, which
amounts to $0.03\%$ in $K_f$. The dependence of the calibration factor on
the operating point itself -- a change of $R_b$ with the measured
coefficients following it -- is derived in Appendix~\ref{app:rb} from the
three-state geometry. It separates into an efficiency term,
$S=d\ln\eta/dR_b$, which the corrected thermistor impedances of
Section~\ref{sec:passivity} place below $10^{-4}~\Omega^{-1}$ in band
mean, and a
mismatch term, $M_R=d\ln(1-|\Gamma_2|^2)/dR_b$, of $1$--$7\times10^{-4}~\Omega^{-1}$,
which follows from the three measured coefficients alone. For a direct
power measurement made at a bridge resistance displaced by $\Delta R$ from
that of the calibration, the error is $\Delta R/R_b+(S+M_R)\,\Delta R$,
dominated by the scale term of the dc substitution ($0.25\%$ for
$0.5~\Omega$); the value of $R_b$ at which a calibration was performed
should therefore accompany it. No cable is moved
between the three states, the adapter being fitted directly at the test port
(Section~\ref{sec:measproc}). The temperature of the laboratory acts on the
reflection coefficient only through the stability of the calibrated system,
which the residual terms already represent. The uncertainty of the
calibration standards is contained in the effective residual terms of the
calibrated system, which is why they are so called; a separate term would
count it twice. No interpolation of measured quantities, and no numerical
integration, is performed at any stage: every measured quantity is evaluated
on the analyzer grid.
The flange connection is bounded in Appendix~\ref{app:flange}.
The length $L_x$ is not fitted but fixed, and its dimensional uncertainty
does not appear in Table~\ref{tab:sources} because its effect is negligible.
Neither does the dc resistance of the two supporting wires, a real term
common to the three states that \eqref{eq:neutral} does not absorb: for
$25~\mu$m platinum wire it is $0.2~\Omega$ per millimeter, the total wire
length between bar and pivot is about $0.3$~mm (a gap of $0.61$~mm,
Section~\ref{sec:scalingA}, less the bead), so it is about $0.06~\Omega$ and,
with the sensitivity above, $0.04\%$ in $K_f$. A change of $\pm10~\mu$m
in $L_x$ moves the calibration factor by at most $8\times10^{-3}\%$ over
the band, two orders of magnitude below the expanded uncertainty. The
reason is structural. In \eqref{eq:skinseries} $L_x$ enters, to the order
that matters, through the internal reactance $j\,\omega\mu_0L_x/8\pi$ in
series with each branch; after the parallel combination \eqref{eq:zth} it
reaches the three thermistor impedances as a reactance that is nearly the
same for the three states, and a common reactance is precisely the neutral
transformation \eqref{eq:neutral}, which leaves $\eta_i$ invariant. Only
the small non-common remainder acts on $K_f$.

Two features distinguish this budget from the coaxial case. There is no
grid-truncation term: the parameter determination is a direct minimization
without a grid (Section~\ref{sec:minimization}), so no such contribution
enters; the conditioning of the minimum is examined in
Appendix~\ref{app:ident}. And the covariance between $F$ and $G$,
estimated in the coaxial work from the trace noise alone, is here obtained
from the complete error model, with $u_F$, $u_G$ and $\mathrm{Cov}(F,G)$
drawn from the same set of replicas; it is moderate to strong, negative and
frequency-dependent, with correlation coefficients reaching
$r\approx-0.8$, and it is retained with its sign in \eqref{eq:budget}. The residual figures of the
calibrated reflectometer that feed $u_G$, $u_{\mathrm{SM}}$ and
$u_{\mathrm{dyn}}$ are those of the WR-42 one-port calibration of
Section~\ref{sec:measproc}.

The strong negative correlation is a consequence of the method. It is not
accidental: the two channels are not
independent, because both are functions of the same measured coefficients.
$G$ depends on them directly, while $F=F(X)$ depends on them through the
parameters, since $X=X(\Gamma_1,\Gamma_2,\Gamma_3)$ is itself obtained from the
minimization over the same three states. A perturbation of a measured
$\Gamma_m$ by \eqref{eq:errmodel} therefore enters $K_f=F\,G$ along two paths at
once, and the two responses are correlated by construction. Because the
coupling is anticorrelated, the errors of the two channels partly cancel:
with $r<0$ the cross term $2\,\mathrm{Cov}(F,G)$ in \eqref{eq:budget} is
negative and reduces the combined uncertainty. The effect is large. Had $F$
and $G$ been treated as independent, the expanded uncertainty of specimen~5
would range from about $1.1\%$ to $1.6\%$ instead of the
$0.5\%$--$1.16\%$ of Table~\ref{tab:budget}; at mid-band the covariance
roughly halves it. The correlation is a property of the three-state
construction and not of the self-consistent determination of $Z_0$:
repeating the propagation with $Z_0$ held fixed at its nominal value, so
that the perturbed coefficients do not re-enter through the thermistor-port
impedance, changes $r$ by at most $0.11$ and $U$ by at most $0.09\%$.

The structure of the functional is what makes the parameter channel
insensitive to the error sources of the reflectometer. Its three terms are
band averages, and the parameters $X$ are common to the whole band: each of
them is supported by every measured point, so a perturbation that varies
across the band -- trace noise, the frequency-dependent part of the residual
terms -- is averaged over the $N_f$ points before it reaches $X$, and its
effect on $F$ is reduced by the band average, in the way the small and nearly
flat $u_F$ of Table~\ref{tab:budget} shows. A perturbation common to the
whole band is not averaged out; it is for these, directivity and tracking,
that the structure of $K_f=F\,G$ decides: directivity cancels identically in
$G$, and tracking enters $G$ in full and is compensated through $F$
(Table~\ref{tab:sources}).

Equation \eqref{eq:budget} propagates the errors of the measurement
through the method: $U$ is the uncertainty of the result given the
thermistor model, and does not contain the error of form of the model
itself. The adequacy of the model is established by the comparison with
the references of Section~\ref{sec:results}, for the specimens tested;
the scope so demonstrated is that of that comparison, to be extended by
the validation of Section~\ref{sec:conclusion}.

Table~\ref{tab:budget} gives the budget of specimen~5, referenced to the
PTB-traceable values; it is the specimen of lowest expanded uncertainty. The
dominant contribution is $u_G$, the reflectometer channel; $u_F$ is nearly
flat across the band, and the source-match term $u_{\mathrm{SM}}$ falls toward
mid-band where the mount is matched. For this specimen the expanded
measurement uncertainty ranges from about $0.5\%$ at
mid-band to $1.16\%$ at the lower edge. The same propagation was
carried out for the other four; the expanded measurement uncertainty ranges from
$0.7\%$ to $1.3\%$ for specimen~4, $0.8$--$1.5\%$ for specimen~1,
$0.9$--$1.8\%$ for specimen~2 and $1.0$--$1.9\%$ for specimen~3, the larger
values of the last two being produced by $u_G$ at the frequencies where the
mount is least well matched and the covariance credit is smallest
($r\approx-0.1$ to $-0.3$); Table~\ref{tab:u1234} lists their values at
the frequencies of the reference tables. The uncertainty depends on the
measured values and on the mount itself, through the match of its three
states, so it cannot be stated a priori for an unknown specimen. What the
laboratory can state is the best measurement uncertainty obtained so far
with the method, in the sense of ISO/IEC~17025 \cite{iso17025}: it is that
of specimen~5, $U\le1.16\%$ ($k=2$) across the band, under the
measurement conditions, with the equipment available and with the present
thermistor model. As an upper bound, the largest expanded uncertainty
obtained in the five mounts is below $2\%$ at every frequency. Neither
figure is an estimate of the error of a particular
determination, which the comparisons of Section~\ref{sec:results} show to
be smaller.

\begin{table}[!t]
\caption{Expanded uncertainty $U$ ($k=2$, per cent) of the calibration
factor of specimens~1--4, from the same propagation as Table~\ref{tab:budget},
at the frequencies of the reference tables; these are the values used in the
normalized errors of Table~\ref{tab:results} and in Table~\ref{tab:etaall}.}
\label{tab:u1234}
\centering
\normalsize
\begin{tabular}{c c c c c}
\hline
$f$ (GHz) & 1 & 2 & 3 & 4\\
\hline
18.0 & 1.28 & 0.91 & 1.39 & 1.33\\
19.0 & 0.79 & 0.94 & 1.29 & 0.93\\
20.0 & 1.21 & 1.44 & 1.71 & 0.68\\
21.0 & 1.48 & 1.41 & 1.51 & 0.99\\
22.0 & 1.29 & 1.10 & 1.40 & 0.84\\
23.0 & 1.54 & 1.52 & 1.91 & 1.01\\
24.0 & 1.53 & 1.33 & 1.59 & 1.12\\
25.0 & 1.00 & 0.97 & 1.02 & 0.76\\
26.5 & 1.11 & 1.83 & 1.68 & 0.87\\
\hline
\end{tabular}
\end{table}

\begin{figure}[!t]
\centering
\includegraphics[width=0.95\textwidth]{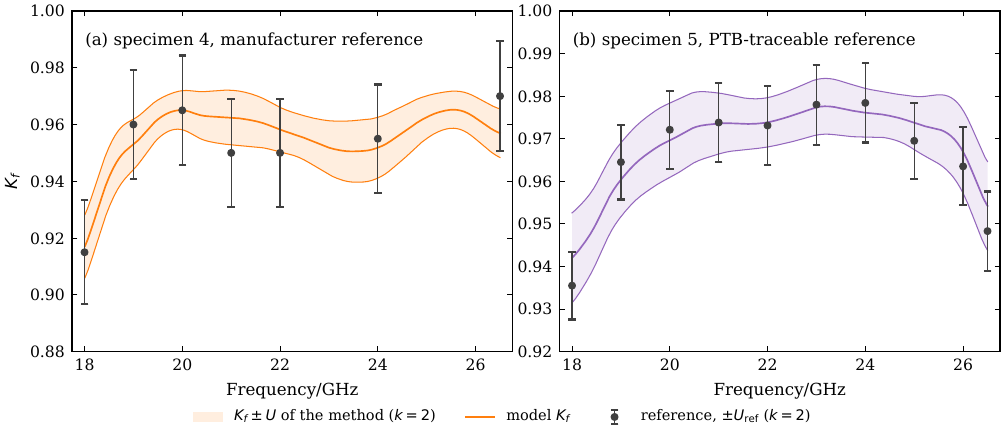}
\caption{Calibration factor with the expanded-uncertainty band of the method.
(a)~Specimen~4 against the manufacturer table ($U=2\%$, $k=2$).
(b)~Specimen~5 against the PTB-traceable reference ($U=0.85$--$0.98\%$, $k=2$).
Solid curve: model $K_f$, smoothed as in Fig.~\ref{fig:kf1234}; shaded band:
$K_f\pm U$ ($k=2$) from \eqref{eq:budget} at every point of the analyzer grid;
points with bars: reference values. Vertical scales differ between the panels.}
\label{fig:kfband}
\end{figure}

Figure~\ref{fig:kfband} shows the calibration factor of specimens~4 and~5
with the expanded-uncertainty band of the method, together with the reference
points and their own bars. This is the complete result of the method for a
specimen: a curve at every point of the analyzer grid with its $U$ at each
point. For specimen~4 the band of the method is narrower than the $U=2\%$
bars of the manufacturer; for specimen~5 it is comparable to the $U=0.85$--$0.98\%$
bars of the PTB-traceable reference, and the two overlap at every frequency.

\section{Transfer Through Waveguide-to-Coaxial Adapters}
\label{sec:adapters}
Waveguide thermistor mounts are frequently used as transfer standards in a
coaxial environment, the mount being connected through a
waveguide-to-coaxial adapter. The quantity of interest is then the
calibration factor of the \emph{assembly} -- mount plus adapter -- seen from
the coaxial port, since that is the object that actually dissipates the
incident power in use. The conventional route obtains it indirectly: the
adapter is characterized separately as a two-port, and its scattering matrix
is de-embedded from a measurement of the assembly, or cascaded onto the
calibration factor of the mount alone. Both forms require an additional
measurement -- the two-port characterization of the adapter -- and propagate
its uncertainty into the result.

The present method obtains the same quantity directly, and without
characterizing the adapter at all. The calibration factor factors as
$K_f = F(X)\,G$, where $F$ carries the model parameters $X$ and $G$ is the
explicit function of the measured reflection coefficients of
\eqref{eq:etameas}. The model factor $F$ is a property of the thermistor
mount: it does not change when an adapter is placed ahead of it, because the
adapter alters only the reflection coefficients presented at the coaxial
port, not the thermistor and its mount. It is therefore computed once, from
the determination of $X$ on the mount alone. The assembly is then measured as
it is used -- the three states $\Gamma_i'$ of mount-plus-adapter, read at the
coaxial port -- and the geometric factor $G'$ is formed from them. The
calibration factor of the assembly is
\begin{equation}
K_f' = F(X)\,G' ,
\label{eq:kftransfer}
\end{equation}
with $F$ unchanged. The multiple reflections between adapter and mount are
contained in the measured $\Gamma_i'$ and enter $G'$ as part of the data; the
scattering matrix of the adapter is never needed, and no de-embedding is
performed. The procedure is thus identical in form to the calibration of the
mount, applied to the assembly.

Although \eqref{eq:kftransfer} is algebraically the same operation as the
calibration of the mount, its uncertainty is not. In the calibration of the
mount, $F$ and $G$ are formed from the \emph{same} measured coefficients, and
the error model \eqref{eq:errmodel} enters both channels along the two paths
of Section~\ref{sec:budget}; the resulting covariance is strongly negative and
reduces the combined uncertainty through the cross term of
\eqref{eq:budget}. In the transfer, $F$ is carried over from the measurement
of the mount while $G'$ is formed from a separate measurement of the assembly,
taken in a different connection. The perturbations of the two channels are then
no longer drawn from the same data; the residuals of two different
calibrations, in waveguide and in 3.5~mm, are taken as independent, so
their covariance is set to zero: the cross
term $2\,\mathrm{Cov}(F,G')$ that offset part of the variance in
\eqref{eq:budget} is lost, and the corresponding term carried formally in
the transfer budget of \cite{greco_coaxial} is zero in this step; it
survives only when $F$ and $G$ are formed from the same measurement. The combined uncertainty of the assembly becomes
\begin{equation}
u^2(K_f') = u_F^2 + u_{G'}^2 + u_{\mathrm{dyn}}^2 ,
\label{eq:budgettransfer}
\end{equation}
with no covariance credit. The reflectometer residuals of the two
measurements are those of two different calibrations, in waveguide and in
3.5~mm, and are not common to them; the instrument terms that are common
enter through $u_{\mathrm{dyn}}$. A residual correlation $r'$ between the
two channels would add $2r'u_Fu_{G'}$ to \eqref{eq:budgettransfer}: at
23~GHz, $r'=\pm0.3$ moves $U$ from $2.20\%$ to $2.45$ or $1.92\%$. This
suppression of the covariance -- not any
excess noise in $G'$ -- is the principal increase of uncertainty incurred in
the transfer.

\begin{figure}[!t]
\centering
\includegraphics[width=0.62\textwidth]{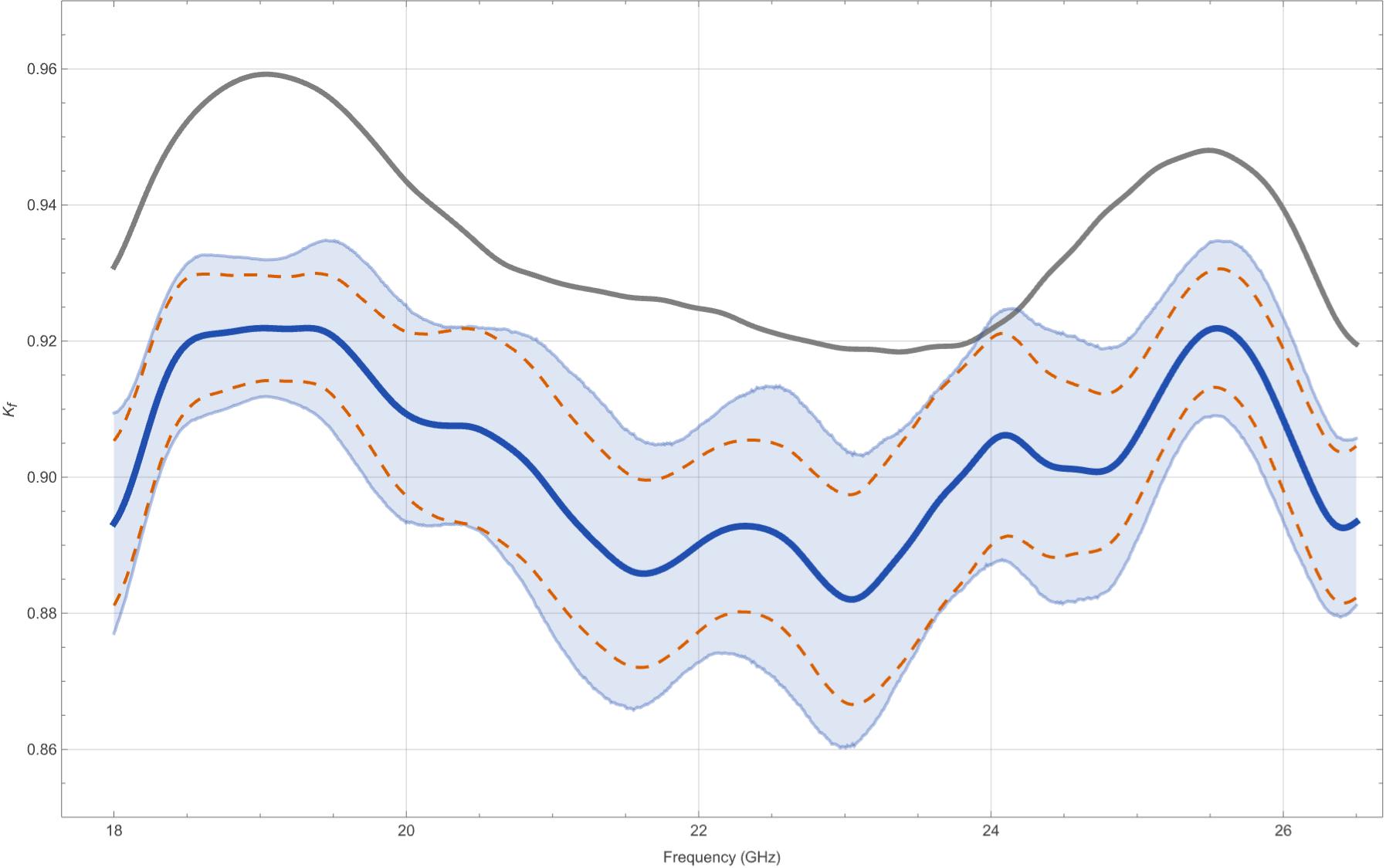}
\caption{Calibration factor of the assembly of specimen~1 and its
waveguide-to-coaxial adapter, measured in 3.5~mm coaxial (blue), with its
expanded-uncertainty band ($k=2$, filled). The dashed lines mark the narrower
band that would follow from the covariance-retaining budget of the mount; the
filled region beyond them is the covariance credit lost in the transfer. The
upper curve (gray) is the calibration factor of the mount alone, in waveguide,
shown for reference: the assembly lies below it by the transmission loss of the
adapter, not by any error. Curves are lightly smoothed for display; the budget
is computed on the unsmoothed data.}
\label{fig:transfer}
\end{figure}

The assembly -- specimen~1 with an HP~K281C WR-42 to 3.5~mm adapter -- is
measured in a 3.5~mm coaxial environment, with the one-port
calibration of the analyzer performed there using a 3.5~mm sliding-load short-open-load-thru (SOLT)
kit (HP~85052B \cite{hp85052b}); the
model factor $F$, and with it $u_F$, is carried over unchanged from the
waveguide determination of the mount. The channel $u_{G'}$ is propagated from
the residual error figures of the coaxial calibration -- effective directivity
$-44$~dB, source match $-31$~dB and reflection tracking $0.006$~dB, taken from
the sliding-load specification of the kit \cite{hp85052b} and applied
conservatively across the band --
with the same structure as in Section~\ref{sec:budget}: the directivity, common
to the three states, cancels in the differences that form $G'$; the source
match survives per state and dominates $u_{G'}$; the trace noise enters as a
random term. The dynamic-accuracy term $u_{\mathrm{dyn}}$ uses the curve of
Table~\ref{tab:dynacc} with the levels and sensitivities of the assembly.
Table~\ref{tab:transferbudget} gives the resulting budget. The combined
expanded measurement uncertainty lies between $1.0\%$ and $2.2\%$ ($k=2$),
$1.7\%$ on average, against the $0.8$--$1.5\%$ of the mount alone; the
increase is
accounted for almost entirely by the loss of the covariance credit, $u_F$
being that carried over from the mount determination and $u_{G'}$ of the
same magnitude as the reflectometer channel of the mount. The variation of
$u_{G'}$ along the band follows the ripple of the match of the assembly,
through the $|\Gamma'_i|^2$ scaling of the source-match term.
Figure~\ref{fig:transfer} shows the calibration factor of the assembly with
its expanded-uncertainty band, together with the narrower band the method
would yield if the covariance were retained; the separation between the two is
the covariance credit that the transfer forgoes.

\begin{table}[!t]
\caption{Uncertainty budget of the transfer, assembly of specimen~1 and its
waveguide-to-coaxial adapter measured in 3.5~mm coaxial. All entries in per
cent. The covariance between $F$ and $G'$ is zero because the two channels
come from separate measurements, unlike the covariance column of
Table~\ref{tab:budget}.}
\label{tab:transferbudget}
\centering
\normalsize
\begin{tabular}{c c c c c c}
\hline
$f$ & $u_F$ & $u_{G'}$ & $u_{\mathrm{dyn}}$ & $\mathrm{Cov}$ & $U_{k=2}$\\
(GHz) & (\%) & (\%) & (\%) & & (\%)\\
\hline
18.00 & 0.462 & 0.659 & 0.225 & 0 & 1.671\\
19.00 & 0.463 & 0.179 & 0.141 & 0 & 1.032\\
20.00 & 0.467 & 0.642 & 0.124 & 0 & 1.607\\
20.99 & 0.473 & 0.904 & 0.103 & 0 & 2.051\\
22.00 & 0.480 & 0.710 & 0.122 & 0 & 1.731\\
23.00 & 0.487 & 0.974 & 0.160 & 0 & 2.201\\
24.00 & 0.491 & 0.737 & 0.111 & 0 & 1.785\\
25.00 & 0.493 & 0.586 & 0.164 & 0 & 1.566\\
26.50 & 0.493 & 0.344 & 0.118 & 0 & 1.225\\
\hline
\end{tabular}
\end{table}

The curve of Fig.~\ref{fig:transfer} shows a ripple that the calibration
factor of the mount alone does not have. The adapter and the mount form a
cascade in which the
multiple reflections between the coaxial transition and the matching
structure are contained in the measured $\Gamma_i'$, and they reach $K_f'$
through $G'$ with their actual frequency structure. The conventional route
reproduces the same ripple when the scattering matrix of the adapter,
measured in a separate connection, is cascaded in full onto the reflection
coefficients of the mount; it does so, however, only to the extent that the
two measurements are mutually consistent in reference planes and
connection, and any inconsistency between them appears as a distorted
ripple. The curve of Fig.~\ref{fig:transfer} is the response of the
assembly as it is actually used, measured as such, with no second
measurement to be reconciled.

A quantitative comparison of this direct transfer against the conventional
two-port characterization of the adapter -- the systematic differences it
introduces and their verification against an independent power reference --
involves considerations beyond the scope of this paper and is left to a
dedicated study.

\section{Further Analysis}
\label{sec:scaling}
\subsection{The Implied Two-Port and the State Correction}
\label{sec:passivity}
The bilinear map \eqref{eq:gamma}, fixed at every frequency by the three
pairs $(\Gamma_{Li},\Gamma_i)$ of the solution, is the scattering
matrix of the matching structure referred to the real impedance $Z_0$ at
the thermistor port. The method does not use it. The expression
\eqref{eq:eta} for the efficiency, although written here with the elements
of $\mathbf S$, is the result of Kerns \cite{kerns1949}: it follows from
the bilinear relation between the reflection coefficients at the two
ports and does not require the scattering matrix to be known; the
calibration factor, the thermistor-port impedance and the efficiency of the
operating state are obtained without it, and have been verified against
the calorimetric references in Section~\ref{sec:results}. The implied
matrix is nevertheless a physical statement about the mount, and it has to
be examined. With the thermistor model of Section~\ref{sec:circuit} the
two-port so obtained is not passive in any of the five specimens: the
largest singular value $\sigma_{\max}$ of $\mathbf S$ exceeds unity over
most of the band, by $12$--$14\%$ at its maximum and $9$--$10\%$ on
average (Table~\ref{tab:passivity}), $S_{22}$ having a magnitude of
$0.1$--$0.4$ and a phase near $90^\circ$ throughout. Equivalently, the map
$T$ of \eqref{eq:gamma}, which contracts the pseudo-hyperbolic distance at
the three measured load points ($\eta_i<1$), expands it elsewhere in the
disk: its Schwarz--Pick factor \eqref{eq:lambdabody}, the efficiency the
structure would show with an arbitrary passive load, exceeds unity on part
of the unit disk.

The origin is in the two degrees of freedom of Section~\ref{sec:invariant}.
The efficiencies of the three states as returned by the model of
Section~\ref{sec:circuit} differ by $5$--$9\%$ in band mean, and the
closure $k_{i,j}=1$ at a real reference requires, by \eqref{eq:kidentity},
the magnification ratio $|T'(\Gamma_{Li})|/|T'(\Gamma_{Lj})|$ to compensate
that spread; with nearly real loads this is produced only by an $S_{22}$ of
a few tenths at a phase near $90^\circ$, and such an $S_{22}$ together with
$|S_{21}|^2\simeq\eta_2$ exceeds the bound $|S_{22}|^2+|S_{21}|^2\le1$. The
defect cannot be removed by anything the calibration factor does not see:
the isometries of Section~\ref{sec:invariant} are lossless elements in
cascade, which leave the passivity of the whole unchanged, and so does a
change of reference, real or complex. What can change it is only a
displacement within $\mathcal M_F$ along its two shape coordinates -- a
modification of the thermistor impedances that differs between the
states, that is, of the bead model, which in Section~\ref{sec:circuit} has
no element depending on the bias beyond $R_b$.

\subsubsection{Formulation}
That displacement is parametrized multiplicatively,
\begin{equation}
Z_{th,i}' = m_i\,Z_{th,i},\qquad m_i=e^{\,s_i+j\theta_i},\qquad i=1,2,3,
\label{eq:mult}
\end{equation}
a scale $s_i$ and a rotation $\theta_i$ of the impedance of each state;
below, $s_i$ is quoted in per cent, $100\,s_i$, so that $s_i=1\%$ means
$|m_i|=e^{0.01}$, a change of modulus of $1\%$ to first order, and
$\theta_i$ in degrees.
The form is the one a change in the operating point of the bead produces:
a change in the properties of the material -- conductivity and loss angle
-- multiplies the impedance of the branch and does not add an element to
it, and a change of loss angle in a medium whose dispersion has constant phase,
as the $\omega^{-1/2}$ law of Section~\ref{sec:circuit}, rotates the impedance by an angle independent
of frequency. In this basis, moreover, the common scale $s_1=s_2=s_3$ is
the real transformer of \eqref{eq:neutral}, the rotations lie almost
entirely within $\mathcal M_F$ (to first order they enter $F$ only through
$\tan\arg Z_{th,i}$, which is $0.06$--$0.32$ for the central state and
below $0.44$ in all), and the remaining
constraint on the scales is linear. An additive form, $Z_{th,i}+\delta
Z_i$, is locally equivalent but would represent an element in series with
the bead; no lossless element, lumped or distributed, has a reactance
constant over a band (Foster's theorem), and the solution below has
rotations constant over the band, which in the additive form would be a
reactance proportional to the impedance itself, not an element.

Nor can the correction be absorbed into a different fixed network for the
bead. A network in which $R_b$ is the only state-dependent element is a
fixed two-port terminated in $R_b$; its input impedance is bilinear in
$R_b$, and three states determine that two-port uniquely at each
frequency, so that any triple of impedances is the image of the three
bias resistances under some fixed network, and the question is only
whether that network is passive. For the impedances of the model of
Section~\ref{sec:circuit} it is, the bead network itself being recovered;
for the corrected impedances it is not, and neither is it for any triple
with $K_f$ fixed and a passive mount over 18--26.1~GHz in the two
specimens examined (Appendix~\ref{app:fixednet}). Passivity can be had in
the bead or in the mount, not in both with fixed elements: the
measurements require an element that changes with the state, which is
what \eqref{eq:mult} represents and what no network with $R_b$ as its only
state-dependent element was found to provide in the specimens examined.

The correction sought is the smallest one that brings the implied two-port
to the floor of passivity while leaving the result of the method
unchanged:
\begin{equation}
\begin{gathered}
\min_{s_i,\theta_i}\ \sum_i\left(s_i^2+\theta_i^2\right)\\
\text{subject to}\quad F(m_iZ_{th,i})=F(Z_{th,i}),\\
\sigma_{\max}\!\left[\mathbf S(m_iZ_{th,i})\right]\le\sqrt{\eta_2}+\delta_\sigma\,,\qquad
\delta_\sigma=10^{-3},
\end{gathered}
\label{eq:sp2}
\end{equation}
solved at every frequency, with $Z_0$ then recomputed from the pair
closure. The equality keeps $K_f$ and $\eta_2$ unchanged, to the
tolerance of the solver (below $3\times10^{-3}\%$ in $K_f$). The square
root of the objective, $\|m-1\|=(\sum_i s_i^2+\theta_i^2)^{1/2}$, is the
logarithmic norm of the correction and the measure of its size used
below; since $|m_i-1|^2=(e^{s_i}-1)^2+4e^{s_i}\sin^2(\theta_i/2)
=(s_i^2+\theta_i^2)(1+s_i)$ up to fourth-order terms, it differs from
$(\sum_i|m_i-1|^2)^{1/2}$ by about $0.1\%$ of its value in band mean for
the corrections found. The
bound on $\sigma_{\max}$ is the floor of what the efficiency of the
operating state allows. Applying the definition of the largest singular
value to the excitation of that state -- $a_1=1$, $b_1=\Gamma_2$,
$b_2=S_{21}/(1-S_{22}\Gamma_{L2})$, $a_2=\Gamma_{L2}b_2$ -- and the power
balance $\eta_2=(|b_2|^2-|a_2|^2)/(1-|\Gamma_2|^2)$,
\begin{equation}
\begin{split}
\sigma_{\max}^2\ \ge\ q_2&\equiv\frac{|\Gamma_2|^2+|b_2|^2}{1+|a_2|^2}\\
&=\eta_2+(1-\eta_2)\,\frac{|\Gamma_2|^2+|a_2|^2}{1+|a_2|^2}\ \ge\ \eta_2 ,
\end{split}
\label{eq:floorbound}
\end{equation}
exactly, for any matrix compatible with the operating state: $\sqrt{q_2}$
is the floor, the lowest value of $\sigma_{\max}$ -- the largest power gain
of the two-port over all excitations -- compatible with the efficiency of
the operating state; at the floor the two-port attenuates every excitation
by the same factor, and the efficiency depends on the load only through
the mismatch factor of \eqref{eq:attlaw} below. Lowering $\sigma_{\max}$
does not lower the dissipation of the structure: a passive matrix with
$\sigma_{\max}=1$ transmits one excitation without loss and others with
more, and what the floor selects is the representation in which the loss
is the same in every direction. The excess
of $\sqrt{q_2}$ over $\sqrt{\eta_2}$ is of the order of
$(1-\eta_2)|\Gamma_2|^2$ -- below $5\times10^{-4}$ in the five specimens,
since the self-consistent $Z_0$ keeps $|\Gamma_{L2}|$ small ($0.05$--$0.19$
in band mean) -- and is covered by the tolerance $\delta_\sigma$ of
\eqref{eq:sp2}. The solutions of \eqref{eq:sp2} reach the floor to that
tolerance, and they take a definite form: for a $2\times2$ matrix
$\sigma_{\max}\sigma_{\min}=|\det\mathbf S|$, and the solutions show
$|\det\mathbf S|=q_2$ to within $2\times10^{-4}$, so that their two
singular values are equal and
\begin{equation}
\mathbf S=\sqrt{q_2}\,\mathbf U,\quad \mathbf U\ \text{unitary},
\label{eq:floor}
\end{equation}
with $q_2=\eta_2$ up to the term of \eqref{eq:floorbound}; in what follows
$\sqrt{\eta_2}$ stands for $\sqrt{q_2}$, the two differing by less than
$5\times10^{-4}$.
At the floor the matching structure is a lossless reciprocal two-port in
cascade with a matched attenuator of power loss $q_2\approx\eta_2$ when its output
reflection vanishes: the reference structure of
Section~\ref{sec:convention} with the attenuator added at its output. To first order in $|S_{11}|^2(1-\eta_2)$ the
map $T$ is an isometry of the disk composed with the radial contraction
$\Gamma_L\to\eta_2\Gamma_L$, and its Schwarz--Pick factor is the efficiency
of a matched attenuator at every load,
\begin{equation}
\eta(\Gamma_L)=\eta_2\,\frac{1-|\Gamma_L|^2}{1-\eta_2^2|\Gamma_L|^2} ,
\label{eq:attlaw}
\end{equation}
with maximum $\eta_2$ at the center. Equal efficiencies of the three
states, up to the mismatch factor of \eqref{eq:attlaw} and to terms of
that order, are thus what the floor of \eqref{eq:sp2} produces. The floor
is a choice: that of the representation in which the efficiency depends
least on the load, the most natural one for a structure of which only
$K_f$ and $\eta_2$ are measured; passivity alone ($\sigma_{\max}\le1$) would be
reached by a smaller correction that does not equalize them -- for the five
specimens, $\|m-1\|$ in band mean is $1.4$--$1.8\%$ with
$\sigma_{\max}\le1$ against $1.5$--$2.1\%$ at the floor, $K_f$ and
$\eta_2$ being unchanged in either case. Nor
can $\sigma_{\max}$ be lowered further: below $\sqrt{q_2}$, by
\eqref{eq:floorbound}, no matrix is compatible with the efficiency of the
operating state.
The form \eqref{eq:floor} refers to the real $Z_0$ of the convention
$k_{i,j}=1$ at the thermistor port, as do the factors $m_i$. Problem
\eqref{eq:sp2} returns, at each frequency, the same solution from every
start; without the
minimum-norm selection the floor would be attained on a two-parameter set
of configurations, which the calibration factor and the passivity of the
two-port do not distinguish and of which the minimum-norm selection picks
one member.

\subsubsection{Results}
\begin{figure}[!t]
\centering
\includegraphics[width=0.95\textwidth]{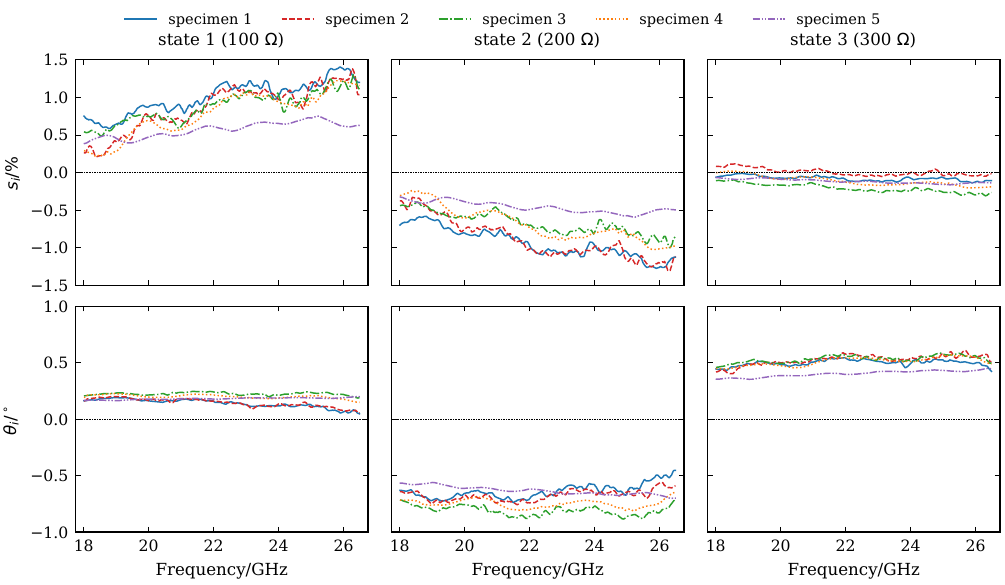}
\caption{State correction \eqref{eq:sp2} for the five specimens: scales
$s_i(f)$ (top) and rotations $\theta_i(f)$ (bottom) of the three states,
with a common scale for each row; lightly smoothed for display (5-point
median and 7-point moving average).}
\label{fig:stheta}
\end{figure}

Table~\ref{tab:passivity} gives, for the five specimens, $\sigma_{\max}$
with the model of Section~\ref{sec:circuit}, the size of the correction
\eqref{eq:sp2}, its components, and the quantities that change and those
that do not. Figure~\ref{fig:stheta} shows $s_i(f)$ and $\theta_i(f)$.
Table~\ref{tab:mbudget} gives the uncertainty budget of the correction for
specimen~5. The error sources of Section~\ref{sec:budget} are propagated
through the solution of \eqref{eq:sp2}: the residuals of \eqref{eq:errmodel}, drawn as in
Section~\ref{sec:budget}, coherent over the band and common to the three
states, together with the trace noise, by Monte Carlo over $N=108$
replicas; the per-state source match, as an independent term
$e_{S,i}\Gamma_i^2$ in each state, by Monte Carlo; the dynamic accuracy of
Table~\ref{tab:dynacc}, as a level error applied to each state in turn
with the responses added in magnitude and divided by $\sqrt3$ as in
\eqref{eq:udyn}, and its phase counterpart, $0.2^\circ$ at the levels of
the three states \cite{hp8510c}, treated in the same way; the bias
resistances, to $2.4\times10^{-4}$ (Section~\ref{sec:results}); the
tolerance of the floor in \eqref{eq:sp2}; and the parameters $X$, through
the replicas of the propagation of Section~\ref{sec:budget} -- their
relative dispersion ($C_x$ $4\%$, $R_{p0}$ $2.3\%$, $\kappa$ $1.8\%$, with
their correlations) applied to the solution of Table~\ref{tab:params},
each replica re-solved for the correction with the measured coefficients.
The quantities reported are averages over frequency, the band means of
Table~\ref{tab:passivity} and the means over thirds of the band in the
lower part of Table~\ref{tab:mbudget}, the natural form for functions
that vary as little with frequency as $s_i$ and $\theta_i$ do; the
residuals of the calibration and the per-state source match, which
dominate the uncertainty at a single frequency point, produce a slow
pattern across the band that averages out, as the directivity and the
tracking do in $G$. The combined standard uncertainties of the band means
are $0.09$, $0.07$, $0.02\%$ and $0.01^\circ$, $0.04^\circ$,
$0.02^\circ$, about a sixth of the values of the scales and a tenth or
better of those of the rotations, and $0.02$--$0.12\%$ and
$0.01$--$0.05^\circ$ for the thirds; the parameter channel displaces the
whole curve $s_i(f)$ without changing its shape. For specimen~4 the same
propagation gives uncertainties of the same size, over thirds in which
the scales are $+0.45$, $+0.87$, $+1.03\%$ ($s_1$) and $-0.43$, $-0.75$,
$-0.88\%$ ($s_2$) and the rotations $-0.73$, $-0.77$, $-0.75^\circ$
($\theta_2$).

\begin{table}[!t]
\caption{The implied two-port with the thermistor model of
Section~\ref{sec:circuit} and with the state correction \eqref{eq:sp2}.
$\sigma_{\max}$: largest singular value of $\mathbf S$, maximum (band
mean); $\|m-1\|$: size of the correction, band mean (maximum);
$\theta_i$ and $s_i$: band means of the rotations (degrees) and scales
(per cent); $\sqrt{\eta_2}$: band mean; Dif: rms closure defect (as a fraction) of the
pairs; $\langle\alpha'\rangle$, $\langle\beta'\rangle$: band averages of
\eqref{eq:abf} with the correction. The calibration factor changes by less
than $3\times10^{-3}\%$ at every frequency.}
\label{tab:passivity}
\centering
\scriptsize
\setlength{\tabcolsep}{2.5pt}
\resizebox{\textwidth}{!}{%
\begin{tabular}{c c c c c c c c c}
\hline
Spec. & $\sigma_{\max}$, model & $\|m-1\|$ (\%) & $\theta_1,\theta_2,\theta_3$ & $s_1,s_2,s_3$ & $\sigma_{\max}$, corrected & $\sqrt{\eta_2}$ & Dif & $\langle\alpha'\rangle,\langle\beta'\rangle$\\
 & max (mean) & mean (max) & ($^\circ$) & (\%) & max (mean) & & model / corr. & \\
\hline
1 & 1.116 (1.086) & 2.01 (2.51) & $+0.14$, $-0.64$, $+0.50$ & $+1.00$, $-0.93$, $-0.08$ & 0.988 (0.981) & 0.980 & 0.070 / 0.037 & 0.999, 0.996\\
2 & 1.123 (1.088) & 1.99 (2.63) & $+0.14$, $-0.68$, $+0.52$ & $+0.87$, $-0.89$, $+0.01$ & 0.987 (0.979) & 0.977 & 0.100 / 0.041 & 0.993, 0.994\\
3 & 1.137 (1.100) & 2.08 (2.78) & $+0.23$, $-0.81$, $+0.53$ & $+0.87$, $-0.67$, $-0.20$ & 0.991 (0.984) & 0.983 & 0.067 / 0.047 & 0.997, 0.996\\
4 & 1.123 (1.100) & 1.96 (2.34) & $+0.20$, $-0.75$, $+0.51$ & $+0.79$, $-0.68$, $-0.11$ & 0.989 (0.984) & 0.983 & 0.049 / 0.028 & 0.999, 0.997\\
5 & 1.118 (1.088) & 1.54 (1.76) & $+0.18$, $-0.63$, $+0.40$ & $+0.57$, $-0.46$, $-0.11$ & 0.992 (0.990) & 0.989 & 0.029 / 0.017 & 0.997, 0.997\\
\hline
\end{tabular}}
\end{table}

\begin{table}[!t]
\caption{Uncertainty budget of the state correction \eqref{eq:sp2} for
specimen~5. Upper part: standard uncertainty of the band means of the
scales $s_i$ (in $10^{-3}$ per cent, i.e. $10^{-5}$ in $|m_i|$) and
rotations $\theta_i$ (in millidegrees) reported in
Table~\ref{tab:passivity}, by source; Monte Carlo terms over $N=108$
replicas, the trace-noise line being contained in the line above it and
not combined again; dynamic-accuracy terms are bounds, magnitudes added
over the three states and divided by $\sqrt3$; the bias resistances are
combined in quadrature over the states; the parameter line is the
dispersion of the replicas of Section~\ref{sec:budget} propagated through
the correction. Lower part: means over thirds of the band with their
combined standard uncertainty, in per cent and degrees.}
\label{tab:mbudget}
\centering
\scriptsize
\setlength{\tabcolsep}{3pt}
\resizebox{\textwidth}{!}{%
\begin{tabular}{l c c c c c c}
\hline
Source & $s_1$ & $s_2$ & $s_3$ & $\theta_1$ & $\theta_2$ & $\theta_3$\\
 & \multicolumn{3}{c}{($10^{-3}$ per cent)} & \multicolumn{3}{c}{(mdeg)}\\
\hline
Residuals \eqref{eq:errmodel} common to the states, with trace noise & 4.3 & 3.2 & 1.1 & 0.6 & 1.8 & 1.2\\
\quad of which trace noise alone & 3.7 & 2.8 & 0.9 & 0.4 & 1.2 & 0.8\\
Source match, per state & 6.8 & 5.3 & 1.5 & 0.7 & 2.5 & 1.5\\
Dynamic accuracy, level (Table~\ref{tab:dynacc}) & 64.7 & 49.4 & 15.4 & 2.4 & 3.5 & 1.8\\
Dynamic accuracy, phase ($0.2^\circ$) & 21.2 & 15.3 & 5.8 & 11.6 & 38.8 & 24.0\\
Bias resistances ($2.4\times10^{-4}$) & 26.4 & 19.8 & 6.6 & 1.2 & 1.3 & 0.5\\
Tolerance of the floor & 8.0 & 4.1 & 1.0 & 3.0 & 3.1 & 0.4\\
Parameters $X$ (replicas of Section~\ref{sec:budget}) & 45.1 & 33.4 & 11.9 & 2.8 & 5.1 & 1.0\\
\hline
Combined (quadrature) & 86.5 & 65.1 & 21.5 & 12.6 & 39.6 & 24.1\\
Band mean (Table~\ref{tab:passivity}) & $+572$ & $-458$ & $-115$ & $+181$ & $-632$ & $+402$\\
\hline
 & \multicolumn{3}{c}{(per cent)} & \multicolumn{3}{c}{(degrees)}\\
18.0--20.8~GHz & $+0.46\pm0.12$ & $-0.37\pm0.09$ & $-0.08\pm0.03$ & $+0.17\pm0.02$ & $-0.59\pm0.05$ & $+0.37\pm0.03$\\
20.8--23.7~GHz & $+0.59\pm0.09$ & $-0.47\pm0.07$ & $-0.12\pm0.02$ & $+0.18\pm0.01$ & $-0.64\pm0.04$ & $+0.41\pm0.02$\\
23.7--26.5~GHz & $+0.67\pm0.10$ & $-0.53\pm0.07$ & $-0.14\pm0.02$ & $+0.19\pm0.01$ & $-0.67\pm0.04$ & $+0.43\pm0.03$\\
\hline
\end{tabular}}
\end{table}
Three results are the same in the five mounts.

First, the correction is small: $\|m-1\|$ is $1.5$--$2.1\%$ in band
mean.
Second, it has the same structure in all specimens. The rotations are
constant across the band, to $\pm0.01$--$0.08^\circ$, and of the same
magnitude in the five mounts: $\theta_1\simeq+0.1$ to $+0.2^\circ$,
$\theta_2\simeq-0.6$ to $-0.8^\circ$, $\theta_3\simeq+0.4$ to $+0.5^\circ$.
The scales carry the frequency dependence: $s_1$ rises across the band and
$s_2$ falls, in opposition and with the same structure -- by
$0.2$--$0.6\%$ between the first and the last third of the band in
specimens~4 and~5, against uncertainties of about $0.1\%$ for those thirds
(Table~\ref{tab:mbudget}) -- while $s_3$
stays near zero; constant factors do not bring $\sigma_{\max}$ to the
floor at every frequency. The problem is solved at
every second point of the analyzer grid. Third, the correction changes nothing that is
measured. By construction $K_f$ does not change; $\eta_2$ changes by less
than $10^{-4}$; $Z_0$ recomputed from the pair closure moves by about
$1~\Omega$ or less in band mean, and the closure defect of the pairs falls by a
factor $1.4$ to $2.4$ (Table~\ref{tab:pairs}). What changes is the internal description: the efficiencies of
the two non-operating states, which come to obey \eqref{eq:attlaw} to
within $4\times10^{-3}$ in band mean, the pair ratios $k_{i,j}$ of Section~\ref{sec:zo}, and
the implied scattering matrix, which becomes that of a structure at the
floor of passivity, of nearly equal efficiency in the three states. These corrected quantities
are the ones reported from here on; the first-step values quoted in
Section~\ref{sec:results} and Appendix~\ref{app:sens} are those of the
model, stated for comparison. The correction is a second step applied
to the solution of the first: the conditions of
Sections~\ref{sec:determination} to~\ref{sec:invariant} are imposed on the
model of Section~\ref{sec:circuit} and fix $X$; the corrected impedances
are not re-entered in them -- the curvature condition, evaluated at the
real reference, is sensitive to the state factors -- and the two steps are
not iterated. The two steps are not interchangeable, and the second cannot
be inverted. The first fixes $F$ and places the three impedances on
$\mathcal M_F$, at a position along the two coordinates that $F$ does not
see which is set by the thermistor model, not by the data. The second is
the projection of that point, by minimum norm, onto the floor of
passivity \eqref{eq:floor}, a two-dimensional subset of $\mathcal M_F$:
the form $\sqrt{\eta_2}\,\mathbf U$ imposes three real conditions on
the five coordinates of $\mathcal M_F$ (the subset on which the efficiency
ratios alone are unity has dimension three, and contains the floor up to
the mismatch factors of \eqref{eq:attlaw}). The projection -- the
minimum-norm solution, reproduced from every start -- is not injective:
the pre-image of a corrected solution is a
three-parameter family of points of $\mathcal M_F$, and the model
impedances that produced it cannot be recovered from it. What the second
step discards is what the data cannot determine -- the position along the
directions that $F$ does not see, which the model had filled with its own
values of $\eta_1/\eta_2$ and $\eta_3/\eta_2$ -- and nothing else, since
$K_f$, $\eta_2$ and $Z_0$ are unchanged. Re-entering $Z'_{th}$ in the
functional would ask the model of Section~\ref{sec:circuit} to reproduce
ratios equal to unity, which it does not produce, and would restore at
each pass what the projection had removed; the result would then depend
on the number of passes.

\begin{figure}[!t]
\centering
\includegraphics[width=0.62\textwidth]{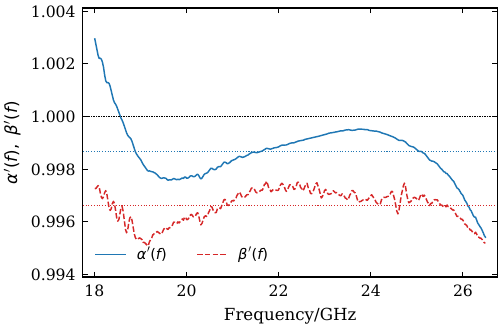}
\caption{Auxiliary variables $\alpha'(f)$ and $\beta'(f)$ of specimen~4 from
\eqref{eq:abf} with the state correction (solid and dashed), and their band
averages (dotted).}
\label{fig:alphabeta}
\end{figure}

\begin{figure}[!t]
\centering
\includegraphics[width=0.8\textwidth]{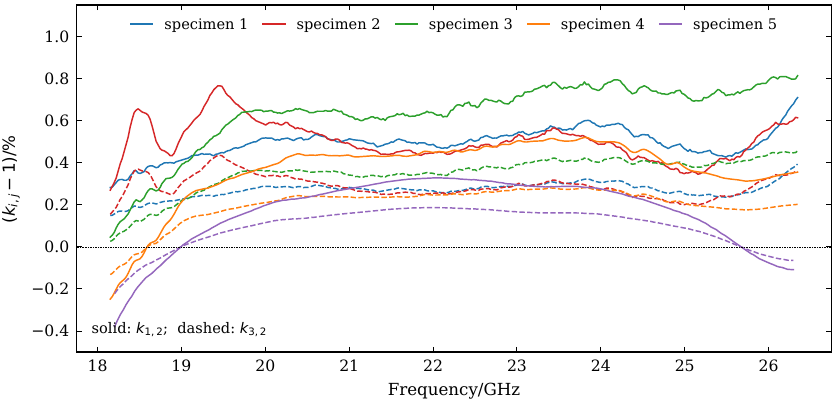}
\caption{Departure of the pair ratios from unity, $k_{i,j}-1$ (per cent),
for the five specimens with the state correction, both pairs evaluated at
the common $Z_0$: solid, pair $(1,2)$; dashed, pair $(3,2)$. Curves lightly
smoothed for display.}
\label{fig:kij}
\end{figure}

The counterparts of the auxiliary variables $\alpha$ and $\beta$ of
\eqref{eq:etaratios} with the corrected impedances, inverted at each
frequency from the efficiencies $\eta_i'$ of the corrected states,
\begin{equation}
\alpha'(f)=\frac{2\eta_1'}{\eta_1'+\eta_3'},\qquad
\beta'(f)=\frac{2\eta_1'\eta_3'}{\eta_2\,(\eta_1'+\eta_3')} ,
\label{eq:abf}
\end{equation}
(with $\eta_2'=\eta_2$, which the correction leaves unchanged), are shown in
Fig.~\ref{fig:alphabeta} for specimen~4: both are within
$\pm0.7\%$ of unity over the band, with band averages of $0.993$--$0.999$
in the five mounts (Table~\ref{tab:passivity}), against $\alpha=0.957$--$0.974$
and $\beta=0.979$--$0.988$ returned by the algorithm with the model of
Section~\ref{sec:circuit} (Table~\ref{tab:params}). The correction sets
the ratios to unity by construction, at every frequency up to the mismatch
factors of \eqref{eq:attlaw}; the first property of
Section~\ref{sec:determination}, the constancy of the ratios across the
band, is preserved, with the value that the floor fixes. The functional uses
only the band averages of $\alpha$, $\beta$, and $K_f$ does not depend on
them (Appendix~\ref{app:ident}); this is why the model of
Section~\ref{sec:circuit} could absorb in $\alpha$, $\beta$ the state
dependence that the correction represents, without any effect on
the result. The pair ratios behave likewise:
Fig.~\ref{fig:kij} shows $k_{i,j}-1$ at the common
$Z_0=(Z_{0,12}+Z_{0,32})/2$ for the two pairs. Both lie between $-0.4$ and
$+0.8\%$ over the whole band, with the same shape in the five mounts and
the pair $(3,2)$ at about half the magnitude of $(1,2)$, which is what the
sensitivities $\mathcal S_{i,2}$ of Appendix~\ref{app:sens} -- the
first-order response of each pair impedance to $k_{i,j}-1$ --
$|\mathcal S_{3,2}|\approx2|\mathcal S_{1,2}|$ -- imply for equal and
opposite displacements of the two pair impedances about their mean. The
convention $k_{i,j}=1$ of Section~\ref{sec:zo} was adopted without
reference to the two invisible degrees of freedom; once these are fixed
by passivity, both factors of \eqref{eq:kidentity} are within $1\%$ of
unity, the convention is exact to that level, and the $Z_0$ it defines is
unchanged within about $1~\Omega$ in band mean. Equal efficiencies are what a passive
matching structure is expected to show, its loss depending weakly on the
load, and are what the coaxial mounts of \cite{greco_coaxial} showed
directly ($\alpha=1.007$, $\beta=0.999$).

\begin{figure}[!t]
\centering
\includegraphics[width=0.62\textwidth]{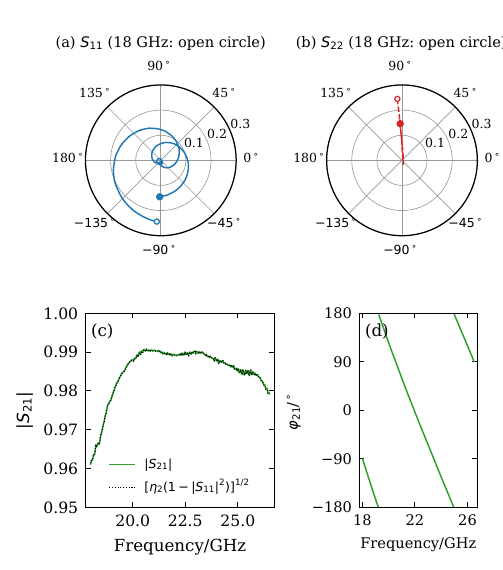}
\caption{Scattering parameters of the matching structure of specimen~5
at the real reference $Z_0$, with the state correction: (a), (b) $S_{11}$
and $S_{22}$ in polar form over 18--26.5~GHz (open circle, 18~GHz; full
circle, 26.5~GHz); (c) $|S_{21}|$, with $[\eta_2(1-|S_{11}|^2)]^{1/2}$ for
comparison; (d) transmission phase $\varphi_{21}$ of
\eqref{eq:delaylength}.}
\label{fig:sparams}
\end{figure}

Figure~\ref{fig:sparams} shows the resulting scattering parameters of
specimen~5. The magnitudes satisfy $|S_{11}|=|S_{22}|$ and
$|S_{21}|^2=\eta_2(1-|S_{11}|^2)$ within $5\times10^{-4}$ at every
frequency, the signature of \eqref{eq:floor} to the order of the term
$(1-\eta_2)|S_{11}|^2$ of \eqref{eq:floorbound}; $S_{22}$ stays close to the imaginary axis (within
$\pm10^\circ$ of $90^\circ$ in this specimen) while its magnitude follows
that of $S_{11}$. The phase of the transmission term,
$\varphi_{21}=\tfrac12\arg(S_{12}S_{21})$, was fitted as
\begin{equation}
\varphi_{21}(f)=\varphi_0-k_z(f)\,\ell_e ,
\label{eq:delaylength}
\end{equation}
with $k_z=(\omega/c_0)\sqrt{1-(f_c/f)^2}$ the propagation constant of the
TE$_{10}$ mode in WR-42 ($c_0$ the speed of light) and $f_c$ its cutoff
frequency; $\ell_e$ is the length of uniform guide with the same delay, and
it gives the electrical position of the thermistor-side reference plane
anticipated in Section~\ref{sec:intro}. Table~\ref{tab:delay} lists the
result for the five specimens. The residual of the fit over the band is
below $1.6^\circ$ rms in every specimen, the length is the same within
1.1~mm in the five mounts, as expected for a common design, and it differs
by less than $0.2$~mm from the value obtained with the impedances of the
model, the correction leaving the transmission phase practically
unchanged. It is an electrical length: it includes the dispersion of the
matching structure and is not the physical distance from the flange to the
thermistor.

\begin{table}[!t]
\caption{Equivalent guide length $\ell_e$ of \eqref{eq:delaylength}, fitted
over the whole band and over 1-GHz windows, with the corrected thermistor
impedances.}
\label{tab:delay}
\centering
\begin{tabular}{c c c}
\hline
Specimen & $\ell_e$, band & $\ell_e$, 1-GHz windows\\
 & (mm) & (mm)\\
\hline
1 & 39.0 & 38.3--41.9\\
2 & 38.8 & 38.2--40.3\\
3 & 39.0 & 38.4--41.2\\
4 & 39.2 & 38.7--40.6\\
5 & 39.9 & 39.5--40.8\\
\hline
\end{tabular}
\end{table}

The corrected model is the complete description of the mount that the
three-state measurement supports: it preserves the calibration factor and
the efficiency of the operating state, it is consistent with the power
balance of a passive structure, and it is the one from which the
quantities beyond $K_f$ and $\eta_2$ are taken -- the efficiencies of the
other two states, the pair ratios, the scattering parameters and the
sensitivity of the efficiency to the bias resistance
(Appendix~\ref{app:rb}). The physical origin of the state dependence is
the subject of continuing work, to be reported separately.

\subsection{Parameter Scaling}
\label{sec:scalingA}
All the results presented so far have been obtained with the thermistor
impedances determined up to the common transformation \eqref{eq:neutral}, to
which the calibration factor and the efficiencies are invariant. It is
nevertheless useful to determine $Z_{th}$ and $Z_0$ more closely. The
calculation itself shows that both are fixed in the same scale. The closed
quadratic \eqref{eq:zoquad} can be written in the compact form
\begin{equation}
Z_{0,i2}^2 + 2\,Z_{0,i2}\,\lambda_a(Z_{th}) + \lambda_b(Z_{th}) = 0,
\label{eq:zoquadcompact}
\end{equation}
where $\lambda_a$ and $\lambda_b$ denote its two coefficients. Under the
transformation $Z_{th}\to a\,Z_{th}$, $a\in\mathbb{R}$, the ratio $\phi_i$ of
\eqref{eq:phi} is unchanged, $\lambda_a$ is homogeneous of degree one and
$\lambda_b$ of degree two in $Z_{th}$, and the equation becomes
\begin{equation}
Z_{0,i2}'^{\,2} + 2\,Z_{0,i2}'\,a\,\lambda_a(Z_{th}) + a^2\lambda_b(Z_{th})
= 0,\qquad Z_{0,i2}'=a\,Z_{0,i2},
\label{eq:zoquadscaled}
\end{equation}
which shows that the method determines the thermistor impedances and the
thermistor-port impedance in the same scale. The state correction of
Section~\ref{sec:passivity} commutes with this scaling: problem
\eqref{eq:sp2} is invariant under $Z_{th}\to aZ_{th}$, since $F$ is and
the implied scattering matrix depends on the loads only through
$\Gamma_{Li}$, in which $Z_0$ scales with them, so the factors $m_i$ are
unchanged by $a$ and the corrected thermistor-port impedance scales with it as
\eqref{eq:zoquadscaled} states.

The purpose of this section is to estimate that scale factor, so as to convert
both quantities into the physical values of the thermistor and of the mount
port. The thermistor parameters, including the bridge resistances $R_{b,i}$ as
they enter the model, are affected by the scaling. The scaled value of
$R_{b,i}$ is a microwave quantity only, and the ratios between the three states
are preserved: the thermistor material is homogeneous and held at a uniform
temperature, so the proportionality between the states is not lost.

\begin{figure}[!t]
\centering
\includegraphics[width=0.62\textwidth]{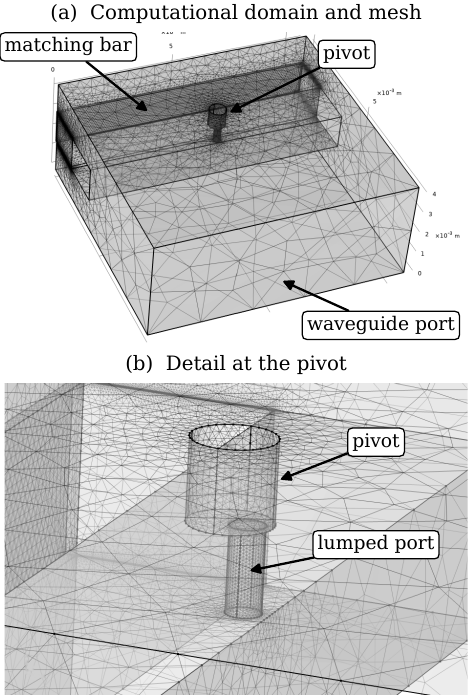}
\caption{Finite-element model of the mount. (a)~Computational domain and
mesh: the input cross-section is the waveguide port; the matching bar spans
the full width of the guide and the pivot hangs from the top wall at its
center. (b)~Detail at the pivot: the small cylinder between the pivot tip and
the top face of the bar is the lumped port that replaces the thermistor.}
\label{fig:femdomain}
\end{figure}

The values obtained in this part are estimates, not traceable results: the
finite-element model represents one surveyed unit, and specimens adjusted at
the factory by small alterations of the matching bar -- bending the center
of the bar changes the impedance seen by the thermistor -- differ from it
individually. They place $Z_{th}$ and $Z_0$ on the physical scale of the
device. The calculation starts from the dimensional survey of a
specimen that was out of service. The small dimensions of the thermistor set
the mesh requirements of the finite-element calculation, and, as noted in
Section~\ref{sec:intro}, the reference plane between the thermistor and the
sensor body is not exactly defined. For these reasons the thermistor is
represented by a cylindrical lumped port,
Fig.~\ref{fig:femdomain}.

The mount is modeled with the finite-element method in the frequency domain,
with perfectly conducting walls, and solved with a finite-element program of
the author's own built on NGSolve \cite{ngsolve}. With the
time convention $e^{j\omega t}$, the electric field in the air-filled interior
$\Omega$ of the mount satisfies the vector wave equation
\begin{equation}
\nabla\times\left(\mu_r^{-1}\,\nabla\times\mathbf{E}\right)
-k_0^2\,\varepsilon_r\,\mathbf{E}=0,\qquad k_0=\frac{\omega}{c_0},
\label{eq:waveeq}
\end{equation}
with $\hat{n}\times\mathbf{E}=0$ on all metal surfaces, where $\mu_r$ and
$\varepsilon_r$ are the relative permeability and permittivity, both unity in
the air-filled interior, $k_0$ is the free-space wavenumber and $c_0$ the speed
of light in vacuum. The field is discretized with curl-conforming
(N\'ed\'elec) elements. The waveguide flange is a wave port carrying the
TE$_{10}$ mode; terminated in its wave impedance, it acts as a perfect
termination.

The lumped port is a cylinder of radius $r_p$ and length $l$ whose volume is
removed from $\Omega$; its lateral surface $\Sigma$ carries the port boundary
condition. Let $\mathbf{p}_0$ and $\mathbf{p}_1$ be the centers of the two end
faces of the cylinder, on the pivot tip and on the top face of the bar,
$l=\|\mathbf{p}_1-\mathbf{p}_0\|$, $\hat{t}=(\mathbf{p}_1-\mathbf{p}_0)/l$, and
$P_\Sigma=|\Sigma|/l$ the perimeter of $\Sigma$. The surface $\Sigma$ plays the role
of the surface that encloses the element in Kerns' formulation
\cite{kerns1949}: the port does not presuppose a transmission line, only that
the fields on $\Sigma$ factor into a voltage-like and a current-like amplitude,
which holds because the gap is electrically small, $k_0 l\le0.34$ over the band
for $l=0.61$~mm. The port voltage is the axial line integral of $\mathbf{E}$
averaged over the perimeter,
\begin{equation}
V=\frac{1}{P_\Sigma}\oint_{\Sigma}\mathbf{E}\cdot\hat{t}\,dS,
\label{eq:portV}
\end{equation}
and the port carries on $\Sigma$ a uniform axial surface current of total
intensity
\begin{equation}
I=\frac{V_s-V}{Z_p},
\label{eq:portI}
\end{equation}
generated by a source of open-circuit voltage $V_s$ and internal impedance
$Z_p$, the reference impedance of the port. In the weak form of
\eqref{eq:waveeq}, \eqref{eq:portI} enters through the boundary term on
$\Sigma$ as the Robin condition
\begin{equation}
\hat{n}\times(\nabla\times\mathbf{E})\big|_{\Sigma}
=\frac{j\omega\mu_0}{P_\Sigma}\,\frac{V-V_s}{Z_p}\,\hat{t},
\label{eq:robin}
\end{equation}
where $\hat{n}$ is the outward normal of $\Omega$ and $V$ is an additional
unknown defined by \eqref{eq:portV}. The perimeter is taken from the meshed
surface, rather than as $2\pi r_p$; with the same $|\Sigma|$ in
\eqref{eq:portV} and \eqref{eq:robin}, the power delivered by the port,
$\tfrac12\mathrm{Re}\{VI^*\}$ with $V$ and $I$ peak phasors, balances
exactly with the power absorbed in the discrete problem; with the same
convention $|a_p|^2-|b_p|^2=\mathrm{Re}\{VI^*\}$ below is twice that
power, a factor common to every power ratio and without effect on them.

The port amplitudes are the power waves referred to $Z_p$
\cite{kurokawa1965},
\begin{equation}
a_p=\frac{V+Z_pI}{2\sqrt{Z_p}},\qquad b_p=\frac{V-Z_pI}{2\sqrt{Z_p}},
\label{eq:powerwaves}
\end{equation}
so that $a_p=V_s/(2\sqrt{Z_p})$: a source $V_s=2\sqrt{Z_p}$ excites the port
with $a_p=1$, and $V_s=0$ terminates it in $Z_p$. At the flange the amplitudes
are those of the TE$_{10}$ mode normalized to unit power with the wave
impedance $Z_{TE}=k_0Z_{F0}/k_z$, where $Z_{F0}$ is the impedance of free space
and $k_z$ the propagation constant of the mode. Solving \eqref{eq:waveeq} once
for each excited port gives the $2\times2$ scattering matrix of the mount
between the flange (port~1) and the thermistor terminals (port~2, reference
$Z_p$). The impedance of the mount seen from the thermistor terminals, with the
flange matched, is then
\begin{equation}
Z_M=Z_p\,\frac{1+S_{22}}{1-S_{22}} .
\label{eq:zmount}
\end{equation}
Although $Z_p$ is required during the calculation, $Z_M$ is independent of it:
a different $Z_p$ changes the scattering matrix but not $Z_M$, which is a
property of the mount alone. The thermistor enters only through $\Gamma_L$ in
the bilinear map \eqref{eq:gamma}, whereas $Z_M(f)$ is the quantity compared
with the self-consistent thermistor-port impedance of Section~\ref{sec:zo}. The model
is lossless, and $|S_{11}|^2+|S_{21}|^2=1$ was verified in every solution.

\begin{figure}[!t]
\centering
\includegraphics[width=0.62\textwidth]{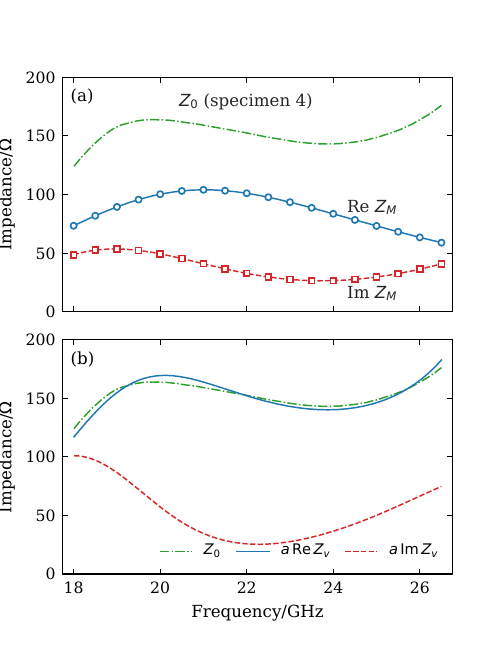}
\caption{(a)~Impedance $Z_M$ of the mount at the lumped port computed by
finite elements (markers: simulated frequencies) and thermistor-port impedance $Z_0$
determined by the method for specimen~4. (b)~$Z_0$ and the impedance $Z_v$ seen
through the line \eqref{eq:zv}, scaled by $a$, with the parameters of
Table~\ref{tab:scaling} for specimen~4.}
\label{fig:zm}
\end{figure}

The impedance $Z_M$ obtained in this way is shown in Fig.~\ref{fig:zm}(a),
together with the thermistor-port impedance $Z_0$ determined by the method for
specimen~4. The two differ not only in scale but also by a rotation of phase.
The lumped port does not coincide exactly with the mounting of the thermistor:
it differs essentially by a short section of transmission line formed by the
thermistor's own wires, not represented in the lumped port. We therefore
compute the impedance seen through a line of characteristic
impedance $Z_c$ and length $\ell_c$,
\begin{equation}
Z_v = Z_c\,\frac{Z_M + j Z_c\tan(k_z\ell_c)}{Z_c + j Z_M\tan(k_z\ell_c)},
\label{eq:zv}
\end{equation}
where the line is given the dispersion of the guide,
\begin{equation}
k_z=\frac{\omega}{c_0}\sqrt{1-\left(\frac{f_c}{f}\right)^{2}},\qquad
Z_c=\frac{Z_{c0}}{\sqrt{1-(f_c/f)^{2}}},
\label{eq:betazc}
\end{equation}
$f_c$ being the cutoff frequency of the TE$_{10}$ mode. With the scale factor
$a$ of \eqref{eq:zoquadscaled}, the set $\{a,\ell_c,Z_{c0}\}$ is found for each
specimen by minimizing
\begin{equation}
\mathcal{J}=\frac{1}{N_f}\sum_{n=1}^{N_f}
\big(Z_{0,n}-a\,\mathrm{Re}\,Z_{v,n}\big)^{2}
\label{eq:jscale}
\end{equation}
over the $N_f$ points of the analyzer grid. The finite-element solution is
computed at 18 frequencies, 0.5~GHz apart, and is interpolated to the analyzer
grid; the measured quantities are not interpolated. The fit uses the
thermistor-port impedance of the first step: the state correction moves it by
about $1~\Omega$ or less in band mean (Section~\ref{sec:passivity}), which
changes $a$ by less than $1\%$, within the spread of
Table~\ref{tab:scaling}.

\begin{table}[!t]
\caption{Scale factor and line parameters obtained by minimizing
\eqref{eq:jscale}, with the rms deviation and the correlation coefficient
between $Z_0$ and $a\,\mathrm{Re}\,Z_v$. Last row: mean and 95\% confidence
interval (Student $t$, four degrees of freedom).}
\label{tab:scaling}
\centering
\setlength{\tabcolsep}{3.5pt}
\begin{tabular}{c c c c c c}
\hline
Specimen & $a$ & $\ell_c$ (mm) & $Z_{c0}$ ($\Omega$) & rms ($\Omega$) & corr\\
\hline
1 & 1.048 & 1.726 & 83.1 & 2.35 & 0.992\\
2 & 1.149 & 2.560 & 88.3 & 7.41 & 0.937\\
3 & 1.227 & 1.683 & 70.3 & 6.07 & 0.960\\
4 & 1.042 & 1.713 & 86.6 & 3.72 & 0.989\\
5 & 1.336 & 1.758 & 75.2 & 5.93 & 0.974\\
\hline
Mean & $1.16\pm0.16$ & $1.89\pm0.47$ & $80.7\pm9.6$ & 5.10 & 0.970\\
\hline
\end{tabular}
\end{table}

The results are listed in Table~\ref{tab:scaling}; the modeled unit is none
of the five specimens. The model reproduces the frequency dependence of $Z_0$
closely in every specimen, with correlation coefficients between 0.937 and
0.992, as illustrated for specimen~4 in Fig.~\ref{fig:zm}(b). Evaluated point
by point, $a(f)=Z_0/\mathrm{Re}\,Z_v$ shows no trend with frequency in any
specimen, the slopes lying within $\pm0.07\%$/GHz. The line length is the most
stable parameter: four specimens lie within 1.68--1.76~mm, whereas specimen~2
departs to 2.56~mm and has the largest rms. The spread of $a$
among specimens is consistent with the individual differences between units
noted above.

The fitted $Z_{c0}$ can be compared with a closed-form estimate from the guide
dimensions alone, $a_g=10.668$~mm and $b_g=4.318$~mm. The power--voltage
impedance ($Z_{PV}$) of the TE$_{10}$ mode, $Z_{PV}=V^2/2P$ with $V$ the
peak voltage across the full height at the center of the guide and $P$ the
transmitted power, is $Z_{PV}=(2b_g/a_g)\,Z_{F0}/\sqrt{1-(f_c/f)^2}$, where
$Z_{F0}=\sqrt{\mu_0/\varepsilon_0}\simeq376.7~\Omega$ is the impedance of
free space introduced above and $f_c$ the cutoff frequency of the mode. With the bar at
mid-height, the gap between the bar and the top wall spans half of that
voltage, so that the impedance seen at the thermistor port is
\begin{equation}
\frac{Z_{PV}}{4}=\frac{b_g}{2a_g}\,
\frac{Z_{F0}}{\sqrt{1-(f_c/f)^{2}}},
\label{eq:zpv4}
\end{equation}
which has the same dispersion as \eqref{eq:betazc}; its asymptotic value
$(b_g/2a_g)\,Z_{F0}=76.2~\Omega$ lies within the confidence interval of
$Z_{c0}$ in Table~\ref{tab:scaling}.

The values of the thermistor and of the thermistor port closer to the physical ones
follow by dividing the impedances by $a$. For the parameters of
Table~\ref{tab:params} this gives
\begin{equation}
R_{p0}'=\frac{R_{p0}}{a},\quad R_{b,i}'=\frac{R_{b,i}}{a},\quad
C_x'=a\,C_x,
\label{eq:scaledparams}
\end{equation}
with $\kappa$, $\alpha$ and $\beta$ unchanged; $L_x$ is a physical length
and is not scaled, so the scaling is exact for the parallel combination and
approximate only in the skin term of \eqref{eq:skin}, whose departure is a
small reactance nearly common to the three states, absorbed by
\eqref{eq:neutral}. The state factors $m_i$ of Section~\ref{sec:passivity}
are dimensionless and apply unchanged to the scaled model: the thermistor
impedances on the physical scale are $m_iZ_{th,i}$ evaluated with the
scaled parameters, and the scaled bridge resistances below are those of
the model before the state factors. Table~\ref{tab:scaledparams} gives
the result for two specimens. Since $a>1$ in the five mounts (Table~\ref{tab:scaling}), the
scaled bridge resistances $R'_{b,i}$ are below the dc values held by the
bridge, by $4$ to $25\%$: this is the expectation stated in
Section~\ref{sec:circuit}, that the sintered material presents a lower
impedance under microwave heating than at dc, made quantitative. It could
not appear in the determination of $K_f$, which is invariant to the scale;
it appears here, where the scale is fixed by the mount. These values are
estimates, not traceable results, for the reason stated at the beginning of
this section; the sintered thermistor is the subject of a model of its own,
left for future work.

\begin{table}[!t]
\caption{Thermistor parameters of Table~\ref{tab:params} scaled by
\eqref{eq:scaledparams}.}
\label{tab:scaledparams}
\centering
\begin{tabular}{c c c c c c}
\hline
Specimen & $C_x'$ (fF) & $R_{p0}'$ ($\Omega$) & $R_{b,1}'$ ($\Omega$)
& $R_{b,2}'$ ($\Omega$) & $R_{b,3}'$ ($\Omega$)\\
\hline
1 & 19.12 & 1375.5 & 95.80 & 191.16 & 286.55\\
5 & 22.75 & 1157.8 & 75.18 & 150.01 & 224.86\\
\hline
\end{tabular}
\end{table}

The fit leaves an imaginary residual: the scaled impedance $a\,Z_v$ retains a
mean reactance between 15 and 51~$\Omega$ in the five specimens
(Fig.~\ref{fig:zm}(b)). It is a finding of the finite-element model; its value
for each specimen is not known, since the model was built from a different
unit. If $Z_0$ were allowed to be complex, the load reflection coefficient
would be written with the power waves of Kurokawa \cite{kurokawa1965},
$\Gamma_{Li}=(Z_{th,i}-Z_0^{*})/(Z_{th,i}+Z_0)$, which with $Z_0=R+jb$
becomes $\big((Z_{th,i}+jb)-R\big)/\big((Z_{th,i}+jb)+R\big)$: an
imaginary part of the reference cannot be distinguished from a reactance
common to the three thermistor impedances, which is \eqref{eq:neutral} with
$a=1$. The residual is therefore not a property that the three measured
reflection coefficients can reveal. It belongs to the finite-element
description of the mount and is not an element of the thermistor impedance
model $Z_{th,i}$ of Section~\ref{sec:circuit}, which contains no reactance
common to the three states; the reference of the model is real by
construction (Section~\ref{sec:zo}). The
calibration factor at given parameters, the efficiencies, the ratio
condition and the hyperbolic term are invariant under a common reactance;
the curvature condition, evaluated at a real reference, is not, and absorbs
such a reactance into the parameters. The residual thus identifies the
element not represented in the model: a reactive output impedance of the
mount at the thermistor port, $+15$ to $+51~\Omega$ inductive as seen by
the finite-element model. It is distinct from the state correction
of Section~\ref{sec:passivity}: that one acts within $\mathcal M_F$ and
is invisible to the calibration factor, this one is a common reactance,
invisible to the calibration factor and to the passivity of the implied
two-port alike, and seen only by the curvature condition.
Representing it -- by a reactive element at the port or by a complex
reference -- is the next step of this work.

\section{Conclusion}
\label{sec:conclusion}
The results for the five K486A specimens can be summarized as follows. The
model reproduces the four manufacturer tables ($U=2\%$) within
$0.84$--$0.98\%$ rms and the PTB-traceable reference ($U=0.85$--$0.98\%$) within
$0.34\%$ rms; over the 38 reference points the normalized error does not
exceed $0.6$. The expanded measurement uncertainty of the method is
$0.5$--$1.16\%$ ($k=2$) for the PTB-traceable specimen and between $0.5\%$
and $1.9\%$ for the five; it is the uncertainty of the result given the
thermistor model, and the adequacy of the model is what the comparison
with the references establishes, for the specimens tested. The comparison
with the traceable reference deserves emphasis: the mount with the traceable
reference is the one the model reproduces most closely. The specimens span
the efficiency range of the mounts in use as traveling standards, as
shown by the efficiencies of the two standards circulated in
CCEM.RF-K28.W \cite{ccem_k28}, which fall within the range of the set and
have close counterparts in it.

The element that the waveguide adds to the method is the self-consistent
determination of the thermistor-port reference impedance. What the data
determine of $Z_0$ is the real reference at which the power balance
between pairs of states closes; that the reference is real is the
convention $k_{i,j}=1$ of Section~\ref{sec:convention}, and the quantity so
obtained is the effective reference impedance of the thermistor port, not
a characteristic impedance in the transmission-line sense. The
determination is a technique in its own right, applicable to other devices
in which the impedance of an inaccessible port must be known from a
standard one-port calibration and a load that can be switched among known
values; it is to be developed beyond the present work, as is the modeling
of the sintered thermistor at high frequencies.

The assembly of a mount and its waveguide-to-coaxial adapter has been
calibrated directly, without characterizing the adapter, with an expanded
measurement uncertainty of about $1.7\%$ ($2.3\%$ total): the model factor
is inherited from the mount and the geometric factor is measured through
the adapter, the cost of the transfer being the loss of the covariance
credit between the two (Section~\ref{sec:adapters}). The method has also
been applied without change to an R486A mount in WR-28 with a
reflectometer of the author's own construction, with $0.92\%$ rms against
the manufacturer table.

The method has thus been presented in its waveguide version. The difficulties
that arise in applying the conditions developed for the coaxial case
\cite{greco_coaxial} to a mount whose thermistor is coupled to the guide
through a matching structure have been stated, and their solutions
developed, leading to a method that requires only three one-port VNA
measurements, with an algorithm largely insensitive to the residual error
terms of the instrument -- the directivity cancels identically in the
reflectometer channel -- and that delivers the calibration factor and
the effective efficiency at every point of the analyzer grid, up to 716
frequencies per sweep, in a few minutes of computation
(Fig.~\ref{fig:flow}). The efficiency so
obtained is that of the mount in the sense of Kerns; its use as the
effective efficiency of a bolometric calibration carries the thermal
equivalence of RF and dc power in the bead assumed in
Section~\ref{sec:method}, which the comparisons with calorimetric
references support at the level of their uncertainties.

Beyond the determination itself, the part of the space of thermistor
impedances that the calibration factor does not see has been characterized
exactly -- two degrees of freedom, the ratios of the efficiencies between
states, whose constancy across the band the functional uses and whose
value it leaves free (Section~\ref{sec:invariant}) -- and the scattering
matrix that the solution implies at the real reference has been found not
to be passive with the thermistor model alone. The smallest correction
along those two coordinates -- a scale and a rotation of about $1\%$ and
$0.5^\circ$ per state, the same in pattern in the five specimens, with
$K_f$ and $\eta_2$ held exactly -- brings its largest singular value to the
floor $\sqrt{\eta_2}$, the matching structure represented as lossless with
a matched attenuator, the efficiencies of the three states equal up to
their mismatch factors, and the pair closure of the thermistor-port
impedance improved by a factor of two (Section~\ref{sec:passivity}). The
same three-state geometry separates the dependence of the calibration
factor on the bias resistance $R_b$ into an efficiency term, given in
closed form along the bilinear continuation of the three states and below
$10^{-4}~\Omega^{-1}$ in band mean once the matching
structure is taken at the floor of passivity, and a mismatch term of
$1$--$7\times10^{-4}~\Omega^{-1}$ that follows from the measured
coefficients alone; together with the scale error of the dc substitution,
$\Delta R/R_b$, they are relevant as a fraction of the uncertainties
obtained by the laboratories, and Appendix~\ref{app:rb} shows how a
departure of $R_b$ -- generally an internal datum of the laboratory, whose
influence on the result is rarely documented -- propagates into a precision
power measurement in which a digital voltmeter reads the balance voltages
of the bridge, as with the HP~432A \cite{keysight2001}. The physical origin
of the state dependence that the correction represents, and the reactive
output element of the mount identified in Section~\ref{sec:scalingA}, are
the subject of continuing work, to be reported in a separate publication.

The procedure does not constitute an independent traceability chain: its
results are verified against references whose traceability is that of their
own certificates, and what it establishes is that a calibration factor
determined from reflection data alone is consistent with those references
within the uncertainties of both.
This work and \cite{greco_coaxial} are complementary: the coaxial paper
establishes the three-state method and the properties on which the
functional rests -- the proportionality of the efficiency ratios, the
relative curvature, the symmetry analysis of the bias resistances -- and
this one adds what the waveguide requires, the self-consistent thermistor-port
impedance and the hyperbolic invariant; the derivations of
\cite{greco_coaxial} are used here without repetition.

As in the coaxial case \cite{greco_coaxial}, it is worth making explicit the
stage this work has reached. The development of a method of this kind
passes through three stages: the formulation of the model and of the
conditions that make its parameters determinable; the verification on a set
of mounts for which a traceable calibration is available; and the
large-scale validation on a number of specimens exceeding the possibilities
of a single laboratory. The present work completes the second stage in
waveguide, with the additional element that the coaxial case did not
require, the self-consistent determination of the thermistor-port
impedance; the verification rests on the comparison with the references,
not on the uncertainty budget, which is a different quantity. The third
stage requires the collaboration of other laboratories, and not only as a
matter of number: the manufacturer tables available for these mounts state
a single figure of two per cent for the whole band, so that the comparison
cannot resolve differences smaller than that, and only the accumulation of
independent results against primary references can take the verification
beyond that limit. For that reason the author offers, as for the coaxial
mounts, to process the reflection coefficients of waveguide mounts from
other laboratories and to return the calibration factor with its
uncertainty budget; the reference value need not be communicated in
advance, and the comparison is completed by reporting it once the result
has been received. The cases accumulated in this way will be the subject of
a later publication, with due acknowledgment to the participating
laboratories. That validation is open to all those interested in
contributing to it, and it is the reason this paper is published in open
access.

\appendix
\section{Identifiability of the Parameter Set}
\label{app:ident}
The analysis follows the corresponding appendix of \cite{greco_coaxial}, to
which we refer for the details, and is carried out for specimen~4. Since
$K_f=F(X)\,G(\Gamma_1,\Gamma_2,\Gamma_3)$ and $G$ does not depend on $X$, the
relative sensitivity of the calibration factor to the parameters is
\begin{equation}
J_{nk}=\left.\frac{\partial\ln K_f(f_n)}{\partial\ln X_k}\right|_{X=X_0}
=\left.\frac{\partial\ln F(f_n)}{\partial\ln X_k}\right|_{X=X_0},
\label{eq:jident}
\end{equation}
with $X=\{C_x,R_{p0},\kappa\}$ and $X_0$ the converged solution. The
auxiliary variables $\alpha$ and $\beta$ do not enter $K_f$, and $L_x$ is
fixed; changing $L_x$ by $0.2$~mm moves the components of $J$ by less than
$1.4\%$. Two different objects are examined in this appendix and should
not be confused: $J$ is the sensitivity of the output to the parameters,
and serves to find the directions of parameter space to which $K_f$ is
least sensitive; the identifiability of $X$ from the data is the question
of the minimized functional, and is examined through its Hessian in
Appendix~\ref{app:hessian}, with $\alpha$, $\beta$ and the implicit
dependence of $Z_0$ on $X$ included.

Local identifiability requires three frequencies at which the rows of $J$
are linearly independent. The three that maximize $|\det J|$ are listed in
Table~\ref{tab:ident}; the matrix they form has full rank and a condition
number of 92, of the same order as that of the whole band, 142. Three suitably
chosen frequencies therefore suffice for the sensitivity matrix to have
full rank; the whole band adds precision, by averaging the measurement
noise over many more points. The identifiability of $X$ from the data is
the question of Appendix~\ref{app:hessian}. The singular value decomposition $J=U\Sigma V^{T}$ of the
unnormalized Jacobian over the 716 points of the band,
Table~\ref{tab:svdident}, gives the directions in the space of the
logarithms of the parameters. The strongest is dominated by $C_x$, the second
by $\kappa$, and the weakest by $R_{p0}$, in agreement with the
$(R_{p0},\kappa)$ direction on which the stability term acts
(Section~\ref{sec:invariant}).

\subsection{The Minimized Functional}
\label{app:hessian}
The Jacobian describes how $K_f$ depends on the parameters; the parameters
themselves are determined by the minimization of \eqref{eq:Ethree}. The
conditioning of that minimum is examined through the Hessian of $E$ at the
solution, over the five unknowns of Section~\ref{sec:minimization} -- the
three parameters $X$ and the auxiliary variables $\alpha$, $\beta$ -- with
the weights of the run, $w_1=2.5\times10^{-4}$, $w_2=0.74$, $w_3=1$; a different
$w_2$ rescales the contribution of the curvature term and leaves the
directions that each term determines unchanged. The Hessian is taken in logarithmic coordinates $\xi_k=\ln(X_k/X_{k,0})$, so that the entries are
curvatures with respect to relative changes and are comparable between
variables. The Hessian is evaluated by centered finite differences. The
point is stationary: the displacement that would cancel the gradient along
each coordinate is below $5\times10^{-4}$ in relative terms.

\begin{table}[!t]
\caption{Components of the logarithmic Jacobian \eqref{eq:jident} of
specimen~4 at the three selected frequencies, multiplied by $10^4$.}
\label{tab:ident}
\centering
\begin{tabular}{c r r r}
\hline
$f$ (GHz) & $C_x$ & $R_{p0}$ & $\kappa$\\
\hline
18.009 & 1251.8 & $-303.2$ & 361.7\\
22.284 & 1355.9 & $-480.3$ & 736.2\\
26.464 & 1503.3 & $-599.7$ & 1161.4\\
\hline
\end{tabular}
\end{table}

\begin{table}[!t]
\caption{Singular values of the logarithmic Jacobian of specimen~4 over the
whole band and composition of the corresponding directions.}
\label{tab:svdident}
\centering
\begin{tabular}{c l l}
\hline
$k$ & $\sigma_k$ & Composition\\
\hline
1 & 4.37 & $C_x$ 69\%, $\kappa$ 22\%, $R_{p0}$ 9\%\\
2 & 0.461 & $\kappa$ 69\%, $C_x$ 28\%, $R_{p0}$ 3\%\\
3 & 0.0309 & $R_{p0}$ 89\%, $\kappa$ 9\%, $C_x$ 3\%\\
\hline
\end{tabular}
\end{table}

\begin{table}[!t]
\caption{Sign of the eigenvalues of the Hessian of each term of
\eqref{eq:Ethree} taken separately, and of the complete functional, at the
solution of specimen~4. An eigenvalue is counted as null when its magnitude
is below $10^{-7}$ of the largest eigenvalue of $E$, i.e. below
$10^{-6}$. The last column gives the directions on which the term acts.}
\label{tab:hsigns}
\centering
\small
\setlength{\tabcolsep}{3pt}
\begin{tabular}{l c c c l}
\hline
Term & neg. & null & pos. & Acts on\\
\hline
Ratio constancy, $\epsilon$ & 0 & 3 & 2 & $\alpha$, $\beta$ only\\
Curvature, $\psi^2$ & 1 & 2 & 2 & one dir.\ in $(\kappa,R_{p0})$\\
Invariant, $\Lambda^2$ & 1 & 2 & 2 & conditioning: $C_x$, $(R_{p0},\kappa)$\\
\hline
Complete functional \eqref{eq:Ethree} & 0 & 0 & 5 & \\
\hline
\end{tabular}
\end{table}

\begin{table}[!t]
\caption{Eigenvalues of the Hessian of \eqref{eq:Ethree} at the solution of
specimen~4, in logarithmic coordinates, and composition of the corresponding
directions.}
\label{tab:heig}
\centering
\normalsize
\begin{tabular}{c l}
\hline
$\lambda$ & Composition\\
\hline
$9.34$ & $\kappa$ 67\%, $R_{p0}$ 30\%, $C_x$ 3\%\\
$9.2\times10^{-4}$ & $\beta$ 84\%, $\alpha$ 16\%\\
$8.5\times10^{-4}$ & $\alpha$ 83\%, $\beta$ 16\%\\
$7.5\times10^{-5}$ & $C_x$ 95\%, $\kappa$ 4\%\\
$1.9\times10^{-6}$ & $R_{p0}$ 70\%, $\kappa$ 29\%\\
\hline
\end{tabular}
\end{table}

Table~\ref{tab:hsigns} gives the sign count of the eigenvalues of each
term taken separately, with the weights of the other two set to zero; an
eigenvalue below $10^{-7}$ of the largest eigenvalue of $E$ in magnitude is
counted as null. The three
terms act on different variables. The ratio-constancy term has two
eigenvalues of order $10^{-3}$ whose directions lie entirely in the
$(\alpha,\beta)$ plane, with no component on $X$, and three eigenvalues below
$10^{-6}$ in magnitude: it determines the auxiliary variables and nothing
else, and conversely $\alpha$ and $\beta$ are determined by nothing else --
the other two terms have exactly null curvature in $\alpha$ and $\beta$. The
curvature term has a single large eigenvalue, $9.34$, along a combination of
$\kappa$ and $R_{p0}$, and curvatures of order $10^{-6}$ elsewhere: $2\times10^{-6}$ along $C_x$
and $-9\times10^{-6}$ along the direction in which $R_{p0}$ and $\kappa$
increase together. The invariant term has eigenvalues of order $10^{-4}$:
$1.1\times10^{-4}$ along a direction dominated by $C_x$,
$1.2\times10^{-5}$ along the $(R_{p0},\kappa)$ direction just mentioned, and
$-1.4\times10^{-4}$ along a direction with components on the three
parameters, which has a projection on the stiff direction of $\psi^2$ and is
dominated by it. No term is of full rank and two of them are indefinite
taken alone; the complete functional,
Table~\ref{tab:heig}, is positive definite. Its five eigenvalues are the
stiff $(\kappa,R_{p0})$ direction from $\psi^2$, the two directions of
$\alpha$ and $\beta$ from $\epsilon$, $C_x$, whose curvature
$7.5\times10^{-5}$ comes almost entirely from $\Lambda^2$, and the
$(R_{p0},\kappa)$ direction. This is the role of the stability term:
without it the functional has no proper minimum along $C_x$ and
$(R_{p0},\kappa)$, where $\psi^2$ alone provides curvatures of order
$10^{-6}$ and $10^{-5}$; with it the minimum is a strict local minimum,
reproduced in the searches of Section~\ref{sec:minimization}; global
uniqueness is not claimed. The calibration
factor changes by less than $10^{-3}\%$ along those two directions
(Table~\ref{tab:svdident}), so the term determines the parameters and not
the result
(Section~\ref{sec:invariant}).

The smallest eigenvalue, $1.9\times10^{-6}$, is positive; its direction
is the $(R_{p0},\kappa)$ combination on which the invariant term supplies
the curvature. That direction is
also the one of the smallest singular value of the Jacobian,
Table~\ref{tab:svdident}, so it is the one to which $K_f$ is least
sensitive: a displacement of $10\%$ in $R_{p0}$ along it (with $\kappa$
from 0.660 to 0.697) changes $K_f$ by $0.06\%$ rms over the band, an order
of magnitude below its expanded uncertainty; the uncertainty budget, which
repeats the minimization in every replica, carries whatever part of the
dispersion along it reaches $K_f$. The same ordering was found in the
coaxial case: the direction least determined by the data is the one on
which $K_f$ depends least.

Compared with the Jacobian, the functional conditions the determination of
$\alpha$, $\beta$ and the stiff $(\kappa,R_{p0})$ direction far better than
$K_f$ would, and $C_x$ comparably; along the $(R_{p0},\kappa)$ direction
of lowest curvature it conditions the parameters to the same degree as
$K_f$ itself.

The same picture is given by the dispersion of the parameters over the
400 replicas of the budget of Section~\ref{sec:budget}, in which
$\{C_x,R_{p0},\kappa\}$ are re-determined in every replica. For
specimen~4 the standard uncertainties are $2.9\%$ in $C_x$, $10.8\%$ in
$R_{p0}$ and $6.4\%$ in $\kappa$, with correlation coefficients $+0.91$
between $R_{p0}$ and $\kappa$, $-0.34$ between $C_x$ and $R_{p0}$ and
$-0.24$ between $C_x$ and $\kappa$. The strong positive correlation of
$R_{p0}$ and $\kappa$ is the $(R_{p0},\kappa)$ direction of
Table~\ref{tab:heig}: the replicas move the two parameters together along
it, which is the direction to which $K_f$ is least sensitive, and this is
why parameter uncertainties of several per cent coexist with a model
channel $u_F$ of a few tenths of a per cent in the calibration factor.

\section{Sensitivity of the Result to the Pair-Closure Approximation}
\label{app:sens}
The reference impedance $Z_0$ obtained in Section~\ref{sec:zo} enters
the result only through the relative curvature $C_{vL}$ of \eqref{eq:cvl};
the calibration factor and the efficiencies follow from \eqref{eq:etameas}
and \eqref{eq:CiZ}, which are independent of $Z_0$. For a lossless matching
structure with real output impedance the convention $k_{i,j}=1$ is exact
(Section~\ref{sec:convention}); for the actual, dissipative structure its
effect is bounded by two properties.

First, the band-averaged $\langle\mathrm{Re}[C_{vL}]\rangle$ is exactly
invariant under a common real scaling $Z\!\to\!aZ$ of $Z_{th,i}$ and $Z_0$,
because the factor $a$ cancels identically in every
$\Gamma_{Li}=(Z_{th,i}-Z_0)/(Z_{th,i}+Z_0)$. The scale ambiguity of $Z_0$
left by the neutral transformation \eqref{eq:neutral} therefore does not
propagate to the result.

Second, the displacement of $Z_0$ produced by a departure of $k_{i,j}$ from
unity is itself of first order but confined to the scale and resistive
directions. Perturbing the pair condition \eqref{eq:zocond} about $k_{i,j}=1$
and differentiating the quadratic \eqref{eq:zoquad} gives, to first order,
\begin{equation}
\frac{\Delta Z_{0,i2}}{Z_{0,i2}} = \mathcal{S}_{i2}\,(k_{i,2}-1),
\label{eq:dzo}
\end{equation}
with the closed-form sensitivity
\begin{equation}
\mathcal{S}_{i2} = -\frac{\phi_i\left[\,2Z_{0,i2}(r_i-r_2) + |Z_{th,i}|^2 - |Z_{th,2}|^2\,\right]}
{2Z_{0,i2}(\phi_i-1)\left[\,Z_{0,i2}(\phi_i-1) + \phi_i r_i - r_2\,\right]},
\label{eq:Ssens}
\end{equation}
where $\phi_i$ is defined in \eqref{eq:phi}. The denominator carries the
factor $(\phi_i-1)$ and the derivative of the quadratic \eqref{eq:zoquad}
with respect to $Z_0$; $\mathcal{S}_{i2}$ is therefore largest when the two outer states become
indistinguishable ($\phi_i\!\to\!1$) or near the double-root condition, and
is otherwise moderate.

The size of the effect on a real mount is obtained by evaluating
\eqref{eq:dzo} with the values of $k_{i,j}$ that the converged solution
yields when each pair is evaluated at the common $Z_0=(Z_{0,12}+Z_{0,32})/2$.
For specimen~4 the sensitivities of \eqref{eq:Ssens} are
$\mathcal{S}_{1,2}\approx+3.5$ and $\mathcal{S}_{3,2}\approx-6.5$ across the
band, and the departures $k_{i,j}-1$ reach $0.9\%$ in magnitude. The sensitivities
vary by less than $10\%$ across the band; the departures $k_{i,j}-1$ do
not, and the displacements follow them. The two pair
impedances $Z_{0,12}$ and $Z_{0,32}$ therefore differ from their mean, in
opposite directions, by about $2.4\%$ rms and up to $3.2\%$
(Table~\ref{tab:pairs}: $4.9\%$ rms and $6.3\%$ between the two pairs).
This spread is the closure defect: the margin within which the three states
fix the thermistor-port impedance once $k_{i,j}=1$ is imposed. It is of the same
sign in the five mounts and ranges from $2.9\%$ rms between pairs
(specimen~5) to $10.0\%$ (specimen~2). The departure of $k_{i,j}$ from
unity reaches $0.9\%$ and the sensitivity is of order five; the
admissibility of the convention rests on the result that follows: the two displacements are opposite and nearly equal, so that
the thermistor-port impedance adopted by the method, their mean, is left almost
unchanged.

The change this leaves in the calibration factor is obtained, as in
\cite{greco_coaxial} (Appendix~A), from the perturbation that the
convention produces in the curvature condition. When the computed
$k_{i,j}$ are imposed in place of unity, each pair impedance moves onto
the common value by \eqref{eq:dzo} -- $Z_{0,12}$ upward by $2.4\%$ and
$Z_{0,32}$ downward by $2.3\%$ for specimen~4 -- and their mean moves by
$0.03\%$ rms and $0.05\%$ at most. The opposition holds at every
frequency, not only in band mean, because $k_{3,2}-1\approx\tfrac12(k_{1,2}-1)$
and $\mathcal S_{3,2}\approx-2\mathcal S_{1,2}$ point by point
(Section~\ref{sec:passivity}); for specimen~2, whose closure defect is the
largest, the displacement of the mean grows from $0.03\%$ at 18~GHz to
$0.5\%$ at the upper edge, $0.16\%$ rms. Evaluated with that displacement and the
model impedances of the solution, the band mean of $\mathrm{Re}[C_{vL}]$
changes by $\Delta\langle\psi\rangle=2.3\times10^{-3}$ for specimen~4,
against $0.16$ and $0.21$ for the displacements of the two pairs taken
separately: the two shifts cancel in the condition to one part in seventy.
The minimization restores $\langle\psi\rangle=0$ by displacing the
parameters along the gradient of the condition, which is the stiff
direction of the Hessian of Appendix~\ref{app:ident} (the curvature term
is of rank one along $\nabla\langle\psi\rangle$), so that the first-order
response $-\mathbf H^{-1}\nabla\Delta E$ reduces to
$\delta\xi=\Delta\langle\psi\rangle\,\nabla\langle\psi\rangle/|\nabla\langle\psi\rangle|^2$
in the logarithmic coordinates of that appendix; with
$|\nabla\langle\psi\rangle|=2.5$ the displacement is $9\times10^{-4}$, and
the calibration factor, whose logarithmic sensitivity along that direction
is $0.07$, changes by $0.006\%$ rms over the band and $0.009\%$ at most.
The same evaluation gives $0.017\%$, $0.014\%$ and $0.003\%$ rms for
specimens 1, 2 and 5, and $0.20\%$ for specimen~3, where the condition is
least sensitive to the parameters ($|\nabla\langle\psi\rangle|=0.55$
against $1.9$--$3.0$ for the others). In every case the effect of the
convention is well below the expanded uncertainty of the specimen
(Section~\ref{sec:budget}). Adopting $k_{i,j}=1$ therefore does not affect
the reported calibration factor within its stated uncertainty.

\section{The Schwarz--Pick Factor Equals the Efficiency}
\label{app:schwarzpick}
The hyperbolic invariant of Section~\ref{sec:invariant} rests on the
contraction of the pseudo-hyperbolic distance under the bilinear map of the
matching two-port. That contraction has an exact physical identity, stated
here: for a reciprocal two-port, the Schwarz--Pick contraction factor of the
map is the efficiency.

The map from the thermistor-port to the input reflection coefficient is
\begin{equation}
\Gamma = T(\Gamma_L) = S_{11} + \frac{S_{21}^2\,\Gamma_L}{1-S_{22}\Gamma_L},
\label{eq:Tmap}
\end{equation}
where reciprocity, $S_{12}=S_{21}$, has been used. Its derivative is
\begin{equation}
T'(\Gamma_L) = \frac{S_{21}^2}{(1-S_{22}\Gamma_L)^2}.
\label{eq:Tprime}
\end{equation}
The Schwarz--Pick factor, the local contraction of the hyperbolic metric at a
point $\Gamma_L$ of the disk, is
\begin{equation}
\lambda(\Gamma_L) = |T'(\Gamma_L)|\,
\frac{1-|\Gamma_L|^2}{1-|T(\Gamma_L)|^2}.
\label{eq:lambdadef}
\end{equation}
The lemma of Schwarz and Pick states that $\lambda\le1$ for any holomorphic
self-map of the disk, with equality only for an automorphism \cite{garnett,
ahlfors}. Substituting \eqref{eq:Tprime} and expanding the denominator
$1-|T(\Gamma_L)|^2$ gives, after reduction,
\begin{equation}
\lambda(\Gamma_L) =
\frac{|S_{21}|^2\,(1-|\Gamma_L|^2)}
{|1-S_{22}\Gamma_L|^2 - |(S_{21}^2-S_{11}S_{22})\Gamma_L + S_{11}|^2},
\label{eq:lambdaeta}
\end{equation}
which is exactly the efficiency $\eta$ of \eqref{eq:eta}. The identity is
algebraic and holds for any reciprocal two-port, lossy or lossless.

The two limits are physically transparent. For a lossless matching structure
the map is a disk automorphism, Schwarz--Pick holds with equality, and
$\lambda=\eta=1$: all the power reaches the thermistor. For a lossy structure
the map strictly contracts, $\lambda=\eta<1$, and the deficit is the
dissipation of the matching bar and pivot. The efficiency, a ratio of powers,
is thus identical to the contraction factor of a geometric metric; this is why
the pseudo-hyperbolic distance is the natural invariant for the problem, and
why $Q_h$ of \eqref{eq:qh} is conserved to the extent that the structure is
lossless.

The bilinear (M\"obius) form of the load-to-input reflection map
\eqref{eq:Tmap} is standard in network theory \cite{orfanidi}, and the
Schwarz--Pick lemma is classical in geometric function theory
\cite{garnett,ahlfors}. The identity established here -- that the Schwarz--Pick
contraction factor of this map coincides exactly with the efficiency of the
reciprocal two-port -- connects the two; it follows in a few lines from
\eqref{eq:eta} and \eqref{eq:Tprime}, and we state it here because it is
what the method uses. Its usefulness here is concrete: it fixes the
scaling of the hyperbolic invariant of Section~\ref{sec:invariant}, whose
measured value is contracted exactly by the efficiency of the mount, a quantity
the method already determines.

\section{Fixed-Network Test}
\label{app:fixednet}
Section~\ref{sec:passivity} states that the state correction cannot be
replaced by a different fixed network for the bead. The statement is
tested here on specimens~4 and~5.

A linear reciprocal two-port terminated in $R_b$ has, in its T form with
elements $Z_a$, $Z_b$, $Z_c$ (series, series and shunt)
\cite{dicke1948}, the input impedance
\begin{equation}
\begin{split}
Z'(R_b)&=Z_a+\frac{Z_c\,(Z_b+R_b)}{Z_b+Z_c+R_b}=A+\frac{B}{R_b+P},\\
A&=Z_a+Z_c,\qquad P=Z_b+Z_c,\qquad B=-Z_c^{2},
\end{split}
\label{eq:tnet}
\end{equation}
a bilinear function of $R_b$ with three complex coefficients
\cite{dicke1948}. Three pairs
$(R_{b,i},Z'_i)$ determine $A$, $P$ and $B$ by a linear system, and
$Z_c=\pm\sqrt{-B}$, $Z_a=A-Z_c$, $Z_b=P-Z_c$: every triple of impedances
is the input impedance of exactly one such two-port, up to the sign of
$Z_c$, which does not affect what follows. The two-port is passive if and
only if the Hermitian part of its impedance matrix is positive
semidefinite \cite{youla1959,belevitch1968},
\begin{equation}
\mathbf H=\tfrac12\left(\mathbf Z_T+\mathbf Z_T^{H}\right)\succeq0,\qquad
\mathbf Z_T=\begin{pmatrix} Z_a+Z_c & Z_c\\ Z_c & Z_b+Z_c\end{pmatrix},
\label{eq:passT}
\end{equation}
and $\lambda_{\min}(\mathbf H)$, in ohm, is its passivity margin: negative, the
network must deliver power for some excitation. The criterion is
invariant under the common scale and reactance of \eqref{eq:neutral},
which multiply $\mathbf H$ by a positive number or leave it unchanged, and
it is necessary for a network of fixed elements at every frequency, so
that a band on which $\lambda_{\min}<0$ excludes such a network over the
whole band. A non-reciprocal two-port, which requires a gyrator
\cite{tellegen1948}, is excluded by the absence of anisotropic material
in the bead, as in Section~\ref{sec:method}.

\begin{figure}[!t]
\centering
\includegraphics[width=0.62\textwidth]{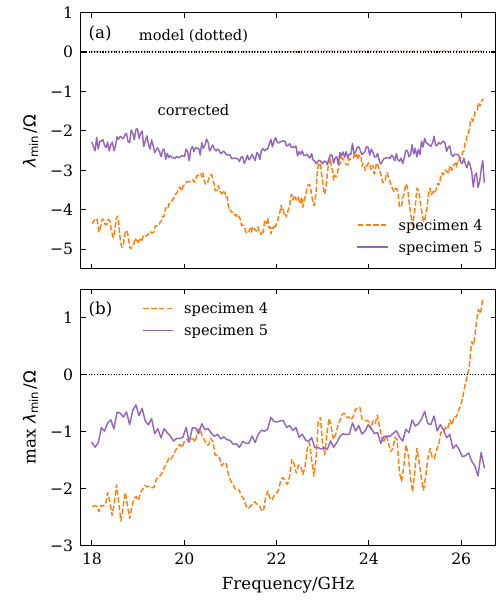}
\caption{Fixed-network test, specimens~4 and~5, at every second (a) and
every fourth (b) point of the analyzer grid. (a) Passivity margin
$\lambda_{\min}$ of the bead two-port \eqref{eq:tnet} for the impedances
of the model (dotted, $+0.01$ to $+0.02$~\capm{\Omega}) and for the corrected
impedances $m_iZ_{th,i}$ (solid and dashed). (b) Largest margin over all triples with
$K_f$ fixed and a passive mount, \eqref{eq:fixsearch}, $x_{\max}=0.3$.}
\label{fig:fixednet}
\end{figure}

Applied to the impedances of the model of Section~\ref{sec:circuit}, the
extraction returns the bead network itself -- $Z_b$ the skin reactance of
the $R_b$ branch, $Z_c$ the series branch, $Z_a$ within $0.1~\Omega$ of
zero, the residue being the second-order skin resistance
$(L_x\mu_0f)^2/48R_b$ of that branch, $0.004$--$0.024~\Omega$, the only
term of the model that is not bilinear in $R_b$ -- with
$\lambda_{\min}=+0.010$ to $+0.022~\Omega$ (Table~\ref{tab:fixednet},
Fig.~\ref{fig:fixednet}(a)). Applied to the corrected impedances
$m_iZ_{th,i}$ of \eqref{eq:sp2}, it returns a network with
$\mathrm{Re}\,Z_a=-6$ to $-25~\Omega$ (specimen~5) and $-17$ to
$-58~\Omega$ (specimen~4) and $\lambda_{\min}=-1.9$ to $-3.4~\Omega$ and
$-1.2$ to $-5.0~\Omega$ at every point of the band.

Whether some other triple admitted by the data is the image of $R_b$ under
a passive fixed network was examined by maximizing the margin over the
five coordinates of $\mathcal M_F$, with the mount required to be passive
but not at the floor:
\begin{equation}
\begin{gathered}
\max_{x\in\mathbb R^6}\ \lambda_{\min}\!\left[\mathbf H\big(Z'(x)\big)\right]\\
\text{subject to}\quad F\big(Z'(x)\big)=F(Z_{th}),\\
\sigma_{\max}\!\left[\mathbf S\big(Z'(x)\big)\right]\le1,\qquad |x_k|\le x_{\max},
\end{gathered}
\label{eq:fixsearch}
\end{equation}
with $Z'_i=Z_{th,i}\,e^{x_i+jx_{3+i}}$ and $\mathbf S$ obtained from the
three pairs $(\Gamma_i,\Gamma_{Li})$ at $Z_0$ recomputed from the pair
closure, as in \eqref{eq:sp2}. The problem was solved by sequential
quadratic programming at every fourth point of the analyzer grid, from
the model, from the minimum-norm solution with $\sigma_{\max}\le1$ and
from random starts, with $x_{\max}=0.3$ ($\pm35\%$ in modulus, $\pm17^\circ$
in phase). The maximum is negative at every point of 18--26.5~GHz for
specimen~5 ($-0.5$ to $-1.8~\Omega$) and up to 26.1~GHz for specimen~4
($-2.6$ to $-0.03~\Omega$), the search running to the edge of the box in
every case ($\|m-1\|=58$--$69\%$), and the sign does not change with
$x_{\max}=0.5$ and twelve random starts at 18--22~GHz. Above 26.1~GHz
specimen~4 admits solutions passive on both sides at the edge of the box,
with changes of $58\%$ in the impedances. A network of fixed elements must
be passive at every frequency; none was found, within the domains and
searches examined, that reproduces the three states over the band
together with a passive mount.

\begin{table}[!t]
\caption{Fixed-network test at selected frequencies. Model: margin of the
bead two-port extracted from the impedances of Section~\ref{sec:circuit}.
Corrected: elements and margin of the two-port extracted from
$m_iZ_{th,i}$ ($Z_c>0$ branch). Search: largest margin over the admissible
triples, \eqref{eq:fixsearch}, and norm of the correction at the optimum.}
\label{tab:fixednet}
\centering
\normalsize
\setlength{\tabcolsep}{5pt}
\begin{tabular}{c c c | c c c c | c c}
\hline
 & & Model & \multicolumn{4}{c|}{Corrected} & \multicolumn{2}{c}{Search}\\
$f$ (GHz) & Spec. & $\lambda_{\min}$ ($\Omega$) & $\mathrm{Re}\,Z_a$ ($\Omega$) & $\mathrm{Re}\,Z_b$ ($\Omega$) & $Z_c$ ($\Omega$) & $\lambda_{\min}$ ($\Omega$) & $\max\lambda_{\min}$ ($\Omega$) & $\|m-1\|$ (\%)\\
\hline
18.0 & 5 & $+0.010$ & $-19.2$ & $+14.8$ & $1275-42j$ & $-2.30$ & $-1.19$ & 68\\
20.0 & 5 & $+0.012$ & $-15.6$ & $+10.4$ & $1204+38j$ & $-2.65$ & $-1.10$ & 65\\
22.0 & 5 & $+0.015$ & $-20.1$ & $+15.8$ & $1172+66j$ & $-2.29$ & $-0.83$ & 65\\
24.0 & 5 & $+0.018$ & $-18.8$ & $+14.2$ & $1119+107j$ & $-2.42$ & $-0.88$ & 64\\
26.5 & 5 & $+0.022$ & $-8.1$ & $+1.5$ & $1010+150j$ & $-3.30$ & $-1.64$ & 63\\
\hline
18.0 & 4 & $+0.010$ & $-25.9$ & $+17.7$ & $1022-192j$ & $-4.36$ & $-2.32$ & 69\\
20.0 & 4 & $+0.012$ & $-42.2$ & $+37.0$ & $1008-192j$ & $-3.37$ & $-1.16$ & 69\\
22.0 & 4 & $+0.015$ & $-39.8$ & $+32.5$ & $950-111j$ & $-4.33$ & $-2.08$ & 67\\
24.0 & 4 & $+0.018$ & $-47.2$ & $+44.0$ & $880-163j$ & $-2.81$ & $-0.82$ & 69\\
26.5 & 4 & $+0.022$ & $-56.4$ & $+58.0$ & $810-205j$ & $-1.21$ & $+1.32$ & 58\\
\hline
\end{tabular}
\end{table}

\section{Dependence of the Efficiency on the Bias Resistance}
\label{app:rb}
The calibration factor depends on the bias resistance $R_b$ at which the
mount is operated, the dc resistance that balances the bridge of the power
meter, through the mismatch factor and, possibly, through the efficiency.
This value is seldom reported with a calibration; a nominal figure is
given, as a formality of compliance with the sensor specification. The
sensitivity of the efficiency to $R_b$ is derived here in closed form from
the three-state measurement, in a way that does not involve the thermistor-port
impedance $Z_0$, and evaluated with the thermistor impedances corrected
as in Section~\ref{sec:passivity}.
The derivation uses the hyperbolic geometry of Section~\ref{sec:invariant},
which is not common in this context, so it is given step by step.

\subsection{Hyperbolic Speed at the Sensor Port}
As the bridge resistance is varied, the measured reflection coefficient
describes a curve $\Gamma(R_b)$ in the unit disk. Its speed in the Poincar\'e
metric of the disk, whose distance is $2\,\mathrm{artanh}\,\rho$ with
$\rho$ the pseudo-hyperbolic distance \eqref{eq:pseudohyp}, is
\begin{equation}
v(R_b)=\lambda_P(\Gamma)\,|\Gamma'|,\qquad
\lambda_P(\Gamma)=\frac{2}{1-|\Gamma|^2},
\label{eq:hspeed}
\end{equation}
where the prime denotes the derivative with respect to $R_b$ and $\lambda_P$
is the density of the metric. We take the logarithmic derivative of the speed,
\begin{equation}
L(R_b)=\frac{d\ln v}{dR_b},\qquad
\ln v=\ln2-\ln\!\big(1-|\Gamma|^2\big)+\ln|\Gamma'| .
\label{eq:hL}
\end{equation}
The two terms are differentiated separately. For the first, with
$|\Gamma|^2=\Gamma\Gamma^{*}$,
\begin{equation}
\frac{d|\Gamma|^2}{dR_b}=\Gamma'\Gamma^{*}+\Gamma\Gamma'^{*}
=2\,\mathrm{Re}[\Gamma^{*}\Gamma'],
\end{equation}
so that
\begin{equation}
-\frac{d}{dR_b}\ln\!\big(1-|\Gamma|^2\big)
=\frac{2\,\mathrm{Re}[\Gamma^{*}\Gamma']}{1-|\Gamma|^2}.
\end{equation}
For the second, $\ln|\Gamma'|=\mathrm{Re}\ln\Gamma'$, whence
\begin{equation}
\frac{d\ln|\Gamma'|}{dR_b}=\mathrm{Re}\!\left[\frac{\Gamma''}{\Gamma'}\right].
\end{equation}
Collecting the two,
\begin{equation}
L(R_b)=\mathrm{Re}\!\left[\frac{\Gamma''}{\Gamma'}\right]
+\frac{2\,\mathrm{Re}[\Gamma^{*}\Gamma']}{1-|\Gamma|^2}.
\label{eq:hLfinal}
\end{equation}

\subsection{Contraction by the Mount}
The curve $\Gamma(R_b)$ is the image of the curve $\Gamma_L(R_b)$ at the
thermistor port under the bilinear map \eqref{eq:gamma} of the mount. By
Appendix~\ref{app:schwarzpick}, that map contracts the Poincar\'e metric at each point
by the Schwarz--Pick factor, which equals the efficiency; the speeds are
therefore related by
\begin{equation}
v(R_b)=\eta(R_b)\,v_{th}(R_b),
\label{eq:hcontr}
\end{equation}
with $v_{th}$ the speed of $\Gamma_L(R_b)$ in the same metric. Taking the
logarithm and differentiating,
\begin{equation}
L(R_b)-L_{th}(R_b)=\frac{d\ln\eta}{dR_b},
\label{eq:hdiff}
\end{equation}
where $L_{th}$ is the logarithmic derivative of $v_{th}$. The sensitivity of
the efficiency to the bias resistance is thus the difference between the
logarithmic rate of the measured curve and that of the thermistor curve.

\subsection{Speed at the Thermistor Port Without $Z_0$}
The speed $v_{th}=\lambda_P(\Gamma_L)|\Gamma_L'|$ can be written without
the reflection coefficient. With $\Gamma_L=(Z_{th}-Z_0)/(Z_{th}+Z_0)$ and
$Z_0$ real,
\begin{equation}
1-|\Gamma_L|^2=\frac{4Z_0\,\mathrm{Re}\,Z_{th}}{|Z_{th}+Z_0|^2},\qquad
\Gamma_L'=\frac{2Z_0\,Z_{th}'}{(Z_{th}+Z_0)^2},
\end{equation}
and the factor $|Z_{th}+Z_0|^2$ and the reference cancel in the quotient,
\begin{equation}
v_{th}=\frac{2\,|\Gamma_L'|}{1-|\Gamma_L|^2}=\frac{|Z_{th}'|}{\mathrm{Re}\,Z_{th}} .
\label{eq:hvth}
\end{equation}
This is the Poincar\'e metric of the right half-plane, $|dZ|/\mathrm{Re}\,Z$,
onto which the Smith map is an isometry for every real $Z_0$. The speed of the
thermistor, and with it the derivative of the efficiency, is therefore
independent of the reference frame; it is also invariant under the neutral
transformation \eqref{eq:neutral}, since $|dZ|/\mathrm{Re}\,Z$ is unchanged by
$Z\to aZ+jb$. Differentiating the logarithm of \eqref{eq:hvth} as before,
\begin{equation}
L_{th}(R_b)=\mathrm{Re}\!\left[\frac{Z_{th}''}{Z_{th}'}\right]
-\frac{\mathrm{Re}[Z_{th}']}{\mathrm{Re}[Z_{th}]},
\label{eq:hLth}
\end{equation}
the exact counterpart of \eqref{eq:hLfinal}: the same first term, and in the
second the denominator $1-|\Gamma|^2$ replaced by $\mathrm{Re}\,Z_{th}$.

\subsection{Three-State Form}
The derivatives are taken along the bilinear continuation of the three
states. For the circuit of Section~\ref{sec:circuit} both curves are
bilinear in $R_b$: $Z_{th}(R_b)$ because $R_b$ is the single element that
varies with the state (up to the second-order skin resistance of
\eqref{eq:skinseries}, $0.004$--$0.024~\Omega$, Appendix~\ref{app:fixednet}),
and $\Gamma(R_b)$ because the map of the mount is bilinear; three states
then determine each curve exactly, and the derivatives at the central
state are those of the bilinear interpolant, with no truncation error for
curves of that form. With the corrected impedances of
Section~\ref{sec:passivity} the load curve is $m(R_b)Z_{th}(R_b)$, of
which only the three states are known, and the same interpolant through
the three corrected points is the continuation adopted; the sensitivity
so obtained is a local prediction of that representation
(Appendix~\ref{app:rbvalues}). With the states
$R_{b,i}=\{R_0-\delta,R_0,R_0+\delta\}$ of \eqref{eq:rbsym}, writing the
interpolant as $(Aw+\Gamma_2)/(Cw+1)$ in $w=(R_b-R_0)/\delta$ and
imposing the values at $w=\pm1$,
\begin{equation}
\frac{\Gamma''}{\Gamma'}=\frac{2}{\delta}\,
\frac{\Gamma_1+\Gamma_3-2\Gamma_2}{\Gamma_3-\Gamma_1},\qquad
\Gamma'=\frac{2}{\delta}\,
\frac{(\Gamma_3-\Gamma_2)(\Gamma_2-\Gamma_1)}{\Gamma_3-\Gamma_1},
\label{eq:hinterp}
\end{equation}
at $R_0$, and likewise for $Z_{th}$. Substituting in \eqref{eq:hLfinal},
\eqref{eq:hLth} and \eqref{eq:hdiff},
\begin{equation}
\frac{d\ln\eta}{dR_b}=
\frac{2}{\delta}\,\mathrm{Re}\!\left[
\frac{\Gamma_1+\Gamma_3-2\Gamma_2}{\Gamma_3-\Gamma_1}
-\frac{Z_{th,1}+Z_{th,3}-2Z_{th,2}}{Z_{th,3}-Z_{th,1}}\right]
+\frac{2\,\mathrm{Re}[\Gamma_2^{*}\Gamma']}{1-|\Gamma_2|^2}
+\frac{\mathrm{Re}[Z_{th}']}{\mathrm{Re}[Z_{th,2}]} ,
\label{eq:hthree}
\end{equation}
with $\Gamma'$ and $Z_{th}'$ the derivatives \eqref{eq:hinterp} at the
central state, in terms of the three measured coefficients and the three
thermistor impedances of the converged solution alone. The first two terms are the
relative curvatures of Section~\ref{sec:zo}, $\Delta^2\Gamma$ and its analogue
in $Z_{th}$, normalized by the secant instead of by the central state; the
remaining two come from the metric, with the derivative of the interpolant
in place of the secant. The sensitivity of the efficiency to
$R_b$ is thus contained in the same three-state geometry that determines the
parameters.

\subsection{Values}
\label{app:rbvalues}
Table~\ref{tab:rb} gives the band averages of $L$, $L_{th}$ and their
difference for the five WR-42 specimens, computed from \eqref{eq:hthree}
with $\delta=100~\Omega$ and the corrected thermistor impedances of
Section~\ref{sec:passivity}, together with the mismatch term $M_R$ of \eqref{eq:sm} below, which involves
the measured coefficients only. $L$ and $L_{th}$ are large and nearly
equal, about $-6\times10^{-3}~\Omega^{-1}$: the speed falls with $R_b$
mainly because the curve approaches the center of the disk. Their
difference, the derivative of the efficiency, is within
$\pm0.4\times10^{-4}~\Omega^{-1}$ in the five mounts, zero within its
dispersion over the band in three of them, that is, below $0.004\%$ per ohm;
it is a derivative evaluated at $R_{b,2}$ from
the three states that surround it, to be used only in the neighborhood of
that point. With the corrected impedances the load curve is $m(R_b)Z_{th}(R_b)$,
which is not bilinear in $R_b$, and \eqref{eq:hthree} is then the exact
derivative of $\ln\eta$ along the bilinear interpolant through the three
corrected points, since the interpolant commutes with the bilinear map of
the mount; the sensitivity so obtained is a local prediction of the
representation adopted. Its smallness has a closed form. For the passive structure
\eqref{eq:floor} the efficiency follows \eqref{eq:attlaw}, whose logarithmic
derivative with respect to $|\Gamma_L|$ is $-2|\Gamma_L|(1-\eta_2^2)/[(1-|\Gamma_L|^2)(1-\eta_2^2|\Gamma_L|^2)]$,
proportional to $|\Gamma_{L2}|$ and to $1-\eta_2^2$: the sensitivity
vanishes with the mismatch of the operating state, which the
self-consistent $Z_0$ keeps small, and with the loss of the mount. With the
thermistor impedances of the model of Section~\ref{sec:circuit} alone,
the same expression \eqref{eq:hthree} returns
$2$--$3\times10^{-4}~\Omega^{-1}$, the efficiency spread between states
that Section~\ref{sec:passivity} traces to the representation.

\begin{table}[!t]
\caption{Band averages of the logarithmic rates \eqref{eq:hLfinal} and
\eqref{eq:hLth}, of their difference \eqref{eq:hthree} -- the sensitivity
$S$ of the efficiency to the bias resistance -- with the corrected thermistor
impedances of Section~\ref{sec:passivity}, and of the mismatch term $M_R$ of
\eqref{eq:sm}. The $\pm$ is the standard deviation over the band.}
\label{tab:rb}
\centering
\setlength{\tabcolsep}{3.5pt}
\begin{tabular}{c c c c c}
\hline
Specimen & $L$ & $L_{th}$ & $S=L-L_{th}$ & $M_R$\\
 & ($10^{-3}/\Omega$) & ($10^{-3}/\Omega$) & ($10^{-4}/\Omega$) & ($10^{-4}/\Omega$)\\
\hline
1 & $-5.67$ & $-5.66$ & $-0.08\pm0.13$ & $+1.1\pm1.5$\\
2 & $-5.69$ & $-5.73$ & $+0.37\pm0.20$ & $+6.7\pm1.5$\\
3 & $-5.69$ & $-5.70$ & $+0.06\pm0.10$ & $+3.3\pm1.4$\\
4 & $-5.67$ & $-5.66$ & $-0.05\pm0.10$ & $+1.2\pm1.4$\\
5 & $-5.58$ & $-5.60$ & $+0.13\pm0.04$ & $+5.5\pm1.5$\\
\hline
\end{tabular}
\end{table}

\subsection{Consequences for Direct Power Measurement}
\label{app:power}
In laboratory use the HP~432A is not read from its panel meter. The
standard practice is the dc-substitution computation with two voltmeters:
a nanovoltmeter (34420A) or an 8.5-digit multimeter (3458A) for the
difference
$V_{\mathrm{comp}}-V_{\mathrm{rf}}$ and a second voltmeter for
$V_{\mathrm{comp}}$, the power being \cite{keysight2001}
\begin{equation}
P_{\mathrm{rf}}=\frac{(V_{\mathrm{comp}}-V_{\mathrm{rf}})
(V_{\mathrm{comp}}+V_{\mathrm{rf}}-2V_0)}{4R_b},
\label{eq:p432}
\end{equation}
with $V_0$ the zero reading. The bridge resistance enters this expression
one to one: a departure $\Delta R$ of the resistance actually held by the
bridge from the value used in \eqref{eq:p432} is a scale error
$\Delta R/R_b$ in the measured power, $0.25\%$ for $0.5~\Omega$. The
accuracy that the manufacturer specifies for this mode of use is
$\pm0.2\%\pm0.5~\mu$W. The relative term is essentially the tolerance of
that resistance. The absolute term is the zero setting: $V_0$ in
\eqref{eq:p432} is the reading with the RF off, and its residual and its
drift between the zero setting and the measurement enter as a power
independent of $P_{\mathrm{rf}}$. That term is not part of the analysis
that follows, which concerns the bridge resistance alone; it adds to the
terms of Table~\ref{tab:power} as an absolute contribution, $0.05\%$ at
1~mW.

The calibration factor depends on $R_b$ as well, through the efficiency
and through the match. The certified $K_f=\eta\,(1-|\Gamma_2|^2)$ already
contains the mismatch factor at the $\Gamma_2$ of the calibration; what
changes when the bridge is displaced is the variation of both factors
between the calibration and the actual resistance,
\begin{equation}
\begin{gathered}
\frac{d\ln K_f}{dR_b}=S+M_R,\\
M_R=\frac{d\ln(1-|\Gamma_2|^2)}{dR_b}
\simeq\frac{1}{2\delta}\ln\frac{1-|\Gamma_3|^2}{1-|\Gamma_1|^2},
\end{gathered}
\label{eq:sm}
\end{equation}
where $S=d\ln\eta/dR_b$ is the sensitivity of \eqref{eq:hthree} and $M_R$
follows from the three measured coefficients alone, without any model.
$M_R$ is the larger of the two by an order of magnitude and changes sign
across the band, since $|\Gamma_2|$ passes through a minimum where the
mount is matched; for specimen~4 it is $-4.0$, $+1.1$ and
$+4.7\times10^{-4}~\Omega^{-1}$ at 18, 22 and 26.5~GHz. The error of a power measurement made at a bridge
resistance displaced by $\Delta R$ from that of the calibration, and
computed with the nominal value, is therefore
\begin{equation}
\frac{\Delta P}{P}\simeq\frac{\Delta R}{R_b}+(S+M_R)\,\Delta R .
\label{eq:dpower}
\end{equation}
Table~\ref{tab:power} evaluates it for specimen~4 and $\Delta R=-0.5~\Omega$.
The scale term dominates by an order of magnitude and does not depend on
the frequency; the match term is of the order of hundredths of a per cent
and of variable sign, and the efficiency term is negligible. Together they
amount to a quarter of a per cent, a quarter of the expanded uncertainty
of the calibration itself, for a displacement that no procedure declares.

\begin{table}[!t]
\caption{Error of a direct power measurement with specimen~4 at a bridge
resistance $0.5$~\capm{\Omega} below that of the calibration, computed with the
nominal resistance, from \eqref{eq:dpower}.}
\label{tab:power}
\centering
\normalsize
\begin{tabular}{l r r r}
\hline
Term & 18~GHz & 22~GHz & 26.5~GHz\\
\hline
Scale, $\Delta R/R_b$ & $-0.250\%$ & $-0.250\%$ & $-0.250\%$\\
Efficiency, $S\,\Delta R$ & $<0.001\%$ & $<0.001\%$ & $<0.001\%$\\
Match, $M_R\,\Delta R$ & $+0.020\%$ & $-0.005\%$ & $-0.024\%$\\
\hline
Total & $-0.230\%$ & $-0.255\%$ & $-0.274\%$\\
\hline
\end{tabular}
\end{table}

Two of the three terms are already compensated in a well-equipped
laboratory. The resistance held by the bridge is a dc quantity, measured with an
8.5-digit multimeter to $2.4\times10^{-4}$ in this work
(Sections~\ref{sec:measproc} and~\ref{sec:results});
used in \eqref{eq:p432} in place of the nominal value, it reduces the scale
term to $0.02\%$, and the remaining contribution of the meter is that of
the voltmeters, below $10^{-4}$. The match term is the variation of the mismatch factor between the
$\Gamma_2$ of the calibration and that at the actual resistance; it is
absorbed whenever the reflection coefficient is measured at the operating
resistance of the bridge and the mismatch correction is computed from it,
which is the practice of a calibration laboratory working with
$\eta_{\mathrm{eff}}$, and it remains when the certified $K_f$ is used as a
whole. The efficiency term $S\,\Delta R$ is the only one that would
require the sensitivity of the efficiency to the bias resistance, a
quantity that neither the microcalorimetric calibration nor the power
meter supplies; the three-state measurement provides it directly, and
places it below $10^{-4}~\Omega^{-1}$ in band mean, so that no correction is needed
for it. Both compensations require that the value of $R_b$ at which the
calibration was performed be known, which is the proposal of this
appendix: with $R_b$ accompanying the calibration, the uncertainty of a
direct power measurement reduces to that of the calibration factor.

\section{Flange Connection}
\label{app:flange}
The mount is mated once per three-state set, so an imperfection of the
flange connection -- an offset, a gap or a tilt between the flanges -- is a
small two-port common to the three states, as the connector was in the
coaxial case \cite{greco_coaxial}. To first order it maps the coefficient at
the flange as $\Gamma_m=\Gamma e^{j\phi}+c_f+\rho_f\,(1-\Gamma^2)$: a
reference-plane shift $\phi$, an additive term $c_f$ and an impedance
discontinuity $\rho_f$. The first two cancel identically in $G$, which depends
on differences of the $\Gamma_i$ only; the discontinuity survives, with a
sensitivity of the reflectometer channel
$S_C=|\Delta K_f/K_f|/|\rho_f|$ whose largest value over the band is 0.41 to
0.53 for the five specimens, independent of $|\rho_f|$ between $10^{-4}$ and
$10^{-2}$ and of its phase.

The repeatability of waveguide flange connections with alignment pins,
measured by repeated connection of a stable device, is better than
$4\times10^{-4}$ in reflection at 75--110~GHz \cite{probst2019}; the
mechanical tolerances of the WR-42 flange are the same in absolute terms on
a guide four times larger, so that figure is a conservative bound for
$|\rho_f|$ here. With $|\rho_f|\le4\times10^{-4}$, $S_C\le0.53$, and the three
connections of Section~\ref{sec:measproc} averaged, the contribution to the
calibration factor is below $0.013\%$ ($k=1$), more than an order of
magnitude below the budget of Section~\ref{sec:budget}; this is why it does not appear
in Table~\ref{tab:sources}. The bound is taken through the reflectometer
channel alone; the re-determination of the parameters compensates part of
it, as it does for the other terms common to the three states.
\ack{The author acknowledges the use of Claude (Anthropic) for assistance
with the English translation and drafting of this manuscript, and for
numerical cross-verification of the derivations. All technical
content, methods, measurements, and results are the author's own. The
author declares no conflicts of interest.}

\funding{This work received no external funding.}

\data{The data that support the findings of this study are available
from the author upon reasonable request.}

\end{document}